\documentclass[12pt,a4paper,fleqn,titlepage,onecolumn,twoside,openright,final]{book}
\usepackage{amsmath,amssymb,amsfonts,amsthm,bbm,bm} 
\usepackage[T1]{fontenc}
\usepackage{lmodern}
\usepackage[utf8]{inputenc} 
\usepackage[protrusion=true,expansion=true]{microtype}
\usepackage{textcomp} 

\usepackage{fancyhdr} 

\usepackage[pdftex,bookmarks,colorlinks,breaklinks]{hyperref}  
\hypersetup{linkcolor=red,citecolor=blue,filecolor=dullmagenta,urlcolor=darkblue} 

\usepackage{tikz-cd} 
\usetikzlibrary{matrix,arrows,decorations.pathmorphing}

\usepackage{pdfsync}

\theoremstyle{plain}

\theoremstyle{definition}

\theoremstyle{remark}

\newtheorem*{remark*}{Remark}
\newtheorem*{example*}{Example}

\newcommand{\reals}{\mathbb R}

\renewcommand{\Re}{\operatorname{Re}}
\renewcommand{\Im}{\operatorname{Im}}

\newcommand{\HH}{\mathcal{H}}

\newcommand{\LL}{\mathcal{L}}
\newcommand{\EE}{\mathcal{E}}

\newcommand{\PP}{\mathcal{P}}

\newcommand{\WW}{\mathcal{W}}
\renewcommand{\SS}{\mathcal{S}}

\newcommand{\CC}{\mathcal{C}}

\newcommand{\UU}{\mathrm{U}}
\newcommand{\SU}{\mathrm{SU}}
\newcommand{\uu}{\mathfrak{u}}

\newcommand{\hh}{\mathfrak{h}}
\newcommand{\sh}{\mathfrak{sh}}

\newcommand{\Ad}{\operatorname{Ad}}
\newcommand{\ad}{\operatorname{ad}}
\newcommand{\SO}{\mathrm{SO}}
\newcommand{\OO}{\mathrm{O}}
\newcommand{\GG}{\mathrm{G}}
\renewcommand{\gg}{\mathfrak{g}}

\newcommand{\1}{\mathbbm{1}}

\newcommand{\J}{\mathbf{J}}
\newcommand{\R}{\mathbf{R}}
\renewcommand{\L}{\mathbf{L}}
\newcommand{\II}{\mathcal{I}}
\newcommand{\JJ}{\mathcal{J}}

\renewcommand{\d}{\operatorname{d}\!}

\newcommand{\dd}[1]{\frac{\operatorname{d}}{\operatorname{d}\!#1}}

\newcommand{\dt}{\operatorname{d}\!t}

\newcommand{\T}{\operatorname{T}\!}
\newcommand{\V}{\operatorname{V}\!}
\renewcommand{\H}{\operatorname{H}\!}

\newcommand{\dist}{\operatorname{dist}}
\newcommand{\distB}{\operatorname{dist}_{\textsc{b}}}
\newcommand{\distWY}{\operatorname{dist}_{\textsc{wy}}}

\newcommand{\distFS}[2]{\operatorname{dist}_{\textsc{fs}}(#1,#2)}
\newcommand{\IWY}{\operatorname{I}_{\textsc{wy}}}

\newcommand{\GHS}{G_{\textsc{hs}}}

\newcommand{\gFS}{g_{\textsc{fs}}}
\newcommand{\gF}{g_{\textsc{f}}}
\newcommand{\gB}{g_{\textsc{b}}}

\newcommand{\gWY}{g_{\textsc{wy}}}

\newcommand{\gC}{g_{\textsc{c}}}

\newcommand{\var}{\operatorname{var}}

\newcommand{\A}{\mathcal{A}}
\newcommand{\AB}{\mathcal{A}_{\textsc{b}}}

\newcommand{\Ac}{\mathcal{A}_{\textsc{c}}}
\newcommand{\F}{\mathcal{F}}
\newcommand{\Hol}{\mathfrak{H}}
\newcommand{\p}{\Pi}

\newcommand{\OHS}{\Omega_{\textsc{hs}}}

\newcommand{\ket}[1]{|{#1}\rangle}
\newcommand{\bra}[1]{\langle #1 |}
\newcommand{\braket}[2]{\langle #1|#2\rangle}
\newcommand{\ketbra}[2]{|#1\rangle\langle #2|}

\newcommand{\length}[1]{\operatorname{Length}[#1]}

\newcommand{\Ker}{\operatorname{ker}}
\newcommand{\Tr}{\operatorname{tr}}

\newcommand{\diag}{\operatorname{diag}}

\newcommand{\texp}{\operatorname{T}\!\exp}

\newcommand{\ttot}{\theta_{\text{tot}}}
\newcommand{\tdyn}{\theta_{\text{dyn}}}
\newcommand{\tgeo}{\theta_{\text{geo}}}
\newcommand{\Geop}{\Phi_{\text{geo}}}

\newcommand{\bfp}{\boldsymbol{p}}
\newcommand{\bfP}{\boldsymbol{P}}
\newcommand{\bfm}{\boldsymbol{m}}

\newcommand{\bfL}{\boldsymbol{\Lambda}}
\newcommand{\bfl}{\boldsymbol{\lambda}}

\newcommand{\g}{\mathfrak{g}}
\newcommand{\eps}{\epsilon}
\newcommand{\bfsigma}{\boldsymbol{\sigma}}
\usepackage{xfrac}
\usepackage{cite}
\usepackage{csquotes}
\usepackage[nottoc]{tocbibind}

\begin{document}
\thispagestyle{empty}
\hfill \includegraphics[width=0.2\textwidth]{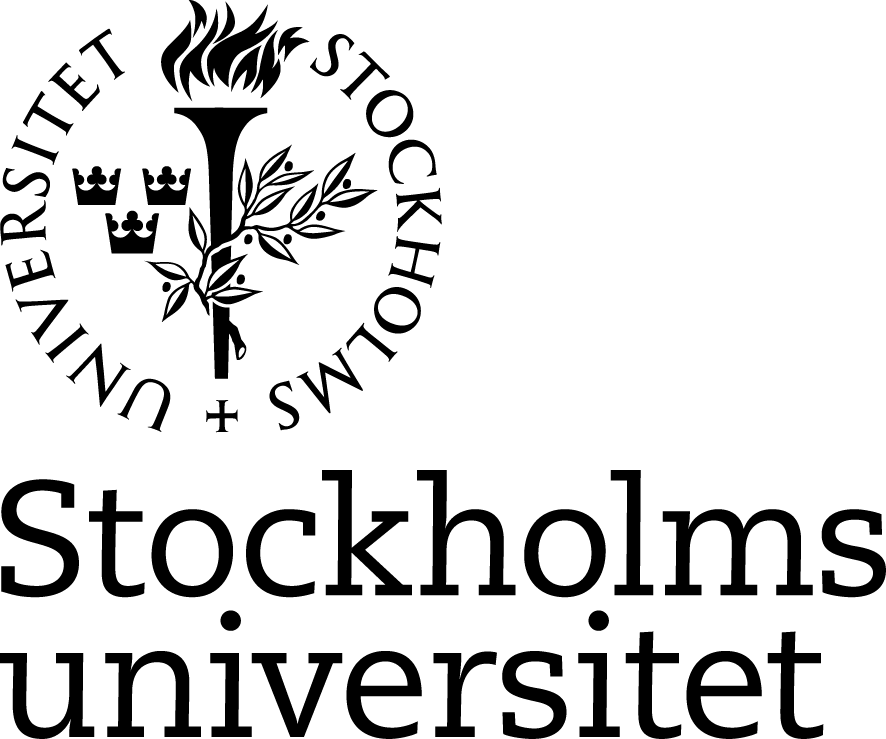} \newline

\vspace{2cm} \noindent {\LARGE Holonomy in Quantum Information Geometry} \newline

\vspace{.5cm} \noindent {\large Ole Andersson} \newline

\vfill \noindent {\small Licentiate Thesis in Theoretical Physics at Stockholm University, Sweden 2018}

\newpage \thispagestyle{empty}
\hfill \vfill 
\noindent 
Thesis for the degree of Licentiate of Philosophy in Theoretical Physics \newline
Department of Physics \newline 
Stockholm University \newline 
Sweden
\pagenumbering{roman}

\pagestyle{empty}
\begin{center}
\includegraphics[width=0.25\textwidth]{logo-svart-svensk_stor_150dpi.png}\vspace{60pt}\\
{\LARGE Holonomy in Quantum Information Geometry}\vspace{40pt}\\
{\Large Ole Andersson}
\vspace{30pt}\\
{\bf Abstract:}
\end{center}
In this thesis we provide a uniform treatment of two non-adiabatic geometric phases for dynamical systems of mixed quantum states, namely those of Uhlmann and of Sj\"{o}qvist \emph{et al.} We develop a holonomy theory for the latter which we also relate to the already existing theory for the former. 
This makes it clear what the similarities and differences between the two geometric phases are. We discuss and motivate constraints on the two phases. Furthermore, we discuss some topological properties of the holonomy of `real' quantum systems, and we introduce higher-order geometric phases for not necessarily cyclic dynamical systems of mixed states. In a final chapter we apply the theory developed for the geometric phase of Sj\"{o}qvist \emph{et al.}\ to geometric uncertainty relations, including some new ``quantum speed limits''.
\vfill
\noindent Akademisk avhandling f\"{o}r avl\"{a}ggande av licentiatexamen vid Stockholms universitet, Fysikum
\vspace{10pt}\\
Licentiatseminariet \"{a}ger rum 21 mars kl 10.15 i sal C5:1007, Fysikum, Albanova universitetcentrum, Roslagstullsbacken 21, Stockholm.

\cleardoublepage

\pagestyle{fancy}
\renewcommand{\chaptermark}[1]{%
        \markboth{#1}{}}
\renewcommand{\sectionmark}[1]{%
        \markright{\thesection\ #1}}
\fancyhf{}  
\fancyhead[LE,RO]{\bfseries\thepage}
\fancyhead[LO]{\bfseries\rightmark}
\fancyhead[RE]{\bfseries\leftmark}
\renewcommand{\headrulewidth}{0.5pt}
\renewcommand{\footrulewidth}{0pt}
\addtolength{\headheight}{0.5pt} 
\fancypagestyle{plain}{%
   \fancyhead{} 
   \renewcommand{\headrulewidth}{0pt} 
}

\addcontentsline{toc}{chapter}{Introduction}
\section*{Introduction}
This thesis deals with the concept of holonomy in quantum mechanics. Since I find this concept, and its consequences, very interesting I have made it central to some of my publications \cite{AnHe2013,AnHe2015,AnBeErSj2016} (but not all \cite{AnBe2017,AnBaBeDuCa2017,AnBaDuCa2018}). And with this thesis I have taken the opportunity to collect my thoughts on the subject.

Holonomy is a phenomenon which has detectable effects on cyclically evolving dynamical systems.
The phenomenon has a purely geometric origin and is not due to the mechanism that drives the evolution.
In quantum mechanics, the most well-known consequence of holonomy is the 
appearance of a geometric relative phase, the so-called Berry phase.
The Berry phase plays a crucial role in several different implementations of quantum mechanics, 
ranging from nuclear physics to condensed matter. 
But the Berry phase is only defined for systems which can be represented by pure states,
and most quantum systems must be represented by mixed states.
A natural question is, therefore, if the definition of the Berry phase can be generalized to also include mixed quantum states.

Armin Uhlmann was, as far as the author is aware, the first to address the issue of a geometric phase for mixed states. 
In a sequence of papers \cite{Uh1986,Uh1989,Uh1991,Uh1992b,Uh1995}, published in the late 80's and early 90's,
Uhlmann developed a holonomy theory for mixed quantum states.
In this theory, the natural notion of geometric phase is an  
extension of Berry's geometric phase.
Moreover, the theory is closely related to one of the most important monotone geometries studied in quantum information---the Bures geometry---which indicates that Uhlmann's phase is closely connected to the probabilistic structure of quantum mechanics. 
Uhlmann's work thus provided a well-motivated geometric phase for mixed states.
However, this does not mean---which is apparent from one of my papers---that Uhlmann's phase is the `correct' geometric phase in all contexts. And indeed, the use of Uhlmann's geometric phase in quantum theory has so far been rather limited. To some extent, this is certainly due to the complexity of Uhlmann's work. But the main reason is probably the lack of satisfying experimental support. This latter fact led 
Erik Sj\"{o}qvist and his collaborators to suggest, in the beginning of the current century, a different definition of geometric phase for mixed states \cite{SjPaEkAnErOiVe2000,ToSjKwOh2004}. 

The definition proposed by Sj\"{o}qvist \emph{et al.}\ was 
extracted directly from well-established quantum interferometric relations and, hence, was almost tautologically certified to be supported by experiments. For pure states, this `interferometric geometric phase' agrees with the Berry phase and, hence, with Uhlmann's phase. 
But for mixed states, as Paul Slater almost immediately pointed out \cite{Sl2002}, the interferometric phase and Uhlmann's phase are in general different. Since its announcement, Sj\"{o}qvist \emph{et al.}'s interferometric geometric phase has been verified in several experiments and, nowadays, this phase seem to be the more popular of the two in applications.

Until now, a holonomy theory similar to the one by Uhlmann but giving rise to Sj\"{o}qvist \emph{et al.}'s geometric phase has been missing.
As a consequence, the relation between the two geometric phases has not been fully investigated. In this thesis we address these issues. After briefly reviewing the work by Uhlmann, we show how Uhlmann's construction can be combined with a well-known technique from symplectic geometry to generate a new holonomy theory for mixed quantum states. 
This new holonomy theory gives rise to Sj\"{o}qvist \emph{et al.}'s geometric phase and, hence, provides an answer to a question raised by Dariusz Chru\'{s}ci\'{n}ski and Andrzej Jamio\l kowski in their well-known book on geometric phases in classical and quantum mechanics:
``\emph{..., what is the relation between the mathematical formulation of Uhlmann and the more ``experimental'' approach of Sj\"{o}qvist et al?}'' \cite{ChJa2004}.

The interferometric geometric phase of Sj\"{o}qvist \emph{et al.}\ is subject to certain constraints and, therefore, it is not always well-defined.
Neither is Uhlmann's geometric phase, but in that case, the constraints are of a different kind. We describe these constraints (which in some cases appear to be limitations and in others are desired) as well as their geometrical origin. We also describe some topological properties of the holonomy and geometric phase of `real systems', and we introduce higher order geometric phases for not necessarily cyclic systems. In a final section we apply the formalism to ``quantum speed limits'', a subject which is very close to my heart (as should be clear from my list of publications \cite{AnHe2014a,AnHe2014b}).
\clearpage

\addcontentsline{toc}{chapter}{Acknowledgments}
\section*{Acknowledgments}
First and foremost, thank you Ingemar for being an excellent supervisor.
Also, a big thank you to all my friends in the KOMKO group.
A special thanks to Hoshang who convinced me to apply to the graduate school.
Last but not least, lots of love to my dear, wonderful family: Annika, Melker, and Anton.
\cleardoublepage

\pagestyle{empty}
\tableofcontents
\cleardoublepage

\pagestyle{fancy}
\renewcommand{\chaptermark}[1]{%
        \markboth{#1}{}}
\renewcommand{\sectionmark}[1]{%
        \markright{\thesection\ #1}}
\fancyhf{}  
\fancyhead[LE,RO]{\bfseries\thepage}
\fancyhead[LO]{\bfseries\rightmark}
\fancyhead[RE]{\bfseries\leftmark}
\renewcommand{\headrulewidth}{0.5pt}
\renewcommand{\footrulewidth}{0pt}
\addtolength{\headheight}{0.5pt} 
\fancypagestyle{plain}{%
   \fancyhead{} 
   \renewcommand{\headrulewidth}{0pt} 
}

\addcontentsline{toc}{chapter}{List of accompanying papers}
\section*{Accompanying papers}

\begin{itemize}
\item[I\;]	
\textbf{Operational geometric phase for mixed quantum states} \\
O. Andersson and H. Heydari \\
New J. Phys. \textbf{15}, 053006, 2013
\item[II\;] 
\textbf{Geometric uncertainty relation for mixed quantum states} \\
O. Andersson and H. Heydari \\
J. Math. Phys. \textbf{55}, 042110, 2014
\item[III\;]
\textbf{Quantum speed limits and optimal Hamiltonians for \\ driven systems in mixed states} \\
O. Andersson and H. Heydari \\
J. Phys. A: Math. Theor. \textbf{47}, 215301, 2014
\item[IV\;]
\textbf{A symmetry approach to geometric phase for quantum \\ ensembles} \\
O. Andersson and H. Heydari \\
J. Phys. A: Math. Theor. \textbf{48}, 485302, 2015
\item[V\;]
\textbf{Geometric phases for mixed states of the Kitaev chain} \\
O. Andersson, I. Bengtsson, M. Ericsson, and E. Sj\"{o}qvist \\
Phil. Trans. R. Soc. A \textbf{374}, 20150231, 2016
\end{itemize}
\clearpage

\addcontentsline{toc}{chapter}{List of additional papers by the author}
\section*{Other papers by the author}
\begin{itemize}
\item[VI\;]
\textbf{Dynamic distance measures on spaces of isospectral mixed quantum states} \\
O. Andersson and H. Heydari \\
Entropy \textbf{15}, 3688, 2013
\item[VII\;]
\textbf{Geometry of quantum evolution for mixed quantum states} \\
O. Andersson and H. Heydari \\
Phys. Scr. T \textbf{160}, 014004, 2014
\item[VIII\;]
\textbf{Geometric uncertainty relation for quantum ensembles} \\
H. Heydari and O. Andersson \\
Phys. Scr. \textbf{90}, 025102, 2015
\item[IX\;]
\textbf{Cliffordtori and unbiased vectors} \\
O. Andersson and I. Bengtsson \\
Rep. Math. Phys. \textbf{79}, 33, 2017
\item[X\;]
\textbf{Self-testing properties of Gisin's elegant Bell inequality} \\
O. Andersson, P. Badzi\c{a}g, I. Bengtsson, I. Dumitru, and A. Cabello \\ 
Phys. Rev. A \textbf{96}, 032119, 2017
\item[XI\;]
\textbf{Device-independent certification of two bits of randomness from one entangled bit and Gisin's elegant Bell inequality} \\
O. Andersson, P. Badzi\c{a}g, I. Dumitru, and A. Cabello \\
Phys. Rev. A \textbf{97}, 012314, 2018
\end{itemize}
\clearpage
\thispagestyle{empty}
\null
\clearpage
\thispagestyle{empty}
\null

\cleardoublepage

\pagenumbering{arabic}

\chapter{The geometric phase}\label{ch: introduction}
A quantum system that goes through a cyclical development may acquire a relative phase.
The emergence of the phase depends in part on the fact that the system  
contains energy. But the phase is also in part a consequence of a geometric phenomenon called holonomy.
In this chapter, which 
serves as an introduction to the thesis,
we review Aharonov and Anandan's approach to holonomy for systems in pure states \cite{AhAn1987}.

\section{The projective Hilbert space}
In quantum mechanics it is assumed that the information about a system in a definite, pure state
can be encoded in a unit vector in a complex Hilbert space. 
If the Hilbert space is $\HH$, the states are thus 
represented by the elements of the unit sphere $\WW$ in $\HH$. 
However, it is also assumed that two unit vectors which differ only by a phase factor 
represent the same state. Therefore, a more precise statement is that the states are parameterized by the elements of the projective Hilbert space $\PP$.
The projective Hilbert space is the space of all unit rank orthogonal projectors on $\HH$.
We have that two unit vectors differ by a phase factor if, and only if, they define 
the same projector:
\begin{equation}
	\ket{\phi}=\ket{\psi}e^{i\theta}\quad \iff \quad \ketbra{\phi}{\phi}=\ketbra{\psi}{\psi}.
\end{equation}

Since two unit vectors differing by a phase factor represents the same quantum state 
one might think that the phase does not contain any physically relevant information.
But this is not always true; it depends on how the phase has emerged.
If $\ketbra{\psi_t}{\psi_t}$ represents the
state of a time-evolving, Hamiltonian quantum system which returns to its initial state at $t=\tau$,
then the relative phase factor between $\ket{\psi_0}$ and 
$\ket{\psi_\tau}$ contains information about the energy of the system as well as information 
about the geometry of the projective Hilbert space. 
To see this we need to connect $\WW$ and $\PP$ by an appropriate fiber bundle
and equip the bundle with a connection.

\subsubsection{About terminology}
Before we proceed, some words should be said about the terminology 
used in this thesis to refer to certain mathematical structures, 
well known to both physicists and mathematicians but under different names.

Central to the subject of this thesis is the notion of a principal fiber bundle.
Most mathematicians know what a principal fiber bundle is, but to physicists 
such bundles are better known as `gauge structures'. (The underlying mathematical structure 
in a gauge theory is a principal fiber bundle.)
In a gauge theory, the symmetry group, or, in physics language, the gauge group, acts on the total space of the   
principal fiber bundle. Moreover, the bundle is equipped with a connection,
which is the same thing as a gauge field. 
Here, we have chosen to use a more mathematics oriented 
terminology. That is, we use the terms ``principal fiber bundle'', ``symmetry group'', and ``connection'', rather than
``gauge structure'', ``gauge group'', and ``gauge field''.
With this said, readers of this thesis, whether it be a physicist or a mathematician,
should not find it difficult to follow the discussion. Otherwise, Chapters 9 and 10 in \cite{Na2003} provide a sufficient introduction to the theory of principal fiber bundles. A more extensive reference is the two volume work \cite{KoNo1996}.

\section{The Hopf bundle}
Define a map $\p$ from $\WW$ to $\PP$ by $\p(\ket{\psi})=\ketbra{\psi}{\psi}$.
This map is onto, i.e., it hits every element of $\PP$, and sends two vectors 
to the same projection operator if, and only if, they differ by a phase factor.
The action by the unitary group $\UU(1)$ on $\WW$,
\begin{equation}
	\big(e^{i\theta},\ket{\psi}\big)\to \ket{\psi}e^{i\theta},
\end{equation}
is thus transitive on the fibers of $\p$.
Consequently, $\p$ is a principal fiber bundle with symmetry group $\UU(1)$. 
The bundle is called the Hopf bundle.%
\footnote{A principal fiber bundle is defined by the complete information about its base space, total space, projection, and the action of the symmetry group. 
Thus, it would be more correct to refer to, e.g., the Hopf bundle as 
$(\SS,\WW,\Pi,\UU(1))$.
But in order to avoid using a too cumbersome notation we will in this thesis refer to fiber bundles by their projections.}%

\subsection{The Berry connection}
The differential of $\p$ maps the tangent spaces of $\WW$ 
onto the corresponding tangent spaces of $\PP$. 
The kernel of the differential at $\ket{\psi}$
is one-dimensional and is called the vertical space at $\ket{\psi}$. 
We write $\V_{\ket{\psi}}\WW$ for the vertical space.
The union of all the vertical spaces, $\V\WW$, is the vertical bundle of $\p$.

A connection on $\p$ is a smooth tangent vector bundle which is complementary to the vertical bundle 
and which is preserved by the symmetry group action. 
We can define a connection on $\p$ as follows.
The real part of the Hermitian product on $\HH$ defines a Riemannian metric $G$ on $\WW$;
if $\ket{\dot{\psi}_1}$ and $\ket{\dot{\psi}_2}$ are tangent vectors at $\ket{\psi}$, then
\begin{equation}
	G\left(\ket{\dot{\psi}_1},\ket{\dot{\psi}_2}\right)
	=\frac{1}{2}\left(\braket{\dot{\psi}_1}{\dot{\psi}_2}+\braket{\dot{\psi}_2}{\dot{\psi}_1}\right).
\end{equation}
Let $\H_{\ket{\psi}}\WW$ be the orthogonal complement of $\V_{\ket{\psi}}\WW$.
The space $\H_{\ket{\psi}}\WW$ is the horizontal space at $\ket{\psi}$,
and the union of all the horizontal spaces is the horizontal bundle $\H\WW$.
Since the action by $\UU(1)$ is by isometries with respect to $G$, 
the action preserves the horizontal bundle.
Therefore, the horizontal bundle is a connection on $\p$.

One can alternatively define $\H\WW$ as the kernel bundle of a connection form on $\WW$.
The orthogonal projection
of a tangent vector $\ket{\dot{\psi}}$ at $\ket{\psi}$ on 
the vertical space $\V_{\ket{\psi}}\WW$ is 
$\ket{\dot{\psi}^v}=\ket{\psi}\braket{\psi}{\dot{\psi}}$.
To see this we first observe that the vertical space is spanned by the unit vector $i\ket{\psi}$.
Then
\begin{equation}
	\ket{\dot{\psi}^v}
	=i\ket{\psi}G\left(i\ket{\psi},\ket{\dot{\psi}}\right)
	=i\ket{\psi}\frac{1}{2}\left(-i\braket{\psi}{\dot{\psi}}+i\braket{\dot{\psi}}{\psi}\right)
	=\ket{\psi}\braket{\psi}{\dot{\psi}}.
\end{equation}
The factor $\braket{\psi}{\dot{\psi}}$ belongs to $\uu(1)$, the Lie algebra of the symmetry group, which equals $i\reals$,
and the assignment 
\begin{equation}
	\A(\ket{\dot{\psi}})=\braket{\psi}{\dot{\psi}}
\end{equation}
defines a $\uu(1)$-valued connection form on $\WW$. This is the
Berry connection form.
Clearly, $\H_{\ket{\psi}}\WW$ is the kernel of the Berry connection form at $\ket{\psi}$.

\subsubsection{Horizontal lifts}
\begin{figure}[t]
	\centering
	\includegraphics[width=0.9\textwidth]{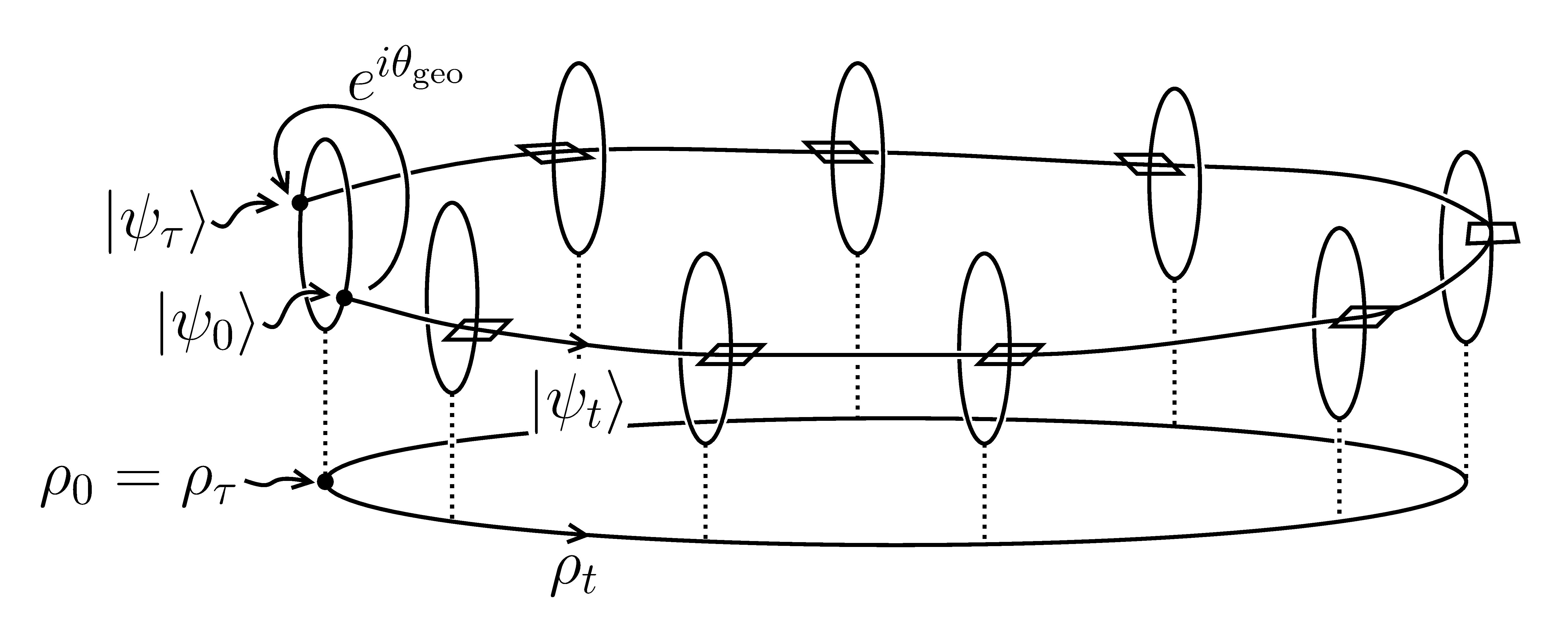}
	\caption{A horizontal lift and the corresponding geometric phase factor of a closed curve in the projective space. Notice that the horizontal lift is everywhere tangential to the horizontal spaces, given by the kernels of the Berry connection. The geometric phase factor is the relative phase factor of the initial and final states of the horizontal lift.}
	\label{fig: horizontal lift}
\end{figure}

The differential of $\p$ maps the horizontal space $\H_{\ket{\psi}}\WW$ 
isomorphically onto the tangent space $\T_\rho\PP$, where $\rho=\ketbra{\psi}{\psi}$.
Therefore, a tangent vector $\dot\rho$ at $\rho$ has a unique preimage $\ket{\dot\psi}$ 
in $\H_{\ket{\psi}}\WW$. The preimage is called the horizontal lift of $\dot\rho$ to $\ket{\psi}$.
More generally, for any curve $\rho_t$ in $\PP$ and any $\ket{\psi_0}$ in the 
fiber of the initial projector $\rho_0$,
there is a unique curve $\ket{\psi_t}$ in $\WW$ which extends from 
$\ket{\psi_0}$, projects onto $\rho_t$ and which
is everywhere horizontal. The latter condition means that, at each instant $t$,
the velocity vector $\ket{\dot{\psi}_t}$ belongs to the horizontal space $\H_{\ket{\psi_t}}\WW$.
The horizontal curve $\ket{\psi_t}$ is the horizontal lift of $\rho_t$ extending from $\ket{\psi_0}$.
If $\rho_t$ is a closed curve, so that $\rho_\tau=\rho_0$ for some $\tau>0$,
the vectors $\ket{\psi_\tau}$ and $\ket{\psi_0}$ belong to the same fiber of $\p$
and hence differ only by a phase factor. This is the geometric phase factor of $\rho_t$. 
The situation is illustrated in Figure \ref{fig: horizontal lift}.

\section{The Aharonov-Anandan geometric phase} 
We introduce the notation $\ket{\dot{\psi}^h}$ for the horizontal projection of the vector $\ket{\dot{\psi}}$.
If $\ket{\dot{\psi}}$ is the velocity vector of a curve generated by a 
Hamiltonian $H$, i.e., if $i\ket{\dot{\psi}}=H\ket{\psi}$, then
\begin{subequations}
\begin{align}
	G\big(\ket{\dot{\psi}^v},\ket{\dot{\psi}^v}\big)
		&= \bra{\psi}H\ket{\psi}^2
		=E_{\ket{\psi}}^2(H),\\
	G\big(\ket{\dot{\psi}^h},\ket{\dot{\psi}^h}\big)
		&=\bra{\psi}H^2\ket{\psi}-E_{\ket{\psi}}^2(H)=\var_{\ket{\psi}}(H).
\end{align}
\end{subequations}
The energy $E=E(H)$ thus governs the drift in the fiber directions and the energy 
uncertainty $\Delta E=\var(H)^{1/2}$ the drift `forward' towards new states.

\subsubsection{The Mandelstam-Tamm quantum speed limit}
We can define a metric on $\PP$ by declaring that the differential of 
$\p$ maps every horizontal space of $\WW$ isometrically onto the 
corresponding tangent space of $\PP$. 
The so obtained metric, $\gFS$, is called the Fubini-Study metric. 
If $\ket{\psi_t}$, $0\leq t\leq \tau$, is the solution to a Schr\"{o}dinger equation, 
$i\ket{\dot\psi_t}=H\ket{\psi_t}$, the projected curve $\rho_t=\ketbra{\psi_t}{\psi_t}$ has the length
\begin{equation}\label{eq: jost}
	\length{\rho_t}
	=\int_{0}^{\tau}\dt\sqrt{\gFS(\dot{\rho}_t,\dot{\rho}_t)}
	=\int_{0}^{\tau}\dt\sqrt{G(\ket{\dot{\psi}^h_t},\ket{\dot{\psi}^h_t})}
	=\tau \Delta E.
\end{equation}
(If $H$ is time-dependent we define $\Delta E$ as the time-average of the 
energy uncertainty along $\ket{\psi_t}$.)
From this observation, and knowledge about the distance function associated with the Fubini-Study metric, 
we can derive a lower bound on the time it takes for the initial state $\ket{\psi_0}$ 
to evolve into a perpendicular state:
If $\braket{\psi_0}{\psi_\tau}=0$, then 
\begin{equation}\label{eq: lengths}
	\length{\rho_t}
	\geq \distFS{\rho_0}{\rho_\tau}
	=\arccos|\braket{\psi_0}{\psi_\tau}|
	=\frac{\pi}{2}.
\end{equation}
Equations \eqref{eq: jost} and \eqref{eq: lengths} yield 
\begin{equation}\label{eq: MT speed limit}
	\tau\geq\frac{\pi}{2\Delta E}.
\end{equation}
The inequality \eqref{eq: MT speed limit} is due to Mandelstam and Tamm \cite{MaTa1945} 
(see also \cite{Fl1973,Bh1983,AnAh1990}). In the literature it goes by the name 
``the Mandelstam-Tamm quantum speed limit''. Quantum speed limits, i.e., fundamental bounds on how fast a quantum state can be transformed into a state with some given properties, have recently attracted much attention.\footnote{The term ``quantum speed limit'' was used for the first time by Margolus and Levitin \cite{MaLe1998}.}  
For more about quantum speed limits 
we refer to the recent reviews \cite{Fr2016} and \cite{DeCa2017}.
See also Sections \ref{UhQSL}, \ref{WYqsl}, and \ref{Oqsl}.

\subsection{The dynamical phase}
We can identify the fiber through $\ket{\psi}$ with $\UU(1)$ by identifying 
$e^{i\theta}$ with $\ket{\psi}e^{i\theta}$. Corresponding to the unit element in $\UU(1)$,
the state $\ket{\psi}$ constitutes a 
natural reference point in the fiber.
But if we look at the fibers along a whole curve of states $\ket{\psi_t}$
it is natural to choose the reference points as close as possible. 
This means that we choose the reference point in the fiber of $\p$ containing $\ket{\psi_t}$ 
to be $\ket{\psi^h_t}$, where $\ket{\psi^h_t}$ is the horizontal lift of the 
projected curve $\rho_t=\ketbra{\psi_t}{\psi_t}$ which extends from $\ket{\psi_0}$. 
The horizontal lift is 
\begin{equation}\label{eq: horiz lift}
	\ket{\psi^h_t}
	= \ket{\psi_t}\exp\Big(\!-\!\int_{0}^{t}\dt\, \A(\ket{\dot{\psi}_t})\Big)
	= \ket{\psi_t}\exp\Big(\!-\!\int_{0}^{t}\dt\,\braket{\psi_t}{\dot{\psi}_t}\Big).
\end{equation}
Now, if $i\ket{\dot{\psi}_t}=H\ket{\psi_t}$, then $\ket{\psi^h_t}=\ket{\psi_t}\exp(iE t)$.
(For a time-dependent Hamiltonian, $E$ is the time-averaged energy along $\ket{\psi_t}$.)
Every inclusion of $\UU(1)$ can be used to pull back $G$ to a metric on $\UU(1)$. However, all the pullback metrics coincide with the 
standard metric on $\UU(1)$. Furthermore, the pullbacks of the vectors $\ket{\psi_t}$ give rise to a curve in $\UU(1)$, namely $t\to \exp(iE t)$.
If $\tau$ is the final time,
the length of this curve is $E\tau$.
The factor $\exp(-iE \tau)$, which is the relative phase factor between $\ket{\psi_\tau}$ and $\ket{\psi^h_\tau}$, is called the dynamical 
phase factor of the projected curve $\rho_t$.
Interestingly, the dynamical phase is connected to another quantum speed limit.

\subsubsection{The Margolus-Levitin quantum speed limit}
There is a lower bound on the time it takes for a state to evolve to an orthogonal state which is 
similar to the Mandelstam-Tamm speed limit but which involves the energy rather than the energy uncertainty.
If $\ket{\psi_t}$ satisfies the Schr\"{o}dinger equation with a positive, time-independent Hamiltonian $H$
and if $\braket{\psi_0}{\psi_\tau}=0$, then 
\begin{equation}\label{eq: Margolus-Levitin quantum speed limit}
	\tau\geq\frac{\pi}{2E}.
\end{equation}
The proof of this inequality is surprisingly simple:
If $E^a$ is the $a$th energy eigenvalue and $\ket{E_a}$ is a corresponding  
energy eigenstate, the time-developed system is represented by
\begin{equation}
	\ket{\psi_t}=\sum_a\ket{E_a}\braket{E_a}{\psi_0}e^{-itE^a}.
\end{equation}
(We assume that the energy eigenstates are normalized and mutually perpendicular.) Then, using that $\cos x\geq 1-2(x+\sin x)/\pi$ for $x\geq 0$,
\begin{equation}
\begin{split}
	\Re\braket{\psi_0}{\psi_t}
	&= \sum_a|\braket{E_a}{\psi_0}|^2\cos(tE^a)\\
	&\geq \sum_a|\braket{E_a}{\psi_0}|^2\Big(1-\frac{2}{\pi}\big(tE^a+\sin(tE^a)\big)\Big)\\
	&=1 - \frac{2tE}{\pi}+\frac{2}{\pi}\Im\braket{\psi_0}{\psi_t}.
\end{split}
\end{equation}
At $t=\tau$, $\Re\braket{\psi_0}{\psi_t}=\Im\braket{\psi_0}{\psi_t}=0$, and the inequality \eqref{eq: Margolus-Levitin quantum speed limit} follows.

The inequality is due to Margolus and Levitin \cite{MaLe1998}
and goes by the name ``the Margolus-Levitin quantum speed limit''. 
Despite being easy to prove, the limit is geometrically puzzling.
Since $H$ is time-independent, the curve $\ket{\psi_t}$
intersects the fibers of $\p$ at a constant angle
\begin{equation}\label{eq: angle}
	\theta=\arctan\left(\frac{\Delta E}{E}\right).
\end{equation}
Then, by Eq.~\eqref{eq: jost},
\begin{equation}
	\tau E 
	= \frac{\tau \Delta E}{\tan\theta}
	= \frac{\length{\rho_t}}{\tan\theta}\geq\frac{\pi}{2\tan\theta}.
\end{equation}
Now, according to the Margolus-Levitin quantum speed limit, if $\rho_t$ 
is close to being a geodesic, $\theta$
can barely exceed $\pi/4$. 
Although it is clear from Eq.~\eqref{eq: angle} that a positive Hamiltonian 
cannot generate a horizontal curve in $\WW$,
it is a bit unclear why the curve has to deviate quite a lot from being horizontal (assuming that the projected curve extends between perpendicular states and is close to being a geodesic).

\subsection{The geometric phase}
\begin{figure}[t]
	\centering
	\includegraphics[width=0.9\textwidth]{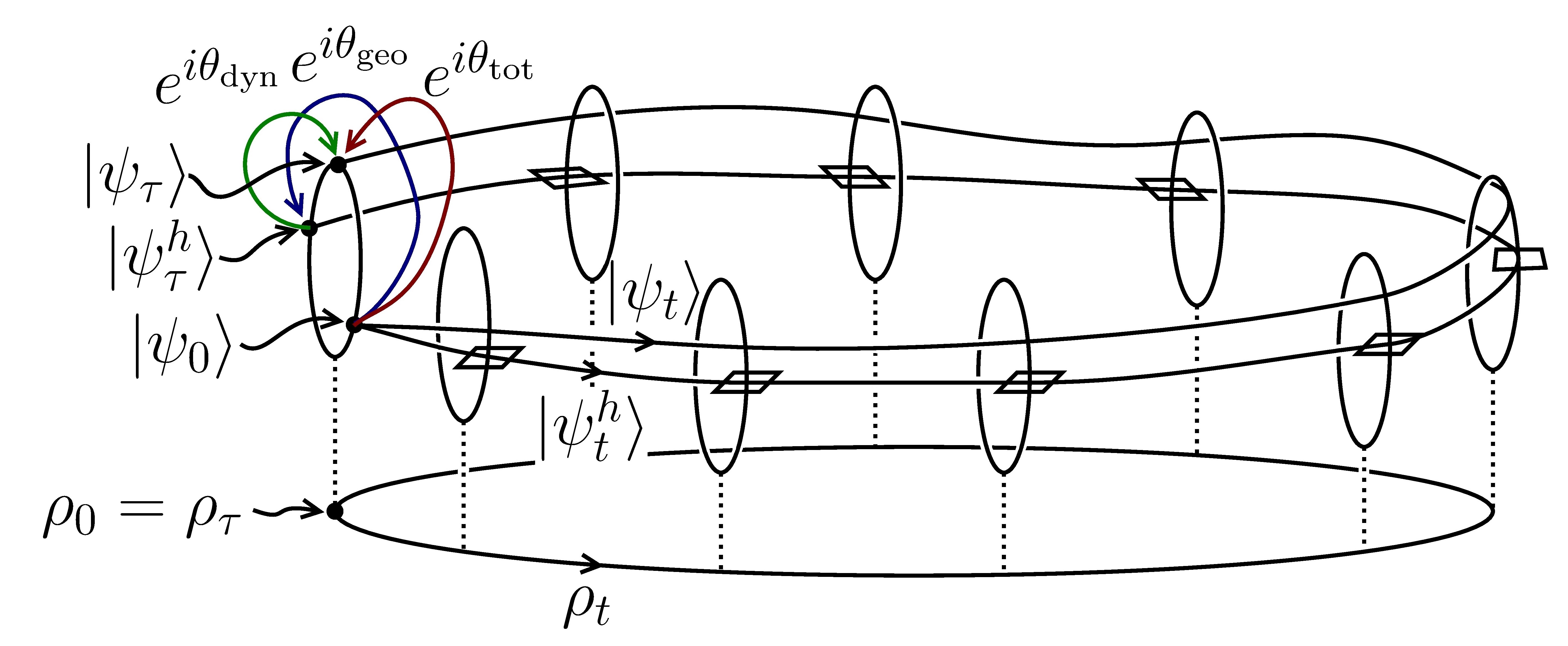}
	\caption{The total, dynamical, and geometric phase factors of a closed curve in the projective space. The total and dynamical phase factors depend on the lift while the geometric phase factor does not. But any pair of them determines the third.}
	\label{fig: the phases}
\end{figure}

Let $\rho_t$ be a closed curve in $\PP$ and let $\ket{\psi_t}$ be any lift of 
$\rho_t$ to $\WW$.
The total, dynamical, and geometric phase factors of $\rho_t$ are defined by
\begin{subequations}\label{eq: phases}
\begin{align}
	e^{i\ttot}&=\braket{\psi_0}{\psi_\tau},\label{eq: global}\\
	e^{i\tdyn}&=\braket{\psi^h_\tau}{\psi_\tau},\label{eq: dynamic}\\
	e^{i\tgeo}&=\braket{\psi_0}{\psi^h_\tau},\label{eq: geometric}
\end{align}
\end{subequations}
where $\ket{\psi^h_t}$ is the horizontal lift of $\rho_t$ extending from $\ket{\psi_0}$. 
The phase factors satisfy 
\begin{equation}\label{eq: cocycle}
	e^{i\ttot} = e^{i\tdyn} e^{i\tgeo} .
\end{equation}
See Figure~\ref{fig: the phases}.
In other words, the total phase is the sum of the dynamical and the geometric phase.
(The phases are only defined modulo $2\pi$.)
Since there are many lifts of $\rho_t$ to $\WW$, the total and the dynamical phases 
are not intrinsic to $\rho_t$. However, the geometric phase \emph{is} intrinsic to $\rho_t$ 
because it does not depend on the initial lift $\ket{\psi_0}$ and, once 
the initial lift is specified, the horizontal lift is unique. The geometric phase factor,
which is the factor in $\UU(1)$ by which $\ket{\psi_0}$ has to be multiplied to become the 
final state of the horizontal lift, is also called the holonomy of $\rho_t$.
Clearly, the holonomy depends on the choice of connection on $\p$ which in turn determines a 
geometry~on $\p$. 

\subsubsection{We give many names to those we love}	
The presentation above is inspired by \cite{AhAn1987}
and, hence, we will refer to the geometric phase \eqref{eq: geometric}
as the Aharonov-Anandan phase. But in different contexts, the same geometric phase is 
attributed to different persons. If the geometric phase results from an adiabatic development it is usually called the Berry phase \cite{Be1984}, or the Berry-Simon phase \cite{Si1983}.
In condensed matter, the phase goes by the name Zak phase \cite{Za1989}. 
The inner product in Eq.~\eqref{eq: geometric} also makes sense for noncyclic evolutions,
and the phase of this product, if nonzero, is a natural generalization of the geometric phase to such evolutions. Samuel and Bhandari \cite{SaBh1988} were among the first to consider geometric phases for noncyclic evolutions of quantum states.
The geometric phase for nonclosed curves has been thoroughly investigated by Mukunda and Simon \cite{MuSi1993I,MuSi1993II}.

\section{Mixed states}
Pure states are not sufficient to describe all phenomena in quantum physics. For example, the properties of a system which is \emph{entangled} with another system cannot be extracted from a pure state, i.e., a single unit rank projector. Neither can the properties a large system in a thermal state. And sometimes, a measurement puts a system in a state which cannot be represented by a projector.
In all these cases, however, the state of the system can be described as a probabilistic ensemble of states.
Such ensembles are called mixed quantum states 
and can be represented by density operators. A unit rank projector, representing a pure state, is 
a special case of a density operator. But a general density operator can have a rank different from 1.
In this thesis we will extend the concept of holonomy to systems in mixed states. 
The extension is complicated by the complexity of the space of 
density operators and the nonabelian and noncanonical character of the holonomy.

\section{Structure of the thesis}
The thesis can be divided into three parts.
In the first part, which consists of Chapters \ref{ch: Quantum State Space} and \ref{ch: Stand pure}, we discuss some topological and 
geometric properties of the space of density operators, 
and we introduce Uhlmann's standard purification bundle.
The standard purification bundle is the common mathematical framework for the different holonomy theories that we will meet in this thesis.
In this first part, many general results are formulated which will be used repeatedly in the subsequent parts.

Chapter \ref{monotone geometries} constitutes the second part of the thesis. 
In Chapter \ref{monotone geometries} we review a construction due to Dittmann and Uhlmann showing that all monotone metrics 
can be obtained as projections of the Hilbert-Schmidt metric via the standard purification bundle. The review is very brief, but the three most important monotone geometries are treated in some detail. These are the Bures geometry, the Wigner-Yanase geometry 
and one which is dedicated to nobody but which we here call the complementary geometry
(for reasons that will be clear). Some authors call this latter geometry the RLD-geometry, where RLD is an acronym for ``Right Logaritmic Derivative''. Why this is so will not be apparent from the current presentation.

The third part consists of Chapters \ref{ch: sjoqvist holonomy} and \ref{apps}. In Chapter \ref{ch: sjoqvist holonomy} we formulate a holonomy 
theory which differs from the theories studied in Chapter \ref{monotone geometries}
 and which gives rise to 
Sj\"{o}qvist \emph{et al.'s} `interferometric geometric phase'. 
Such a holonomy theory has been lacking until now and, therefore, 
Chapter \ref{ch: sjoqvist holonomy} is a bit more technical and more detailed than the other chapters.
In the last chapter, Chapter \ref{apps}, we 
consider three different applications of the theory developed in Chapter \ref{ch: sjoqvist holonomy}.
For example, we derive two quantum speed limits. 
As we have seen, a quantum speed limit is a fundamental
lower bound on the time it takes to 
perform a certain quantum processing task.
Some of the existing quantum speed limits are of a geometric nature
and are, as we will see, more-or-less immediate consequences of the theory derived in this thesis.
\vspace{7pt}\\ 

\hfill{Enjoy!}

\chapter{Quantum state space}\label{ch: Quantum State Space}

A mixed quantum state, or quantum ensemble, can be represented by a density operator on a Hilbert space. 
By definition, a density operator is a self-adjoint, positive semi-definite, trace-class operator with unit trace. 
Here we will only consider quantum systems which can be modeled on finite 
dimensional Hilbert spaces.
The assumption that density operators are trace-class is then redundant;
a density operator on a finite dimensional Hilbert space is
simply a Hermitian operator whose eigenvalues are nonnegative and sum up to $1$.

In the current chapter we will discuss some topological properties of
the space of density operators on a general finite dimensional Hilbert space.
Specifically, we will describe three stratifications of this space.
The strata are made up of manifolds of density operators which share certain spectral properties.
Towards the end of the chapter we will also introduce some important 
geometries on the space of density operators,
the properties of which we will explore in later chapters.
Special attention will be given to the space  
of faithful density operators.
Here, ``faithful'' is shorthand for having full rank.

\section{Stratifications}
Let $\HH$ be a complex $n$-dimensional Hilbert space
and let $\SS$ be the space of density operators on $\HH$.
The space $\SS$ is an $(n^2-1)$-dimensional compact convex subset of 
$\hh(\HH)$, the space of Hermitian operators on $\HH$. 
We assume that $\hh(\HH)$ carries its canonical topology as a real vector space and we
equip $\SS$ with the subspace topology.

\subsection{Stratification with respect to rank}
The space $\SS$ 
is in general not a smooth manifold  \cite{GrKuMa2005}.\footnote{The case when $\HH$ is two-dimensional is the only exception. See Section \ref{sec: qubits}. In higher dimensions, the boundary of $\SS$ has `low-dimensional corners'.}
But each of its subspaces of density operators which have a fixed common rank is a smooth manifold. 
Recall that the rank of an operator is the complex dimension of its support.
We will write $\SS^k$ for the space of all density operators on $\HH$ which have rank $k$.
The dimension of $\SS^k$ is $2nk-k^2-1$.

The space $\SS^1$, containing the unit rank density operators, is the projective Hilbert space that we became acquainted with in Chapter~\ref{ch: introduction}. 
The spaces $\SS^k$ for $2\leq k\leq n-1$
have rather complicated topologies which we will explore in the next section. (See also \cite{BeZy2017}, especially Sec.~8.5.)
The space $\SS^n$, which consists of the faithful density operators on $\HH$,
is the topological interior of $\SS$.
Being the interior of an $(n^2-1)$-dimensional compact convex set, 
this space is a convex manifold 
diffeomorphic to an open $(n^2-1)$-dimensional ball.

\subsubsection{Spectra of density operators}
By the weight spectrum of a density operator $\rho$ we mean the 
sequence $\bfp=(p^1,p^2,\dots,p^k)$ of the operator's \emph{positive} eigenvalues arranged in decreasing order of magnitude. 
We assume that the eigenvalues are repeated in accordance with their degeneracy. 
The length $k$ of the weight spectrum equals the rank of $\rho$.
We call two density operators isospectral if they have the same weight spectrum.
When we want to list only the distinct eigenvalues of $\rho$ we use capital letters.
Thus, we write $\bfP=(P^1,P^2,\dots,P^l)$ for the distinct, positive eigenvalues of
$\rho$ arranged in decreasing order of magnitude. The spectrum $\bfP$ is the eigenvalue spectrum of $\rho$.

By the degeneracy spectrum of $\rho$ we mean the sequence of positive integers $\bfm=(m_1,m_2,\dots,m_l)$ where $m_a$ is the degeneracy of the eigenvalue $P^a$.
We call two density operators isodegenerate if they share the same degeneracy spectrum.
Finally, we define the eigenprojector spectrum of $\rho$ to be the 
sequence $\bfL=(\Lambda_1,\Lambda_2,\dots,\Lambda_l)$ where $\Lambda_a$ is the orthogonal projection onto the eigenspace of $\rho$ corresponding to 
$P^a$. We can now represent $\rho$ as 
\begin{equation}\label{eq: spec deco}
	\rho=\sum_aP^a\Lambda_a.
\end{equation}
This representation is the spectral decomposition of $\rho$. 
(We will explicitly specify the range of a summation index only when the range is not obvious.)
The normalization requirement $\Tr\rho=1$ is equivalent to $\sum_aP^am_a=1$.

\subsection{Unitary orbits}
The density operators which represent the instantaneous states of an evolving closed quantum system 
all belong to the same orbit of the left adjoint representation of the unitary group $\UU(\HH)$ on $\hh(\HH)$,
\begin{equation}
	\UU(\HH)\times\hh(\HH)\ni(U,A)\to \Ad_U\!A=UAU^\dagger\in\hh(\HH).
\end{equation}
The unitary orbit of a density operator $\rho$ contains all and only those density operators 
which are isospectral to $\rho$. We write $\SS_{\bfp}$
for the common unitary orbit of the density operators which have weight spectrum $\bfp$.

The unitary orbits are homogeneous spaces and, hence, closed manifolds. 
Indeed, for any density operator $\rho$ with weight spectrum $\bfp$, the mapping
\begin{equation}\label{factor map}
	\UU(\HH)/\UU_\rho \ni U\UU_\rho \mapsto U\rho U^\dagger \in \SS_{\bfp}
\end{equation}
is a diffeomorphism. Here, $\UU_\rho$ is the group of unitary operators which commute with $\rho$. This group is usually called the isotropy group of $\rho$.
One can also describe $\SS_{\bfp}$ as the orbit of 
$\rho$ of the left 
adjoint representation of the special unitary group $\SU(\HH)$ on $\hh(\HH)$. 
And if $\SU_\rho$ is the group of special unitary operators which commute with $\rho$, i.e., the group of unitary operators in 
$\UU_\rho$ which have determinant equal to $1$, then 
\begin{equation}\label{special factor map}
	\SU(\HH)/\SU_\rho \ni U\SU_\rho \mapsto U\rho U^\dagger \in \SS_{\bfp}
\end{equation}
is a diffeomorhism. 

If $\rho$ has rank $k$, the orbit of $\rho$ is completely contained in $\SS^k$.
The unitary orbits therefore partition the $\SS^k$s and, hence, $\SS$ into closed manifolds of 
isospectral density operators. We will refer to this partition as the unitary stratification. 
Notice that not all orbits in the unitary stratification have the same topology,
even if the orbits belong to the same $\SS^k$.
From \eqref{factor map} follows that the orbits of two density operators are homeomorphic if, and only if, the 
density operators are isodegenerate. The dimension of $\SS_{\bfp}$ is 
$2nk-k^2-\sum_am_a^2$, where the $m_a$s are the degeneracies of the eigenvalues in $\bfp$.

\subsection{Classical manifolds}
Two adjacent isospectral density operators do not commute. For
if $\rho$ is a density operator of rank $k$ and $\CC^k_\rho$
is the space of all the density operators in $\SS^k$ which commute with $\rho$,
\begin{equation}
	\CC^k_\rho = \{ \sigma\in\SS^k : \sigma\rho=\rho\,\sigma \},
\end{equation}
then, as we will see in the next section, $\CC^k_\rho$ is a manifold which 
intersects the unitary orbit of $\rho$ in a complementary manner in $\SS^k$ at $\rho$.
We call $\CC^k_\rho$ the commutative manifold of $\rho$. The dimension of $\CC^k_\rho$ is $\sum_am_a^2-1$,
where, again, the $m_a$s are the degeneracies of the positive eigenvalues of $\rho$.

Since `being commuting' is not a transitive relation among density operators, even if we fix the rank, a commutative manifold is not \emph{the} commutative manifold of each of its members. Consequently, the commutative manifolds do not partition the $\SS^k$s. It will prove useful to reduce the commutative manifolds 
to `fully commutative manifolds' that \emph{do} partition the 
$\SS^k$s and, hence, $\SS$,
and which, on top of that, intersect any unitary orbit at, at most, one density operator.
To this end, for every eigenprojector spectrum $\bfL$ we write $\SS_{\bfL}$ for the space of
all the density operators which have the eigenprojector spectrum $\bfL$.
If $\bfL$ has length $l$, $\SS_{\bfL}$ is a manifold
of dimension $l-1$.
Indeed, $\SS_{\bfL}$ is canonically diffeomorphic to the open simplex
\begin{equation}\label{simplex}
	\Delta_{\bfm}\equiv\big\{(x^1,x^2,\dots,x^l)\in\reals^l : x^a>x^{a+1}> 0\text{ and } \sum_a x^am_a=1\big\},
\end{equation}
where the $m_a$s are the ranks of the operators in $\bfL$.
We have that $\SS_{\bfL}$ is contained in $\CC^k_\rho$ if $\rho$ has degeneracy spectrum $\bfL$.

We will call $\SS_{\bfL}$ a classical manifold.
The density operators in a classical manifold are simultaneously diagonalizable and, hence, can be simultaneously identified with classical probability distributions; hence the name ``classical manifold''.
Any two different operators in the same classical manifold have the same degeneracy spectrum but not 
the same eigenvalue spectrum. Therefore, a unitary orbit can intersect a classical manifold at, at most, one operator. 
To be precise, the unitary orbit of the density operator $\rho$ and the classical manifold containing $\rho$ intersect only at $\rho$.
Since `having the same eigenprojector spectrum' \emph{is} an equivalence relation on each $\SS^k$, 
the classical manifolds partition the $\SS^k$s and $\SS$. We will refer to this partition as the classical stratification. 
   
\section[Hermitian representations of tangent vectors]{Hermitian representations of tangent\\ vectors}\label{hermitian rep}
The space of density operators is contained in the hyperplane of unit trace operators in $\hh(\HH)$. 
This hyperplane is parallel to 
the subspace of traceless Hermitian operators $\sh(\HH)$.
Tangent vectors to the constant rank strata $\SS^k$ can thus be canonically 
identified with traceless Hermitian operators. In information geometry, 
the canonical identification is called the mixture representation.
Other, equally important, identifications are the exponential
representations. 
These are defined only on the space of faithful density operators $\SS^n$.

\subsection{The mixture representation}
\begin{figure}
	\centering
	\includegraphics[width=0.9\linewidth]{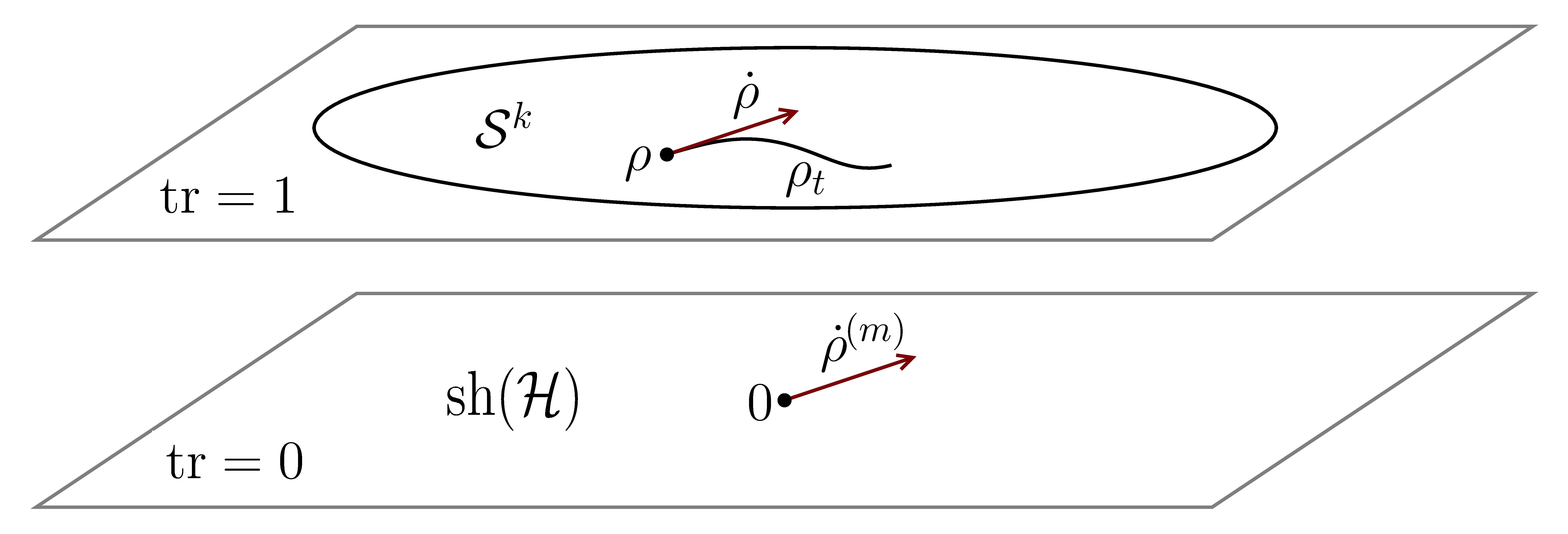}
	\caption{The mixture representation of a tangent vector.
	Since $\SS^k$ is parallel to the hyperplane of traceless Hermitian operators, the tangent vectors of $\SS^k$ can be identified with traceless Hermitian operators, simply by `parallel translating' them in the ambient space of Hermitian operators. This identification is called the mixture representation.}
	\label{fig: mixture representation}
\end{figure}

The mixture representation identifies tangent vectors of $\SS^k$  
with Hermitian operators which have a vanishing trace;
the tangent vector $\dot\rho$ at $\rho$ is identified with the operator $\dot\rho^{(m)}$ in $\sh(\HH)$ defined by
\begin{equation}
	\dot\rho^{(m)}=\lim_{t\to 0}\frac{1}{t}(\rho_t-\rho).
\end{equation}
Here $\rho_t$ is any curve in $\SS^k$ that extends from $\rho$ with velocity $\dot\rho$, see Figure~\ref{fig: mixture representation}.
We split the tangent space at $\rho$ into two subspaces:
\begin{subequations}
\begin{align}
	 \T^u_\rho\SS^k &= \{ \dot\rho \in \T_\rho\SS^k: \dot\rho^{(m)}=-i[A,\rho] \text{ for some }  A\in\sh(\HH) \},\label{unitary tangent}\\
	 \T^c_\rho\SS^k &= \{ \dot\rho \in \T_\rho\SS^k : [\dot\rho^{(m)},\rho]=0 \}.
\end{align}
\end{subequations}
The spaces $\T^u_\rho\SS^k$ and $\T_\rho^{c}\SS^k$ are
the tangent spaces of the unitary orbit and the commutative 
manifold of $\rho$, respectively.
That is, $\T^u_\rho\SS^k=\T_\rho\SS_{\bfp}$ where $\bfp$ is the weight spectrum of $\rho$ and $\T^c_\rho\SS^k=\T_\rho\CC^k_\rho$.
Moreover, the spaces are complementary in the sense that 
$\T_\rho\SS^k=\T^u_\rho\SS^k \oplus \T_\rho^{c}\SS^k$.
That the sum is direct is obvious from the following characterizations
of the two spaces: Let $\bfL=(\Lambda_1,\Lambda_2,\dots,\Lambda_l)$ be the eigenprojector spectrum of $\rho$. Then
\begin{subequations}
\begin{align}
	 \T^u_\rho\SS^k &= \big\{ \dot\rho \in \T_\rho\SS^k: \sum_{a}\Lambda_a\dot\rho^{(m)}\Lambda_a=0\big\},\\
	 \T^c_\rho\SS^k &= \big\{ \dot\rho \in \T_\rho\SS^k :\sum_{a}\Lambda_a\dot\rho^{(m)}\Lambda_a=\dot\rho^{(m)} \big\}.
\end{align}
\end{subequations}
Notice that if $\dot\rho^{(m)}=-i[A,\rho]$, we can assume that $\Lambda_aA\Lambda_a=0$
for all $a$.  
The tangent space of the classical manifold of $\rho$, which is a subspace of $\T^c_\rho\SS^k$, can be described as
\begin{equation}
	\T_\rho\SS_{\bfL}=\big\{ \dot\rho\in\T_\rho\SS^k : \dot\rho^{(m)}=\sum_a x^a\Lambda_a\text{ where } \sum_{a} x^am_a=0\big\}.
\end{equation}

\subsection{Exponential representations} 
On the space of faithful density operators we can use generalized covariances to 
identify tangent vectors with traceless Hermitian operators.
Generalized covariances also give rise to Riemannian metrics called quantum Fisher metrics. 
Some of the more important metrics in quantum information geometry are quantum Fisher metrics \cite{Ha2006}.

A generalized covariance is a field $\kappa$ of real-valued inner products on $\hh(\HH)$ which is smoothly parameterized by the faithful density operators. The field is required to be unitarily invariant and classically adapted:
\begin{subequations}
\begin{align}
	&\kappa_{U\rho U^\dagger}(UAU^\dagger,UBU^\dagger)=\kappa_\rho(A,B)
	\text{ for every unitary $U$,}\\
	&\kappa_\rho(A,B)=\Tr(\rho AB)\text{ if $A$ commutes with $\rho$}.
\end{align}
\end{subequations}
Important examples of generalized covariances include the symmetric
and the Bogolubov-Kubo-Mori generalized covariance:
\begin{subequations}
\begin{align}
	\kappa^\textsc{s}_\rho(A,B)&=\frac{1}{2}\Tr\big( \rho \,\{A,B\}\big),\label{symmen}\\
	\kappa^\textsc{bkm}_\rho(A,B)&=\int_0^1 d\lambda \Tr\big( \rho^{\lambda} A \rho^{1-\lambda} B \big).
\end{align}
\end{subequations}
The curly brackets in \eqref{symmen} is the skew-commutator, $\{A,B\}=AB+BA$.
Given a generalized covariance $\kappa$,
the exponential representation of a tangent vector $\dot\rho$ at $\rho$  is
the Hermitian operator $\dot\rho^{(e)}$ defined by
\begin{equation}
	\kappa_\rho\big(\dot\rho^{(e)},A\big)=\Tr\big(\dot\rho^{(m)}A\big).
\end{equation}
In general, the exponential representation depends on the generalized covari-ance.~But for every vector $\dot\rho$ in $\T_\rho \CC^n_\rho$,
the exponential representation is 
\begin{equation}\label{classical log}
	\dot\rho^{(e)}= \dot\rho^{(m)}\rho^{-1},
\end{equation}
irrespective of what is the generalized covariance. 

\subsubsection{Metrics associated with generalized covariances}
To every generalized covariance $\kappa$ we can associate a Riemannian metric $g_\kappa$ on $\SS^n$.
The metric is at $\rho$ defined as 
\begin{equation}\label{tjosan}
	g_{\kappa}(\dot\rho_1,\dot\rho_2) = \kappa_\rho\big(\dot\rho_1^{(e)},\dot\rho_2^{(e)}\big).
\end{equation}
The metric associated with the symmetric generalized covariance
is of special importance in quantum information geometry and estimation theory \cite{He1976,Ha2006,Pa2009}. It is called the quantum Fisher information metric.
We will denote it by $\gF$. 

The symmetric logarithmic derivative of a tangent vector $\dot\rho$ at $\rho$ is the Hermitian operator 
$L_{\dot\rho}^{\textsc{s}}$ defined by 
\begin{equation}\label{SLD}
	\dot\rho^{(m)}=\frac{1}{2}\{\rho, L_{\dot\rho}^{\textsc{s}}\}.
\end{equation}
Equivalently, we can define the symmetric logarithmic derivative as
\begin{equation}\label{symlog}
	L_{\dot\rho}^{\textsc{s}}= 2\int_0^\infty \d{t}\, e^{-\rho t} \dot\rho^{(m)} e^{-\rho t}.
\end{equation}
For the symmetric generalized covariance, the exponential representation of a tangent vector coincides 
with the symmetric logarithmic derivative. We thus have that
\begin{equation}\label{Fisher metric}
	\gF(\dot\rho_1,\dot\rho_2)=\frac{1}{2}\Tr\big(\rho\,\{L_{\dot\rho_1}^{\textsc{s}},L_{\dot\rho_2}^{\textsc{s}}\}\big).
\end{equation}

Bogolubov's logarithmic derivative
of a tangent vector $\dot\rho$ at $\rho$ is  
\begin{equation}
	L^{\textsc{bkm}}_{\dot\rho}=\dd{t}\log\rho_t\big|_{t=0}
\end{equation}
where $\rho_t$ is any curve extending from $\rho$ with initial velocity $\dot\rho$. Then
\begin{equation}
	\dot\rho^{(m)} = \int_0^1  \d\lambda\, \rho^{\lambda} L_{\dot\rho}^{\textsc{bkm}} \rho^{1-\lambda}.
\end{equation}
The Bogolubov-Kubo-Mori metric is
\begin{equation}
	g_{\textsc{bkm}}(\dot\rho_1,\dot\rho_2)
	= \kappa^\textsc{bkm}_\rho\big(\dot\rho_1^{(e)},\dot\rho_2^{(e)}\big)
	= \int_0^1 d\lambda\Tr\big(\rho^{\lambda} L_{\dot\rho_1}^{\textsc{bkm}} \rho^{1-\lambda} L_{\dot\rho_2}^{\textsc{bkm}}\big).
\end{equation}

\section{Generally on notation and terminology}
We will henceforth not distinguish between a tangent vector 
and its mixture representation, unless it is necessary 
for clarity. We will thus skip the cumbersome ${}^{(m)}$-notation.
Moreover, relative to a generalized covariance we will write $L_{\dot\rho}$ rather than $\dot\rho^{(e)}$ for the exponential representation 
of a tangent vector $\dot\rho$, and we will call $L_{\dot\rho}$ the logarithmic derivative of $\dot\rho$. In fact, only the symmetric generalized covariance will play a role in this thesis and, hence,
$L_{\dot\rho}$ will denote the symmetric logarithmic derivative of 
$\dot\rho$.

By \emph{a curve} of operators we mean a piecewise smooth one-parameter family of 
operators.  For every curve $A_t$, the parameter is
assumed to take its values in an unspecified closed interval $0\leq t\leq \tau$,
unless otherwise is explicitly stated. (The parameter may, but need not, represent time.)
The operators $A_0$ and $A_\tau$ are the initial and final operators of the curve
and we say that the curve is closed if $A_0=A_\tau$.

We write $\dot A_{t}$ for the velocity vector of the curve at 
$t$. The velocity vector is well defined only at smooth values of $t$.
At every nonsmooth value of $t$---the number of which we assume 
is finite---we assume that unique unidirectional velocity vectors exist. 
For a closed curve we neither assume nor exclude
that the initial and final velocities are the same.

\chapter{Standard purification and holonomy}\label{ch: Stand pure}
A quantum state is called pure if it can be represented by a unit rank density operator. 
In quantum information theory,  
purification refers to the fact that every density operator 
can be thought of as representing the reduced state of some 
pure state.
Let $\HH^k$ be a $k$-dimensional Hilbert space.
One can show that for each density operator $\rho$ 
on $\HH$ which has rank at most $k$, there exists a unit vector $\ket{\psi}$ in 
$\HH\otimes\HH^k$ such that $\rho=\Tr_{\HH^k}\!\ketbra{\psi}{\psi}$.
An equivalent statement is that there exists a linear function $\psi$ from $\HH^k$ to $\HH$ such that $\rho=\psi\psi^\dagger$.
We will call such a linear function an amplitude for $\rho$. A key observation is that every amplitude
of $\rho$ has the form $\psi U$ where $U$ is a unitary operator on $\HH^k$.

\section{Standard purification}\label{sec: standard purification}
Let $\LL$ be the space of linear maps from $\HH^k$ to $\HH$ equipped with the Hilbert-Schmidt Hermitian product:
$\braket{\psi_1}{\psi_2} = \Tr(\psi_1^\dagger\psi_2)$.
Furthermore, let $\LL^k$ be the space of linear maps from 
$\HH^k$ to $\HH$ which have rank $k$ and let $\WW^k$ be the
space of linear maps $\psi$ in $\LL^k$ which have a unit Hilbert-Schmidt norm: 
\begin{equation}
	\WW^k = \big\{ \psi \in \LL^k : \Tr(\psi^\dagger\psi)=1 \big\}.
\end{equation}
The standard purification bundle over $\SS^k$ is 
the principal fiber bundle $\Pi$ from $\WW^k$ onto $\SS^k$ defined by 
\begin{equation}
	\p(\psi) = \psi\psi^\dagger.
\end{equation}
The symmetry group of $\Pi$ is $\UU(k)$, the group of unitary operators on $\HH^k$.
It acts from the right on $\WW^k$ by operator precomposition: $\R_U(\psi)=\psi U$.
We will in this thesis use terminology introduced by Uhlmann~\cite{Uh1986,Uh1991}
and call $\WW^k$ the purification space and the elements in $\WW^k$ amplitudes. 

\subsection{Vertical tangent vectors}
The tangent space of $\WW^k$ at $\psi$ consists of those linear operators in $\LL^k$ 
which are Hilbert-Schmidt perpendicular to $\psi$:
\begin{equation}
	\T_\psi\WW^k = \big\{ \dot\psi \in\LL^k : \Tr(\psi^\dagger \dot\psi + \dot\psi^\dagger\psi) = 0 \big\}.
\end{equation} 
Moreover, the differential $d\p$ of the standard purification bundle projection
sends a tangent vector $\dot\psi$ at $\psi$ to $\dot\psi\psi^\dagger + \psi \dot\psi^\dagger$.
We say that $\dot\psi$ is vertical if it is annihilated by $d\p$. This is equivalent to 
$\dot\psi=\psi\xi$ for some skew-Hermitian operator $\xi$ on $\HH^k$, i.e., an element in the Lie algebra $\uu(k)$ of $\UU(k)$.
We write $\V_\psi\WW^k$ for the space of all vertical tangent vectors at $\psi$:
\begin{equation}\label{eq: vert space}
	\V_\psi\WW^k = \{ \psi \xi : \xi\in\uu(k) \}.
\end{equation}
This is the vertical space at $\psi$.
The vertical space coincides with the 
tangent space at $\psi$ to the fiber over $\psi\psi^\dagger$.
Furthermore, the vertical spaces at all amplitudes combine to a smooth vector 
subbundle $\V\WW^k$ of the tangent bundle $\T\WW^k$ 
of $\WW^k$. We call this bundle the vertical bundle of $\p$.
The symmetry group action preserves the vertical bundle: 
\begin{equation}
	\V_\psi\WW^k \ni \psi \xi \xrightarrow{d\R_U} \psi U\Ad_{U^\dagger}\!\xi \in \V_{\psi U}\WW^k.
\end{equation}

\subsection{Horizontal tangent vectors}\label{sec: horiz tang vect}
A horizontal bundle is a smooth subbundle $\H\WW^k$ of $\T\WW^k$ 
which is everywhere complementary to the vertical bundle.
If $\H\WW^k$ is also preserved by the symmetry group action it is called 
an Ehresmann connection.
The horizontal bundle $\H\WW^k$ thus qualifies as an Ehresmann connection provided that 
$\T_\psi\WW^k = \V_\psi\WW^k \oplus \H_\psi\WW^k$ and $d\R_U(\H_\psi\WW^k) = \H_{\psi U}\WW^k$
for every amplitude $\psi$ and unitary $U$. 
The tangent vectors in $\H\WW^k$ are called horizontal vectors.

Ehresmann connections, which unlike the vertical bundle are neither unique nor canonical,
can be obtained from piecing together the kernels of a connection form.
A connection form on $\p$ is a smooth $\uu(k)$-valued differential form $\A$ on $\WW^k$ such that
$\A_\psi(\psi \xi) = \xi$ and $\A_{\psi U} d\R_U = \Ad_{U^\dagger}\!\A_\psi$
for all $\psi$ in $\WW^k$, $\xi$ in $\uu(k)$, and $U$ in $\UU(k)$.
The kernel bundle of $\A$ is an Ehresmann connection on $\p$, 
\begin{equation}
	\H_\psi\WW^k = \Ker\A_\psi = \{ \dot\psi\in\T_\psi\WW^k : \A_\psi(\dot\psi)=0 \}.
\end{equation}

Conversely, every 
Ehresmann connection is the kernel bundle of a connection form.
In fact, there is a one-to-one correspondence between Ehresmann connections and connection forms. 
Therefore, one usually does not distinguish between the two concepts and calls both 
Ehresmann connections and connection forms ``connections''. 

\subsubsection{Horizontal lifts and projected metrics}
The differential of $\p$ maps $\H_\psi\WW^k$ isomorphically onto 
$\T_\rho\SS^k$, where $\rho=\psi\psi^\dagger$. 
Therefore, if $\dot\rho$ is any tangent vector at $\rho$,
there is a unique horizontal vector $\dot\psi^h$ at $\psi$ such that $d\Pi(\dot\psi^h)=\dot\rho$. 
If $\dot\psi$ is any tangent vector at $\psi$ which gets projected onto $\dot\rho$,
the horizontal vector is 
\begin{equation}
	\dot\psi^h=\dot\psi - \psi\A(\dot\psi).
\end{equation}
The vector $\dot\psi^h$ is called the horizontal lift of $\dot\rho$ to $\psi$.
Since the connection is invariant under the symmetry group action,
each $d\R_U$ sends horizontal lifts to horizontal lifts. That is, 
if $\dot\psi^h$ is the horizontal lift of $\dot\rho$ at $\psi$, then $\dot\psi^h U$ is the horizontal lift of $\dot\rho$ at $\psi U$.

Suppose that the purification space carries a right invariant Riemannian metric $G$, i.e., a metric with respect to which the 
symmetry group acts by isometries.
We can then define a metric $g$ on $\SS^k$ as follows. For any pair of tangent 
vectors $\dot\rho_1, \dot\rho_2$ at $\rho$ let $\dot\psi_1^h, \dot\psi_2^h$ be the horizontal lifts to 
any amplitude of $\rho$. Then define 
\begin{equation}
	g(\dot\rho_1,\dot\rho_2) = G\big(\dot\psi_1^h,\dot\psi_2^h\big).
\end{equation}
This is a Riemannian metric on $\SS^k$ called the projection of $G$. 
We notice that if $G$ is also invariant under the left action by $\UU(\HH)$, i.e., $G$ is bi-invariant,
then the projection of $G$ is adjoint invariant: 
$g_{U\rho U^\dagger}(U\dot\rho_1U^\dagger,U\dot\rho_2U^\dagger) = g_\rho(\dot\rho_1,\dot\rho_2)$.

\subsection{The mechanical connection}\label{le mech con}
Let $G$ be a right invariant Riemannian metric on $\WW^k$.
We can then take the orthogonal complement of the vertical bundle as the horizontal bundle. 
The identity $d\R_U(\H_\psi\WW^k) = \H_{\psi U}\WW^k$ holds since $\R_U$ preserves the orthogonal direct sum.
The associated connection form is defined by the requirement that 
$\psi\A_\psi(\dot\psi)$ is the orthogonal projection of $\dot\psi$ onto $\V_\psi\WW^k$.
Alternatively, we can define the connection form using 
the moment of inertia tensor $\II$ and metric momentum $\JJ$.
These tensors are defined as follows.
Let $\uu(k)^*$ be the space of real-valued linear functions on $\uu(k)$ and 
at each amplitude $\psi$ define
\begin{subequations}
\begin{alignat}{2}
	\II_\psi&:\uu(k)\to \uu(k)^*,\quad &  &\II_{\psi}(\xi)\zeta=G(\psi\xi,\psi\zeta),\\
	\JJ_\psi&:\T_\psi\WW^k\to \uu(k)^*,\quad & &\JJ_{\psi}(\dot\psi)\zeta=G(\dot\psi,\psi\zeta).
\end{alignat}
\end{subequations}
Then $\A_{\psi}=\II_{\psi}^{-1}\JJ_{\psi}$.
We will adopt the terminology of classical mechanics and call $\A$
the mechanical connection associated with the right invariant metric~$G$.

\subsubsection{Horizontal lifts of geodesics}
Let $g$ be the projection of $G$
and let $\rho_t$ be a geodesic relative $g$ in $\SS^k$.
Recall that a characterizing property of geodesics is that they are made up of 
`shortest length segments'. This means that for any two close enough 
arguments $t_1 < t_2$, the segment $\rho_t$, $t_1\leq t\leq t_2$, is the 
unique shortest curve connecting $\rho_{t_1}$ to $\rho_{t_2}$
(up to reparametrization). 

Uniqueness of a shortest curve connecting two density operators is lost 
if the density operators are `far apart'. Nevertheless, a curve whose length equals the infimum 
of the lengths of all curves connecting the same two density operators is necessarily a geodesic. See, e.g, \cite[Sec.\,II, Prop.\,2.6]{Sa1996}.
It is this infimum which, by definition, is the geodesic distance between the two density operators.
We will write $\dist(\rho_1,\rho_2)$ (or $\dist_g(\rho_1,\rho_2)$) for the geodesic distance 
between the density operators $\rho_1$ and $\rho_2$. Next we will prove that geodesics in $\SS^k$ 
lift to horizontal geodesics in $\WW^k$. The proof relies on two observations about geodesics in $\WW^k$.

The first observation is that geodesics in $\WW^k$ have a conserved metric momentum. 
To see this assume that $\psi_t$ is a geodesic in $\WW^k$.
Then choose any $\xi$ in $\uu(k)$ and consider the variation 
$\psi_{\eps;t}=\psi_t\exp(\eps\xi)$. 
The curves $\psi_{\eps;t}$ all have the same speed since 
the symmetry group acts by isometries. Consequently,
\begin{equation}
\begin{split}
	\dd{t} \JJ_{\psi_{t}}(\dot{\psi}_{t})\xi
	&= G\big(\nabla_{\dot{\psi}_{t}}\dot{\psi}_{t},\psi_t\xi\big)	
		+ G\big(\dot{\psi}_{t},\nabla_{\dot{\psi}_{t}}\psi_t\xi\big)\\
	&= G\big(\dot{\psi}_{t},\nabla_{\psi_t\xi}\dot{\psi}_{t}\big)\\
	&=\frac{1}{2} \dd{\eps} G\big(\dot{\psi}_{\eps;t},\dot{\psi}_{\eps;t}\big)\Big|_{\eps=0}\\
	&=0.
\end{split}
\end{equation}
Here, $\nabla$ is the Levi-Civita connection of $G$,
see \cite[Sec.\,II]{Sa1996}.
It follows from this observation that if a geodesic extends perpendicularly from a fiber of $\Pi$,
then it will penetrate all the fibers it passes perpendicularly.
In other words, a geodesic which is horizontal at some point is horizontal everywhere.

The second observation is that a curve in $\WW^k$ which is a shortest 
curve between the fibers of its endpoints have to 
exit the initial fiber, and enter the final fiber, perpendicularly.
This follows immediately from \cite[Sec.\,III, Prop.\,2.4]{Sa1996}.
Notice that by `being shortest' we do not mean that it is unique but only that its 
length is minimal among the lengths of all the curves that connect the two fibers.
It now follows from the first observation that a shortest curve 
between two fibers of $\Pi$ is a horizontal geodesic.

Finally, let $\psi_t$ be a horizontal lift of a geodesic $\rho_t$.
Since $\rho_t$ can be partitioned into shortest length segments,
the lift admits a corresponding partition into segments which are 
shortest curves between the fibers containing respective segments endpoints.
The segments of $\psi_t$ are thus horizontal geodesics and, accordingly, 
so is $\psi_t$.

\begin{remark*}
We have seen that the metric momentum, in a sense, measures the 
angles by which a curve penetrates the fibers of $\Pi$.
One might wonder if the moment of inertia tensor has a geometric interpretation as well.
This is indeed the case. If we fix an amplitude $\psi$, then $\L_\psi(U)= \psi U$ is an embedding of the 
symmetry group onto the fiber of $\p$ containing $\psi$. 
We can pull back $G$ via this embedding to a metric $\L_\psi^* G$ on $\UU(k)$.
The pull-back metric is right invariant and, thus, is determined by its values on the Lie algebra 
of $\UU(k)$. On the Lie algebra, the pull-back metric is given by the moment of inertia tensor: $\L_\psi^* G(\xi_1,\xi_2) = \II_{\psi}(\xi_1)\xi_2$.
\end{remark*}

\subsection{The Hilbert-Schmidt metric}\label{sec: the HS metric}
A most important example (in this thesis \emph{the} most important example) 
of a bi-invariant Riemannian metric on $\WW^k$ is given by the real part of the Hilbert-Schmidt Hermitian product
\begin{equation}\label{ghs}
	\GHS(\dot\psi_1,\dot\psi_2)=\frac{1}{2}\Tr\big(\dot\psi_1^\dagger \dot\psi_2 + \dot\psi_2^\dagger \dot\psi_1\big).
\end{equation}
The orthogonal complement of the vertical bundle 
is an Ehresmann connection and the projection of $\GHS$ is adjoint invariant.
In the next section we will see that the projection 
of $\GHS$ on the space of faithful density operators is proportional to the quantum Fisher information metric. We will from now on refer to $\GHS$ as the Hilbert-Schmidt metric.

\subsubsection{Geodesics of the Hilbert-Schmidt metric}
The purification space $\WW^k$ is an open, dense subset of the unit sphere in $\LL^k$, and 
geodesics of the Hilbert-Schmidt metric on $\WW^k$ are `great arcs'. 
In other words, up to reparameterization, every geodesic $\psi_t$ in $\WW^k$
is of the form 
\begin{equation}\label{great arc}
	\psi_t = \frac{(\tau-t)\psi_0 + t\psi_\tau}{\|(\tau-t)\psi_0 + t\psi_\tau\|_{\textsc{hs}}}.
\end{equation}
The denominator is the Hilbert-Schmidt norm of the numerator,
i.e., the square root of the Hilbert-Schmidt product of the numerator with itself.
A straightforward calculation shows that the curve $\psi_t$ is horizontal according to the mechanical connection associated with 
the Hilbert-Schmidt metric if, and only if, $\psi_0^\dagger\psi_\tau=\psi_\tau^\dagger\psi_0$.

\subsubsection{The Hilbert-Schmidt symplectic form}
Had we defined $\GHS$ as four times the real part of the Hilbert-Schmidt product, the
projection onto the faithful density operators would have been exactly 
the quantum Fisher information metric \eqref{Fisher metric}.
The reader might wonder why we did not define $\GHS$ so that this relation holds. Well, the choice of proportionality factor made here is for `geometrical convenience'.
For example, when $\WW^k$ is equipped with the metric \eqref{ghs},
great circles have length $2\pi$, as in the Euclidian case, and the area of $\WW^k$ agrees with the Euclidean area of the $(2nk-1)$-dimensional unit sphere, namely $2\pi^{nk}/(nk-1)!$. 

Interestingly, another choice of proportionality factor is suggested by the most fundamental equation in quantum mechanics: 
Any positive multiple of the imaginary part of the Hilbert-Schmidt product is a symplectic 
form on $\LL^k$. If we select the symplectic form which is \emph{twice} the imaginary part,
\begin{equation}\label{ohs}
	\OHS(\dot\psi_1,\dot\psi_2)=-i\Tr\big(\dot\psi_1^\dagger \dot\psi_2 - \dot\psi_2^\dagger \dot\psi_1\big),
\end{equation}
then for every observable $A$, the flow lines of the Hamiltonian vector field associated with the expectation value function $E_\psi(A)=\Tr(\psi^\dagger A\psi)$ satisfy the Schr\"{o}dinger equation $i\dot\psi=A\psi$. Thus \eqref{ohs} seems like a natural choice of symplectic structure on $\LL^k$. Now, the metric 
which is compatible with $\OHS$, i.e., which together with $\OHS$ and $i$ form a 
K\"{a}hler structure (see \cite[Ch.\,8]{Na2003} and Section~\ref{sec: KKS}) is the one 
given by twice the real part of the Hilbert-Schmidt product.
We will refer to $\OHS$ as the Hilbert-Schmidt symplectic form.

\section{Parallel transport and holonomy}
\begin{figure}[t]
	\centering
	\includegraphics[width=0.9\textwidth]{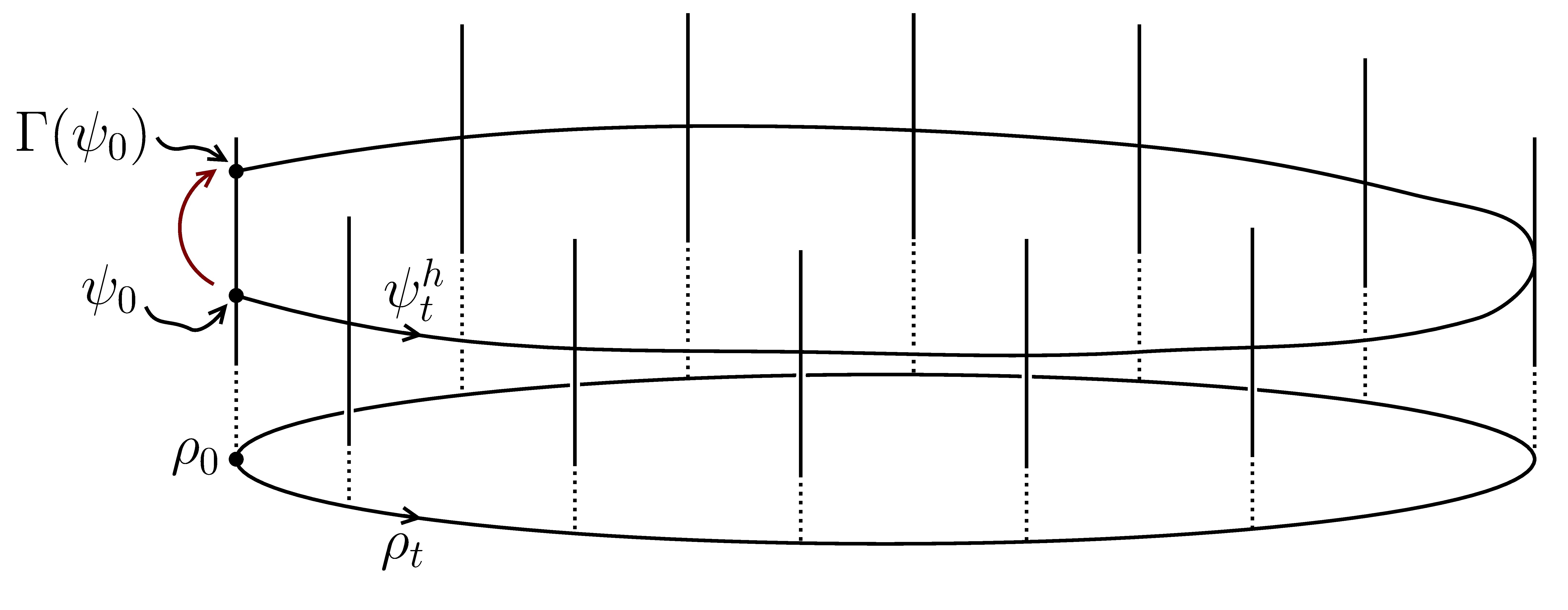}
	\caption{Parallel transport in the standard purification bundle.
	The parallel transport operator connects the initial and the final (and, in fact, all the intermediate) fibers of $\rho_t$ in a way that is determined by $\rho_t$ and the connection. Thus, it encodes geometric properties $\rho_t$.
	For a closed $\rho_t$, the holonomy at $\psi_0$ is defined by $\Gamma(\psi_0)=\psi_0 \Hol_{\psi_0}$.}
	\label{fig: horizontal lift of curve}
\end{figure}

If $\rho_t$ is a curve of density operators having rank $k$ 
and if $\psi_0$ is an amplitude of the initial density operator, there is a unique 
curve of amplitudes $\psi_t$ in $\WW^k$ which extends from $\psi_0$, projects onto $\rho_t$
and which is everywhere horizontal: $\dot{\psi}_t\in\H_{\psi_t}\WW^k$. 
See \cite[Ch.\,II, Sec.\,3, Prop.\,3.1]{KoNo1996}. This curve 
is the horizontal lift of $\rho_t$ extending from $\psi_0$.
The existence and uniqueness of horizontal lifts can be abstracted in an 
operator $\Gamma$ which sends each $\psi_0$ in the fiber over $\rho_0$
to the final amplitude in the horizontal lift of $\rho_t$ that extends from $\psi_0$,
see Figure~\ref{fig: horizontal lift of curve}.
The operator $\Gamma$ is the parallel transport operator associated with $\rho_t$. (We will write $\Gamma[\rho_t]$ when we want to emphasize that $\Gamma$ depends on $\rho_t$.)
The parallel transport operator commutes with the symmetry group action: 
\begin{equation}
	\Gamma\R_U = \R_U \Gamma.
\end{equation}
This is so because $\R_U$ sends each horizontal lift of $\rho_t$ 
to another horizontal lift of $\rho_t$.

If $\rho_t$ is closed, the parallel transport operator maps the fiber over $\rho_0$
onto itself. We define the holonomy of $\rho_t$ based at the amplitude $\psi_0$ 
to be the unitary $\Hol_{\psi_0}$ in $\UU(k)$ defined by
$\Gamma(\psi_0)=\psi_0 \Hol_{\psi_0}$, see Figure~\ref{fig: horizontal lift of curve}.
As indicated, the holonomy depends on the initial amplitude.
(Of course, the holonomy also depends on the curve $\rho_t$. And when we want to emphasize this, or distinguish it from the holonomy of another curve, we will write $\Hol_{\psi_0}[\rho_t]$.) Moreover, the holonomy transforms 
covariantly under changes of initial amplitude:
\begin{equation}
	\Hol_{\psi_0 U}= U^\dagger \Hol_{\psi_0} U.
\end{equation} 

Given that we know the connection form and that we have an explicit 
lift, we can directly write down a horizontal lift; if $\psi_t$ is a lift 
of $\rho_t$, then
\begin{equation}\label{hor push}
	\psi^h_{t}=\psi_t \texp\Big(\!-\!\int_0^t\dt\, \A(\dot\psi_t)\Big)
\end{equation}
is a horizontal lift of $\rho_t$. Here $\texp$ is the positively time-ordered exponential.
The parallel transport of the initial amplitude is, thus,
\begin{equation}\label{pto}
	\Gamma(\psi_0)=\psi_\tau\texp\Big(\!-\!\int_0^\tau\dt\, \A(\dot\psi_t)\Big).
\end{equation}
And the holonomy is 
\begin{equation}\label{the holonomy}
	\Hol_{\psi_0}=(\psi_0^\dagger\psi_0)^{-1}\psi_0^\dagger\psi_\tau \texp\Big(\!-\!\int_0^\tau\dt\, \A(\dot\psi_t)\Big).
\end{equation}

Since the holonomy is not invariant under changes of initial amplitude,
it is not measurable, even indirectly. However, by applying an adjoint invariant function to the holonomy one can assign a quantity to a closed curve of density operators which is measurable, at least in principle.
Examples of such functions are the trace and the determinant;
the scalars $\Tr \Hol_{\psi_0}$ and $\det \Hol_{\psi_0}$ do not depend on the initial amplitude $\psi_0$. Next we will introduce another 
quantity related to holonomy which does not depend on the initial amplitude, namely the geometric phase. 

\subsection{Geometric phase}\label{sec: geometric phase}
The geometric phase factor of a curve of density operators $\rho_t$ is 
\begin{equation}\label{gmtrc phase}
	\Geop[\rho_t] 
	= \Tr(\psi_0^\dagger\Gamma(\psi_0)).
\end{equation}
Here, $\psi_0$ is any amplitude of the initial density operator.
The argument of the geometric phase factor is called the geometric phase of $\rho_t$:
\begin{equation}
	\tgeo[\rho_t] = \arg\Geop[\rho_t].
\end{equation} 
(We will often leave out the reference to the curve and simply write $\Geop$ and $\tgeo$ for the geometric phase factor and geometric phase of $\rho_t$.) 
The geometric phase is only defined if the geometric phase factor does not vanish, and then only up to addition of integer multiples of $2\pi$.
If $\rho_t$ is a closed curve, the geometric phase factor and the geometric phase can be expressed in terms of the holonomy of the curve:
\begin{subequations}
\begin{align}
	\Geop &= \Tr(\psi_0^\dagger\psi_0\Hol_{\psi_0}),\label{eq: geometric phase factor}\\
	\tgeo &= \arg\Tr(\psi_0^\dagger\psi_0\Hol_{\psi_0}).\label{eq: geometric phase}
\end{align}
\end{subequations}
Obviously, the geometric phase factor contains less information about the curve $\rho_t$
than the holonomy. On the other hand, it is invariant under changes of initial amplitude, and, therefore, is potentially measurable, if only indirectly. The literature on geometric phase and its usage in physics is overwhelming. Some collective references
of relevance for the current exposition are \cite{ShWi1989,BoMoKoNiZw2003,ChJa2004}.

\subsection{The holonomy group}
Let $\rho_t$ and $\sigma_t$ be two closed curves in $\SS^k$ with
a common initial density operator $\rho_0=\sigma_0$.
The product of $\rho_t$ and $\sigma_t$ is the curve $(\rho\ast\sigma)_t$ defined as
\begin{equation}
	(\rho\ast\sigma)_t = 
	\begin{cases} 
		\rho_{2t} & 0\leq t\leq \tau/2,\\
		\sigma_{2t-\tau} & \tau/2\leq t\leq \tau.
	\end{cases}
\end{equation}
The product curve is also a closed curve at $\rho_0=\sigma_0$
and, hence, its parallel transport operator sends the fiber over the common initial density operator onto itself. Uniqueness of solutions to ordinary differential equations implies that
$\Gamma[(\rho\ast\sigma)_t]=\Gamma[\sigma_t]\Gamma[\rho_t]$.
Then, commutativity of the parallel translation operators with the 
right action of the symmetry group yields
\begin{equation}\label{group property}
	\Hol_{\psi_0}[(\rho\ast\sigma)_t]=
	\Hol_{\psi_0}[\sigma_t]\Hol_{\psi_0}[\rho_t].
\end{equation}
The group of all unitaries that can be realized as holonomies at $\psi_0$ of closed curves at $\psi_0\psi_0^\dagger$ is the holonomy group at $\psi_0$. We denote this group $\operatorname{Hol}(\psi_0)$.
The holonomy group is a subgroup of $\UU(k)$.
A shift of the initial amplitude changes the holonomy group 
into a conjugate subgroup of $\UU(k)$. 

\section{Complementary purification}\label{comp pur}
For any amplitude $\psi$ in $\WW^k$, the product $\psi^\dagger\psi$ is 
a faithful density operator on $\HH^k$.
Let $\SS^c$ be the space of faithful density operators on $\HH^k$.
The comple\-mentary purification bundle is the principal fiber bundle 
$\p^c$ from $\WW^k$ onto $\SS^c$ defined by 
\begin{equation}
	\p^c(\psi)=\psi^\dagger\psi.
\end{equation}
The symmetry group of $\p^c$ is $\UU(\HH)$, the group  of unitary operators on $\HH$,
which acts from the left on $\WW^k$ by operator postcomposition, $\L_U(\psi)=U\psi$.
The complementary purification bundle will play different roles 
in subsequent chapters.
In Chapter~\ref{monotone geometries} we will analyze the geometry of the complementary purification bundle to a considerable extent
in the case of faithful density operators.
Here we restrict ourselves to introducing some notation which will be useful in Chapter~\ref{ch: sjoqvist holonomy}.

Let $\bfp$ be a weight spectrum of length $k$ for density operators on $\HH$
and let $\bfP=(P_1,P_2,\dots,P_l)$ and $\bfm=(m_1,m_2,\dots,m_l)$
be the corresponding eigenvalue and degeneracy spectrum, respectively.
An eigenprojector spectrum $\bfl=(\lambda_1,\lambda_2,\dots,\lambda_l)$ for 
density operators on $\HH^k$ is called compatible with $\bfp$ and $\bfm$
if the rank of each $\lambda_a$ equals the degeneracy $m_a$.
Given such a compatible eigenprojector spectrum we write 
$\WW_{\bfp;\bfl}$
for the fiber of $\p^c$ over $\sum_aP^a\lambda_a$. That is,
\begin{equation}
	\WW_{\bfp;\bfl}=\big\{\psi \in\WW^k : \psi^\dagger\psi =  \sum_aP^a\lambda_a\big\}.
\end{equation}
A key observation which will be investigated in greater detail in Chapter~\ref{ch: sjoqvist holonomy},
and which will culminate in a holonomy theory for isospectral mixed states, is that
the standard purification bundle restricts to a fiber bundle $\p_{\bfp;\bfl}$ from $\WW_{\bfp;\bfl}$
onto $\SS_{\bfp}$. The symmetry group of $\p_{\bfp;\bfl}$ is the 
group $\UU_{\bfl}$ of all the unitary operators on $\HH^k$ which commute with the projections in $\bfl$.

By varying the eigenprojector spectrum $\bfl$ we obtain different but isomorphic
bundles over the unitary orbit $\SS_{\bfp}$.
One might wonder what happens if we keep the eigenprojector spectrum $\bfl$ fixed but vary the 
eigenvalue spectrum among the ones compatible with $\bfl$. One then obtains a fiber bundle over 
the space $\SS_{\bfm}$ of all the density operators which have degeneracy spectrum $\bfm$.
More precisely, if $\SS_{\bfl}^c$ is the space of density operators on $\HH^k$ which have 
eigenprojector spectrum $\bfl$, and $\WW_{\bfl}$ is the preimage of $\SS^c_{\bfl}$ 
under the under the complementary bundle, i.e.,
\begin{equation}
	\WW_{\bfl}=\big\{\psi \in\WW^k : \psi^\dagger\psi \in \SS^c_{\bfl}\big\},
\end{equation} 
the standard purification bundle restricts to a principal fiber bundle $\p_{\bfl}$
from $\WW_{\bfl}$ onto $\SS_{\bfm}$. The symmetry group of $\p_{\bfl}$ is again $\UU_{\bfl}$. 

\chapter{Monotone geometries}\label{monotone geometries}
In quantum mechanics, the admissible transformations of states
are called quantum channels \cite{Ha2006}. Examples of quantum channels are unitary transformations, partial traces, and state reductions due to measurements.
The formal definition of a quantum channel is the following.
Let $\HH_1$ and $\HH_2$ be Hilbert spaces.
A quantum channel $\EE$ from $\HH_1$ to $\HH_2$ is a linear map from the space of operators on $\HH_1$ to the space of operators on $\HH_2$ which is 
completely positive and trace preserving.
Complete positivity is the requirement that for every Hilbert space $\HH$ 
and every positive operator $A$ on $\HH_1\otimes \HH$, the operator 
$(\EE\otimes\1)(A)$ is a positive operator on $\HH_2\otimes \HH$. 
That $\EE$ is trace preserving means that $\Tr\EE(B)=\Tr B$ for every operator 
$B$ on $\HH_1$. These two properties guarantee that $\EE$ takes density operators on $\HH_1$ to density operators $\HH_2$. 

In quantum mechanics it is assumed that a quantum channel cannot 
make two quantum states more distinguishable.
This is equivalent to the assumption that information processing cannot increase the amount of information \cite{Ha2006}.
Thus, when a system passes through a channel, information about its state can be lost but not gained.
Therefore, a distance function which is assumed to measure the 
distinguishability between two states must be monotone.
A distance function $\dist$ on density operators
is monotone if for every quantum channel $\EE$,
$\dist(\EE(\rho_1),\EE(\rho_2))\leq \dist(\rho_1,\rho_2)$
whenever the two sides of the inequality makes sense.
If the distance function comes from a Riemannian metric $g$,
the distance function is monotone if, and only if, $g$ is monotone.
That is, if $g(d\EE(\dot\rho_1),d\EE(\dot\rho_2))\leq g(\dot\rho_1,\dot\rho_2)$.
We call the geometry of a monotone metric a monotone geometry.
It turns out that every monotone metric on the space faithful density operators $\SS^n$ can be obtained as the 
projection of the Hilbert-Schmidt metric relative an appropriate
choice of connection in the standard purification bundle \cite{DiRu1992,Pe1996,DiUh1999}. Here we will review this result.
Special attention will be given to the three most important monotone geometries. In this chapter we restrict our study to the space of faithful density operators. Moreover, we assume that the purification amplitudes have domain and support in the same Hilbert space.

\section{Bures geometry}
The quantum Fisher information metric \eqref{Fisher metric} is the most important metric in quantum information geometry.
The quantum Fisher information metric appears, e.g., in the fundamental quantum Cramer-Rao bound:
Consider a system whose state depends on an unknown, continuous parameter $\lambda$
whose value we want to infer by measuring a suitable observable---an ``estimator''. 
A theorem in quantum estimation theory states that the 
variance of the estimator 
is lower bounded by the quantum Cramer-Rao bound $1/(M\F(\lambda))$, see, e.g., \cite[Ch.\,VIII, Sec.\,4]{He1976}.
$M$ is the number of times the measurement is repeated 
and $\F(\lambda)$ is the quantum Fisher information.
If the state of the system is $\rho_\lambda$, the 
quantum Fisher information is the squared speed of $\rho_\lambda$
in the quantum Fisher information geometry:
\begin{equation}
	\F(\lambda)=\gF(\dot\rho_\lambda,\dot\rho_\lambda).
\end{equation}

In Section \ref{sec: the HS metric} we mentioned that the 
quantum Fisher information metric is proportional to the projection of $\GHS$ when the connection is the mechanical connection of $\GHS$.
The projection is called the Bures metric (which, thus, is essentially the same as the quantum Fisher information metric). In this section we will review some basic properties of the Bures geometry.

\subsection{The Bures metric}
Let the connection of the standard purification bundle be 
the mechanical connection associated with the Hilbert-Schmidt metric $\GHS$.
That is, let the horizontal bundle be the orthogonal complement of the vertical bundle
with respect to $\GHS$. Then, a tangent vector $\dot\psi$ at $\psi$ is horizontal if and only if 
\begin{equation}\label{eq: Bures horizontal}
	\dot\psi^\dagger\psi=\psi^\dagger \dot\psi.
\end{equation}
This follows immediately from the requirement that 
$\GHS(\dot\psi,\psi\xi)=0$ for every skew-Hermitian operator $\xi$.
Every solution of Eq.~\eqref{eq: Bures horizontal} is of the form 
$\dot\psi=\frac{1}{2}L\psi$ where $L$ is a Hermitian operator.
Furthermore,
$L$ must have a vanishing expectation value at
$\rho=\psi\psi^\dagger$ for otherwise $\dot\psi$ will not be 
tangential to $\WW^n$. The projection of $\dot\psi$ is
\begin{equation}\label{proj SLD}
	d\Pi(\dot\psi)=\dot\psi\psi^\dagger + \psi \dot\psi^\dagger = \frac{1}{2}\{\rho,L\}.
\end{equation}
Hence, $L$ is the symmetric logarithmic derivative of  $d\p(X)$,
c.f., Eq.~\eqref{SLD}.

Let $\gB$ be the projection of $\GHS$.
According to Eq.~\eqref{proj SLD}, the horizontal lift of a tangent vector $\dot\rho$ at $\rho$ to the amplitude $\psi$ is $\frac{1}{2}L_{\dot\rho}\psi$ where $L_{\dot\rho}$
is the symmetric logarithmic derivative of $\dot\rho$. Consequently,
\begin{equation}\label{eq: bures metric}
	\gB(\dot\rho_1,\dot\rho_2)
	=\GHS\big(\tfrac{1}{2}L_{\dot\rho_1}\psi,\tfrac{1}{2}L_{\dot\rho_2}\psi\big)
	=\frac{1}{8}\Tr\big(\rho\,\{L_{\dot\rho_1},L_{\dot\rho_2}\}\big).
\end{equation}
A comparison with Eq.~\eqref{Fisher metric} tells us that
the Bures metric is one quarter of the quantum Fisher information metric.
By Eqs.~\eqref{tjosan} and \eqref{SLD}, alternative formulas for the Bures metric are
\begin{equation}\label{alt bures}
	\gB(\dot\rho_1,\dot\rho_2)
	=\frac{1}{4}\Tr(\dot\rho_1 L_{\dot\rho_2})
	=\frac{1}{2}\Tr\big(\dot\rho_1 (\L_\rho+\R_\rho)^{-1}\dot\rho_2\big).
\end{equation}
Since the Bures and the quantum Fisher information metrics are proportional, the corresponding geometries are essentially the same. 

The commutative manifold and the unitary orbit of a density operator $\rho$ are perpendicular relative to the Bures metric. 
To see this let $\dot\rho_c$ and $\dot\rho_u$
be tangent vectors in $\T^c_\rho\SS^n$ and $\T^u_\rho\SS^n$, respectively. According to Eq.~\eqref{classical log}, the symmetric logarithmic derivative of $\dot\rho_c$ is $\dot\rho_c\rho^{-1}$ and, hence, the horizonal lift of 
$\dot\rho_c$ to $\sqrt{\rho}$ is $\frac{1}{2}\dot\rho \rho^{-1/2}$.
Furthermore, by Eq.~\eqref{hor push}, writing $\dot\rho_u=-i[A,\rho]$ for some Hermitian $A$, the horizontal lift of $\dot\rho_u$ to $\sqrt{\rho}$ is 
$-iA\sqrt{\rho}-\sqrt{\rho}\AB(-iA\sqrt{\rho}\,)$. A straightforward calculation shows that 
\begin{equation}
	\GHS\big(\tfrac{1}{2}\dot\rho \rho^{-1/2},-iA\sqrt{\rho}-\sqrt{\rho}\AB(-iA\sqrt{\rho}\,)\big)=0.
\end{equation}

\subsubsection{Bures distance}
In Section \ref{sec: the HS metric} we proved that
if the standard purification bundle is equipped with the mechanical connection of 
some right invariant metric, the geodesics of the projected metric lift to horizontal 
geodesics in the purification space. This result applies to the situation considered here; Bures geodesics lift to horizontal Hilbert-Schmidt geodesics. 

Recall that a horizontal Hilbert-Schmidt geodesic in $\WW^n$ has, possibly after a reparameterization, the form in Eq.~\eqref{great arc}
for some amplitudes $\psi_0$ and $\psi_\tau$ which satisfy $ \psi_0^\dagger\psi_\tau=\psi_\tau^\dagger\psi_0$.
The shortest geodesic connecting the fibers over $\rho_0$ and $\rho_\tau$ must be such that $\|\psi_0-\psi_\tau\|_{\textsc{hs}}^2=2-2\Tr(\psi_0^\dagger\psi_\tau)$ is minimal.
The minimal norm is attained if
\begin{equation}\label{extern}
	\Tr(\psi_0^\dagger\psi_\tau)=|\Tr(\psi_0^\dagger\psi_\tau)|=\max_U\big\{|\Tr(\psi_0^\dagger\psi_\tau U)|\big\}.
\end{equation}
The maximum is taken over all unitaries on $\HH$.
A theorem of Uhlmann \cite{Uh1976} states that the right hand side of Eq.~\eqref{extern}
equals the fidelity of $\rho_0$ and $\rho_\tau$:
\begin{equation}\label{fidelity}
	F(\rho_0,\rho_\tau)=\| \sqrt{\rho_0} \sqrt{\rho_\tau} \|_{\Tr}=\Tr\sqrt{\sqrt{\rho_\tau}\rho_0\sqrt{\rho_\tau}}.
\end{equation}
The norm in this definition is the trace norm, defined by $\| A \|_{\Tr}=\Tr\sqrt{A^\dagger A}$.
We can now write down a formula for the Bures distance between $\rho_0$ and $\rho_\tau$:
\begin{equation}\label{Buresavstand}
\distB(\rho_0,\rho_\tau)=\arccos\Re\Tr(\psi_0^\dagger\psi_\tau)=\arccos F(\rho_0,\rho_1).
\end{equation}

\subsection{Uhlmann holonomy and geometric phase}
Let $\dot\psi$ be a tangent vector at $\psi$ in $\WW^n$.
The tangent vector has a unique, orthogonal decomposition $\dot\psi=\frac{1}{2}L\psi+\psi\xi$ where
where $L$ is Hermitian and $\xi$ is skew-Hermitian. The first term is the 
horizontal part of $\dot\psi$ and the second term is the vertical part. 
The Bures connection form 
is the $\uu(\HH)$-valued one-form $\AB$ defined by $\AB(\dot\psi)=\xi$,
and the operator $L$ is the symmetric logarithmic derivative of the projection 
$d\Pi(\dot\psi)=\dot\psi\psi^\dagger+\psi \dot\psi^\dagger$. Thus, by Eq.~\eqref{symlog},
\begin{equation}
	\AB(\dot\psi)
	=\psi^{-1}\Big(\dot\psi-\int_0^\infty\dt\, e^{-t\psi\psi^\dagger}\big(\dot\psi\psi^\dagger+\psi \dot\psi^\dagger\big)e^{-t\psi\psi^\dagger}\psi\Big).
\end{equation}
An alternative formula can be derived from the observation
\begin{equation}\label{con}
	\{\psi^\dagger\psi,\AB(\dot\psi)\}=\psi^\dagger \dot\psi-\dot\psi^\dagger\psi.
\end{equation}
The superoperator $\L_{\psi^\dagger\psi}+\R_{\psi^\dagger\psi}$ is invertible and, hence,
\begin{equation}\label{Burres}
\begin{split}
	\AB(\dot\psi)
	&=(\L_{\psi^\dagger\psi}+\R_{\psi^\dagger\psi})^{-1}(\psi^\dagger \dot\psi-\dot\psi^\dagger\psi)\\
	&=(\L_{\psi^\dagger\psi}+\R_{\psi^\dagger\psi})^{-1} 
	\Big(\L_{\psi^\dagger\psi}\big(\psi^{-1} \dot\psi\big) 
	- \R_{\psi^\dagger\psi}\big((\psi^{-1}\dot\psi)^\dagger\big)\Big).
\end{split}
\end{equation}

Let $\rho_t$ be a curve of faithful density operators.
According to Eq.~\eqref{pto}, the parallel transport operator associated with $\rho_t$ is 
\begin{equation}
	\Gamma(\psi_0)=\psi_\tau\texp\Big(\!-\!\int_0^\tau\dt\, \AB(\dot{\psi}_t)\Big),
\end{equation}
where $\psi_t$ is any lift of $\rho_t$ which extends from $\psi_0$.
If $\rho_t$ is closed, the holonomy of $\rho_t$ at $\psi_0$ is thus given by
\begin{equation}\label{Uh holonomy}
	\Hol_{\psi_0}^{\textsc{uh}}
	=\psi_0^{-1}\psi_\tau\texp\Big(\!-\!\int_0^\tau\dt\, \AB(\dot{\psi}_t)\Big).
\end{equation}
We will call this the Uhlmann holonomy of $\rho_t$ \cite{Uh1986,Uh1991}.
Hence the superscript ``UH''.
The Uhlmann geometric phase factor and geometric phase are \cite{Uh1989,Uh1995}
\begin{subequations}
\begin{align}
	\Geop^{\textsc{uh}}
	&= \Tr\big(\psi_0^\dagger\Gamma(\psi_0)\big)
	= \Tr\big(\psi_0^\dagger\psi_0\Hol_{\psi_0}^{\textsc{uh}}\big),\\
	\tgeo^{\textsc{uh}}
	&= \arg\Geop^{\textsc{uh}}
	= \arg\Tr\big(\psi_0^\dagger\psi_0\Hol_{\psi_0}^{\textsc{uh}}\big).
\end{align}
\end{subequations}

\subsubsection{H\"{u}bner's formula}
The curve $\rho_t$ has a canonical closed lift to $\WW^n$, namely the lift which at each instant $t$ is given by the square root of $\rho_t$.
If we expand $\rho_t$ as an incoherent orthonormal ensemble of pure states
$\rho_t=\sum_a p^a_t \ketbra{\psi_{a;t}}{\psi_{a;t}}$,
the square root lift has the expansion
$\sqrt{\rho_t}=\sum_a \sqrt{p^a_t} \ketbra{\psi_{a;t}}{\psi_{a;t}}$.
Then, according to Eq.~\eqref{con},
\begin{equation}
\begin{split}
	\bra{\psi_{a;t}}\{\rho_t,\AB(\tfrac{\d}{\d t}\sqrt{\rho_t}\,)\} \ket{\psi_{b;t}}
	= \bra{\psi_{a;t}}[\sqrt{\rho_t},\tfrac{\d}{\d t}\sqrt{\rho_t}\,]\ket{\psi_{b;t}}.
\end{split}
\end{equation}
From this we can deduce H\"{u}bner's formula \cite{Hu1993}:
\begin{equation}\label{hubner}
	\AB\big(\tfrac{\d}{\d t}\sqrt{\rho_t}\,\big)=\sum_{a,b}\frac{1}{p^a_t+p^b_t}\ketbra{\psi_{a;t}}{\psi_{a;t}}[\sqrt{\rho_t},\tfrac{\d}{\d t}\sqrt{\rho_t}\,]\ketbra{\psi_{b;t}}{\psi_{b;t}}.
\end{equation}

\subsection{Uhlmann's quantum speed limit}\label{UhQSL}
Consider a curve of faithful density operators $\rho_t$ generated by a Hamiltonian:\!
\begin{equation}\label{vonNeumann}
 	i\dot\rho_t=[H,\rho_t].
\end{equation}
We can lift $\rho_t$ to a solution $\psi_t$ satisfying the Schr\"{o}dinger equation $i\dot\psi_t=H\psi_t$.
The solution is not Bures horizontal. For if we plug the velocity of $\psi_t$ into the horizontality equation
\eqref{eq: Bures horizontal}, it reduces to $2i\psi^\dagger_t H\psi_t = 0$ which, since $\psi_t$ is invertible, implies that $H=0$.
According to Eq.~\eqref{hor push}, a horizontal lift of $\rho_t$ is given by
\begin{equation}\label{denna}
	\psi_{t}^h=\psi_t\texp\Big(\!-\!\int_0^t\dt\,\AB(\dot\psi_t)\Big).
\end{equation}
We define $\sigma_t$ to be the complementary projection of the horizontal lift, i.e.,
$\sigma_t=\psi_{t}^{h\dagger}\psi_{t}^h$. We further define the Hermitian operator $A$ as
\begin{equation}
	A=i\texp\Big(\!-\!\int_0^t\dt\,\AB(\dot\psi_t)\Big)^\dagger\AB(\dot\psi_t)\texp\Big(\!-\!\int_0^t\dt\,\AB(\dot\psi_t)\Big).
\end{equation}
Differentiation of Eq.~\eqref{denna} then yields $i\dot\psi_{t}^h =(H\psi_{t}^h - \psi_{t}^h A)$. Furthermore, 
since the velocity field of the horizontal lift satisfies Eq.~\eqref{eq: Bures horizontal}, 
\begin{equation}\label{singsing}
	2\psi_{t}^{h\dagger} H \psi_{t}^h = \{\sigma_t,A\}.
\end{equation}
It follows that
\begin{equation}
\begin{split}
	\Tr(\dot\psi_{t}^{h\dagger} \dot\psi_{t}^h) 
	&= \Tr(\rho_t H^2)+ \Tr(\sigma_t A^2) - \Tr(2\psi_{t}^{h\dagger} H \psi_{t}^h A)\\
	&= \Tr(\rho_t H^2) - \Tr(\sigma_t A^2).
\end{split}
\end{equation}
Note that the left hand side is the square of the Bures speed of $\rho_t$.
Taking the trace of both sides of \eqref{singsing} shows that $\Tr(\rho_t H)=\Tr(\sigma_t A)$.
Thus,
\begin{equation}\label{uppskattning}
\begin{split}
	\var_{\rho_t}(H) 
	&=  \Tr(\rho_t H^2) - \Tr(\rho_t H)^2\\
	&= \gB(\dot\rho_t,\dot\rho_t) + \Tr(\sigma_t A^2) - \Tr(\sigma_t A)^2\\
	&\geq \gB(\dot\rho_t,\dot\rho_t).
\end{split}
\end{equation}
We conclude that 
\begin{equation}
	\distB(\rho_0,\rho_\tau)
	\leq \int_0^\tau \dt \sqrt{\gB(\dot\rho_t,\dot\rho_t)}
	\leq \tau \Delta E,
\end{equation}
where $\Delta E$ is the time-average of the energy uncertainty along $\rho_t$.
The inequality 
\begin{equation}\label{Uhqsl}
	\tau\geq \frac{\distB(\rho_0,\rho_\tau)}{\Delta E}
\end{equation}
is ``the Uhlmann quantum speed limit'' \cite{Uh1992}.

\section{Wigner-Yanase geometry}
The standard purification bundle over the faithful density operators 
is trivial; a trivialization from $\SS^n\times \UU(\HH)$ onto $\WW^n$ is given by
$(\rho,U)\to\sqrt{\rho} U$. The domain of the trivialization is the total space of the trivial bundle $(\rho,U)\to\rho$. 
The tangent bundle of $\SS^n\times \UU(\HH)$ canonically splits into the direct sum of $\T\SS^n$ and $\T\UU(\HH)$. The latter is the vertical bundle of the trivial bundle projection. We can then take $\T\SS^n$ as the horizontal bundle. This bundle is an Ehresmann connection. The image of $\T\SS^n$ under the differential of the trivialization is thus a connection in the standard purification bundle. For reasons that will soon be clear we call this connection the Wigner-Yanase connection.

The section $\rho\to(\rho,\1)$ is a global horizontal lift of $\SS^n$ into the total space of the trivial bundle. Therefore, the square root section $s(\rho)=\sqrt{\rho}$ is a global horizontal lift of $\SS^n$ into $\WW^n$. This means that if $\rho_t$ is a curve of faithful density operators, then $\sqrt{\rho_t}$ is a horizontal lift of $\rho_t$.
Since $\sqrt{\rho_t}$ is a closed curve if $\rho_t$ is a closed curve, the holonomy groups of the Wigner-Yanase connection are all trivial.
Nevertheless, we can obtain an interesting geometry on $\SS^n$ by projecting down the Hilbert-Schmidt metric.

\subsection{The Wigner-Yanase metric}
The Wigner-Yanase metric $\gWY$ on $\SS^n$ is the projection of the Hilbert-Schmidt metric relative to the Wigner-Yanase connection. The horizontal lifts of the tangent vectors $\dot\rho_1$ and $\dot\rho_2$ at $\rho$ to $\sqrt{\rho}$ are $ds(\dot\rho_1)$ and $ds(\dot\rho_2)$, respectively.
We hence have that 
\begin{equation}\label{WY pullback}
	\gWY(\dot\rho_1,\dot\rho_2)=\GHS(ds(\dot\rho_1),ds(\dot\rho_2)).
\end{equation}

Recall that the tangent space at $\rho$ splits into the direct sum of the tangent spaces of the unitary orbit and the commutative manifold of $\rho$. For a tangent vector $\dot\rho_c$ which is tangential to the commutative manifold at $\rho$ we have that $ds(\dot\rho_c)=\frac{1}{2}\dot\rho_c\rho^{-1/2}$ and, hence, that
\begin{equation}\label{classical WY}
	\gWY(\dot\rho_c,\dot\rho_c)
	=\frac{1}{4}\Tr\big((\dot\rho_c)^2\rho^{-1}\big)
	=\frac{1}{4}\Tr\big(\dot\rho_c L_{\dot\rho^c}\big).
\end{equation}
If $\dot\rho_u$ is tangential to the unitary orbit at $\rho$, and $\dot\rho_u=-i[A,\rho]$ with $A$ Hermitian, then $ds(\dot\rho_u)=-i[A,\sqrt{\rho}\,]$. Hence,  
\begin{equation}\label{unitary WY}
	\gWY(\dot\rho_u,\dot\rho_u)
	=-\Tr\big([A,\sqrt{\rho}\,]^2\big).
\end{equation}
The tangent spaces of the commutative manifold and the unitary orbit are perpendicular relative $\gWY$. 
This follows immediately from the observation
\begin{equation}
	\GHS\big(\tfrac{1}{2}\dot\rho \rho^{-1/2},-i[A,\sqrt{\rho}\,]\big)=0.
\end{equation}
Consequently, if we decompose a tangent vector $\dot\rho$ at $\rho$ into its commutative and orbital parts, $\dot\rho=\dot\rho_c+\dot\rho_u$, and $\dot\rho_u=-i[A,\rho]$, we find that
\begin{equation}\label{Why}
	\gWY(\dot\rho,\dot\rho)
	=\frac{1}{4}\Tr\big(\dot\rho_c L_{\dot\rho_c}\big)
	-\Tr\big([A,\sqrt{\rho}\,]^2\big).
\end{equation}
We recognize the second term in this expansion as twice the Wigner-Yanase skew-information of $\rho$ with respect to the observable 
represented by $A$. Hence the name ``Wigner-Yanase metric''.
The Wigner-Yanase skew-information was introduced by Wigner and Yanase \cite{WiYa1963}
as a measure of the noncommutativity of $\rho$ and $A$.
We write $\IWY(\rho,A)$ 
for the Wigner-Yanase skew-information:
\begin{equation}
	\IWY(\rho,A)
	=-\frac{1}{2}\Tr\big([A,\sqrt{\rho}\,]^2\big)
	=\Tr(\rho A^2)-\Tr(\sqrt{\rho}A\sqrt{\rho}A).
\end{equation}

\subsubsection{The Wigner-Yanase distance function}
Equation \eqref{WY pullback} says that the Wigner-Yanase metric is the pull back of the Hilbert-Schmidt metric by the square root section.
Therefore, the geodesic distance between $\rho_0$ and $\rho_\tau$
is the infimum of the Hilbert-Schmidt lengths of curves 
\emph{in} the image of the square root section which extend from $\sqrt{\rho_0}$ to  $\sqrt{\rho_\tau}$.  In general, this type of `constrained' variational 
problem is difficult to solve. But, luckily, our knowledge of spherical geometry helps us at this point.  

We know that, up to reparameterization, the shortest curve \emph{in} $\WW^n$ connecting $\sqrt{\rho_0}$ to 
$\sqrt{\rho_\tau}$  is of the form 
\begin{equation}\label{great rot arc}
	\psi_t = \frac{(\tau-t)\sqrt{\rho_0} + t\sqrt{\rho_\tau}}{\|(\tau-t)\sqrt{\rho_0} + t\sqrt{\rho_\tau}\|_{\textsc{hs}}}.
\end{equation}
The distance between $\rho_0$ to  $\rho_\tau$ is thus bounded from below by the Hilbert-Schmidt length of $\psi_t$. The initial and final amplitudes in $\psi_t$
are, by construction, the square roots of their projections. But so are also all the intermediate amplitudes (i.e, every $\psi_t$ is a positive semi-definite, Hermitian operator). In other words, each $\psi_t$ is contained in the image of the square root section. We conclude that the Wigner-Yanase distance between 
$\sqrt{\rho_0}$ and $\sqrt{\rho_\tau}$ equals the Hilbert-Schmidt length of $\psi_t$. That is,
\begin{equation}\label{Wigneravstand}
	\distWY(\rho_0,\rho_\tau)
	=\arccos\big(\Tr(\sqrt{\rho_0}\sqrt{\rho_\tau}\,)\big).
\end{equation}
Notice that the trace is nonnegative since $\sqrt{\rho_0}\sqrt{\rho_\tau}$ is positive semi-definite.
The quantity $A(\rho_1,\rho_2)=\Tr(\sqrt{\rho_1}\sqrt{\rho_2}\,)$ is called the \emph{affinity} of $\rho_1$ and $\rho_2$.
\subsection{A quantum speed limit involving the Wigner-Yanase skew-information}\label{WYqsl}
A quantum speed limit can be derived from Eq.~\eqref{Why}.
Suppose that $\rho_t$ satisfies Eq.~\eqref{vonNeumann}.
The final time is then bounded from below:
\begin{equation}
	\tau\geq\frac{\distWY(\rho_0,\rho_\tau)}{\big\langle \sqrt{2\IWY(\rho_t,H)}\,\big\rangle}.
\end{equation}
The denominator is the time-average of the square root of twice the  Wigner-Yanase skew-information 
over the interval $0\leq t\leq \tau$.
For qubits this is a weaker speed limit than that of Uhlmann which we derived in Section \ref{UhQSL},
see \cite{PiCiCeAdSP2016}.
But if this is the case in general is not known.

\section{The complementary geometry}
A third monotone geometry can be obtained from the observation that the kernel
bundle of the mechanical connection on the complementary purification bundle 
is also a connection in the standard purification bundle.

Recall from Section \ref{comp pur} that the complementary purification bundle over 
$\SS^n$ is the left $\UU(\HH)$-principal bundle $\p^c$ which sends the 
amplitude $\psi$ in $\WW^n$ to $\psi^\dagger\psi$ in $\SS^n$.
The complementary vertical space at $\psi$ is the space $\V^c_\psi\WW^n$
of all the tangent vectors at $\psi$ which get annihilated by the differential of $\p^c$.
We have, in analogy with Eq.~\eqref{eq: vert space}, that
\begin{equation}
	\V^c_\psi\WW^n = \{ \xi\psi : \xi\in\uu(\HH)\}.
\end{equation}
We define the horizontal space $\H^c_{\psi}\WW^n$ as the orthogonal complement of $\V^c_\psi\WW^n$ with respect to the Hilbert-Schmidt metric. A vector $\dot\psi$ at $\psi$ is then horizontal if 
$\psi \dot\psi^\dagger = \dot\psi\psi^\dagger$.
This condition is equivalent to $\dot\psi=\frac{1}{2}\psi R$ for some Hermitian operator $R$. The operator must satisfy $\Tr(\psi^\dagger\psi R)=0$.
We will next show that the union of all the horizontal spaces is an Ehresmann connection for $\Pi$.

The  first requirement is that $\H^c_\psi\WW^n$ should be complementary to $\V_\psi\WW^n$, the vertical space of $\p$ at $\psi$. 
This follows from the observations that the dimensions of $\V_\psi\WW^n$ and $\H^c_\psi\WW^n$ sum up to the dimension of $\WW^n$ and
that $\V_\psi\WW^n$ and $\H^c_\psi\WW^n$ only have the zero-vector in common 
(since $\uu(\HH)$ and $\hh(\HH)$ only have the zero-operator in common).
The second requirement is that the horizontal bundle should be preserved by the action of the symmetry group. This is also the case. For if $\dot\psi$ belongs to $\H^c_\psi\WW^n$ and $U$ is any unitary, then
$d\R_U(\dot\psi)=\dot\psi U$ is a horizontal vector at $\psi U$:
\begin{equation}
	(\psi U)(\dot\psi U)^\dagger - (\dot\psi U)(\psi U)^\dagger = \psi \dot\psi^\dagger - \dot\psi\psi^\dagger = 0.
\end{equation}
We conclude that $d\R_U$ sends $\H^c_\psi\WW^n$ into $\H^c_{\psi U}\WW^n$. 

\subsection{The complementary metric} 
The horizontal lift of a tangent vector $\dot\rho$ at $\rho$
to an amplitude $\psi$ is 
\begin{equation}\label{holift}
	\dot\psi^h=\tfrac{1}{2}\dot\rho(\psi^\dagger)^{-1}.
\end{equation}
Let $\gC$ be the projection of the Hilbert-Schmidt metric. Then, for $\dot\rho_1$ and $\dot\rho_2$ at $\rho$ having horizontal lifts $\dot\psi_1^h$ and $\dot\psi_2^h$, respectively,
\begin{equation}\label{comp metric}
	\gC(\dot\rho_1,\dot\rho_2)=\GHS(\dot\psi_1^h,\dot\psi_2^h)=\frac{1}{8}\Tr\big( \rho^{-1} (\dot\rho_1^\dagger \dot\rho_2+\dot\rho_2^\dagger \dot\rho_1) \big).
\end{equation}
If, in particular, $\dot\rho_1$ is tangential to the commutative manifold of $\rho$ and $\dot\rho_2$ is tangential to the unitary orbit of $\rho$,
so that $\dot\rho_2=-i[A,\rho]$ for some Hermitian $A$, then
the horizontal lifts of $\dot\rho_1$ and $\dot\rho_2$ to $\sqrt{\rho}$
are $\frac{1}{2}\dot\rho\rho^{-1/2}$ and $\frac{1}{2}[A,\rho]\rho^{-1/2}$, respectively.
By a straightforward calculation,
\begin{equation}
	\gC(\dot\rho_1,\dot\rho_2)
	=\GHS\big(\tfrac{1}{2}\dot\rho\rho^{-1/2},\tfrac{1}{2}[A,\rho]\rho^{-1/2}\big)=0.
\end{equation}
Thus, the commutative manifold and the unitary orbit of $\rho$ are perpendicular at $\rho$ with respect to the complementary metric.

\subsection{Complementary connection form}
Every tangent vector $\dot\psi$ at $\psi$ has a unique decomposition
$\dot\psi=\dot\psi^h+\dot\psi^v$ where $\dot\psi^h$ is horizontal and $\dot\psi^v$ is vertical.
The complementary connection $\Ac$ is defined by $\dot\psi^v=\psi \Ac(\dot\psi)$.
According to Eq.~\eqref{holift}, the horizontal component~is 
\begin{equation}
	\dot\psi^h=\frac{1}{2}\big(\dot\psi\psi^\dagger + \psi \dot\psi^\dagger\big)(\psi^\dagger)^{-1}
	=\frac{1}{2}\big(\dot\psi + \psi \dot\psi^\dagger(\psi^\dagger)^{-1}\big).
\end{equation}
Consequently, the complementary connection is
\begin{equation}\label{Cannes}
	\Ac(\dot\psi)=\psi^{-1}(\dot\psi-\dot\psi^h)
	=\frac{1}{2}\big(\psi^{-1}\dot\psi - (\psi^{-1}\dot\psi)^\dagger\big).
\end{equation}
In particular, along the square root lift,
\begin{equation}
	\Ac\big(\tfrac{\d}{\dt}\sqrt{\rho_t}\big)
	=\frac{1}{2}\big[(\sqrt{\rho_t}\,)^{ -1},\tfrac{\d}{\dt}\sqrt{\rho_t}\,\big].
\end{equation}
A counterpart to H\"{u}bner's formula \eqref{hubner} is
\begin{equation}\label{hubners}
\begin{split}
	\Ac\big(\tfrac{\d}{\d t}\sqrt{\rho_t}\,\big)
	&=\frac{1}{2}\sum_{a,b}\ketbra{\psi_{a;t}}{\psi_{a;t}}
		\big[(\sqrt{\rho_t}\,)^{-1},\tfrac{\d}{\d t}\sqrt{\rho_t}\,\big]\ketbra{\psi_{b;t}}{\psi_{b;t}}\\
	&=\frac{1}{2}\sum_{a,b}\bigg(\frac{1}{\sqrt{p^a_t}}-\frac{1}{\sqrt{p^b_t}}\bigg)\ketbra{\psi_{a;t}}{\psi_{a;t}}\tfrac{\d}{\d t}\sqrt{\rho_t}\ketbra{\psi_{b;t}}{\psi_{b;t}}.
\end{split}
\end{equation}

\section{Qubits}\label{sec: qubits}
In this section we illustrate the theory we have developed so far
in the simplest nontrivial case, namely that of a system in a mixed qubit state. The geometrical intuition is greatly enhanced by the fact that 
the space of density operators on a two-dimensional Hilbert space can be identified with the three-dimensional Euclidean ball, named the Bloch ball
by quantum mechanics.\footnote{Colleagues.} 

\subsection{The Bloch ball}
Given an orthonormal basis $\{\ket{0},\ket{1}\}$ for the two-dimensional Hilbert space, 
an explicit diffeomorphism from $\SS^2$ onto the interior of the Bloch ball is
\begin{equation}\label{diffeomorphism}
	\rho\to \mathbf{r}
	=\big(\Tr(\rho\sigma_x),\Tr(\rho\sigma_y),\Tr(\rho\sigma_z)\big).
\end{equation}
The operators $\sigma_x$, $\sigma_y$ and $\sigma_z$ are the Pauli operators defined by
\begin{subequations}
\begin{align}
	\sigma_x &= \ketbra{1}{0}+\ketbra{0}{1},\\
	\sigma_y &= i(\ketbra{1}{0}-\ketbra{0}{1}),\\
	\sigma_z &= \ketbra{0}{0}-\ketbra{1}{1}.
\end{align}
\end{subequations}
We notice that the maximally mixed state is sent to the origin 
of the Bloch ball. And the more pure the state is, the closer to the boundary of the Bloch ball it ends up. In fact, the diffeomorphism \eqref{diffeomorphism} extends to the pure states.
These get sent to the surface of the Bloch ball, called the Bloch sphere.

The inverse diffeomorphism is 
\begin{equation}\label{diffeomorphismen}
	\mathbf{r}=(x,y,z)\to\rho=\frac{1}{2}(\1+\mathbf{r}\cdot\bfsigma).
\end{equation}
Here $\mathbf{r}\cdot\bfsigma$ is shorthand for $x\sigma_x+y\sigma_y+z\sigma_z$.
We will use \eqref{diffeomorphismen} to pull back the Bures, the Wigner-Yanase, and the complementary metrics to the Bloch ball.
The pull-back metrics will also be denoted by $\gB$, $\gWY$, and $\gC$, respectively.

\subsubsection{Eigenvalues, eigenstates, inverse and square root}
Writing $r=\sqrt{x^2+y^2+z^2}$,
the eigenvalues and eigenvectors
of a density operator given by the right hand side of 
\eqref{diffeomorphismen} are
\begin{subequations}\label{estate}
\begin{align}
	p^0&=\frac{1}{2}(1+r), & \ket{\psi_{0}}
	&= \frac{e^{i\theta_0}}{\sqrt{2r(r+z)}}\big((r+z)\ket{0}
	+ (x+iy)\ket{1}\big),\label{nollande}\\
	p^1&=\frac{1}{2}(1-r), & \ket{\psi_{1}}
	&= \frac{e^{i\theta_1}}{\sqrt{2r(r+z)}}\big((x-iy)\ket{0}
	- (r+z)\ket{1}\big).\label{ettande}
\end{align}
\end{subequations}
The phase factors $e^{i\theta_0}$ and $e^{i\theta_1}$ are arbitrary.
Clearly, the formulas for the eigenstates are valid only if $z\ne -r$.
For $z=-r$,
\begin{subequations}
\begin{align}
	\ket{\psi_{0}} &= \ket{1}e^{i\theta_0},\\
	\ket{\psi_{1}} &= \ket{0}e^{i\theta_1}.
\end{align}
\end{subequations}
Using the formulas in Eq.~\eqref{estate}, one can easily write down formulas for the inverse and the square root of a faithful density operator in terms of the Euclidean coordinates; if $\rho$ is given by the right hand side of \eqref{diffeomorphismen}, then
\begin{subequations}
\begin{align}
	\rho^{-1} 
	&= \frac{2}{1-r^2}(\1-\mathbf{r}\cdot\bfsigma),\label{invers}\\
	\sqrt{\rho} 
	&= \frac{\sqrt{1+r}+\sqrt{1-r}}{2\sqrt{2}}\Big(\1
	+	\frac{1}{1+\sqrt{1-r^2}}\,\mathbf{r}\cdot\bfsigma\Big).\label{rot}
\end{align}
\end{subequations}

\subsubsection{Unitary orbits and classical manifolds}
\begin{figure}[t]
	\centering
	\includegraphics[width=0.65\textwidth]{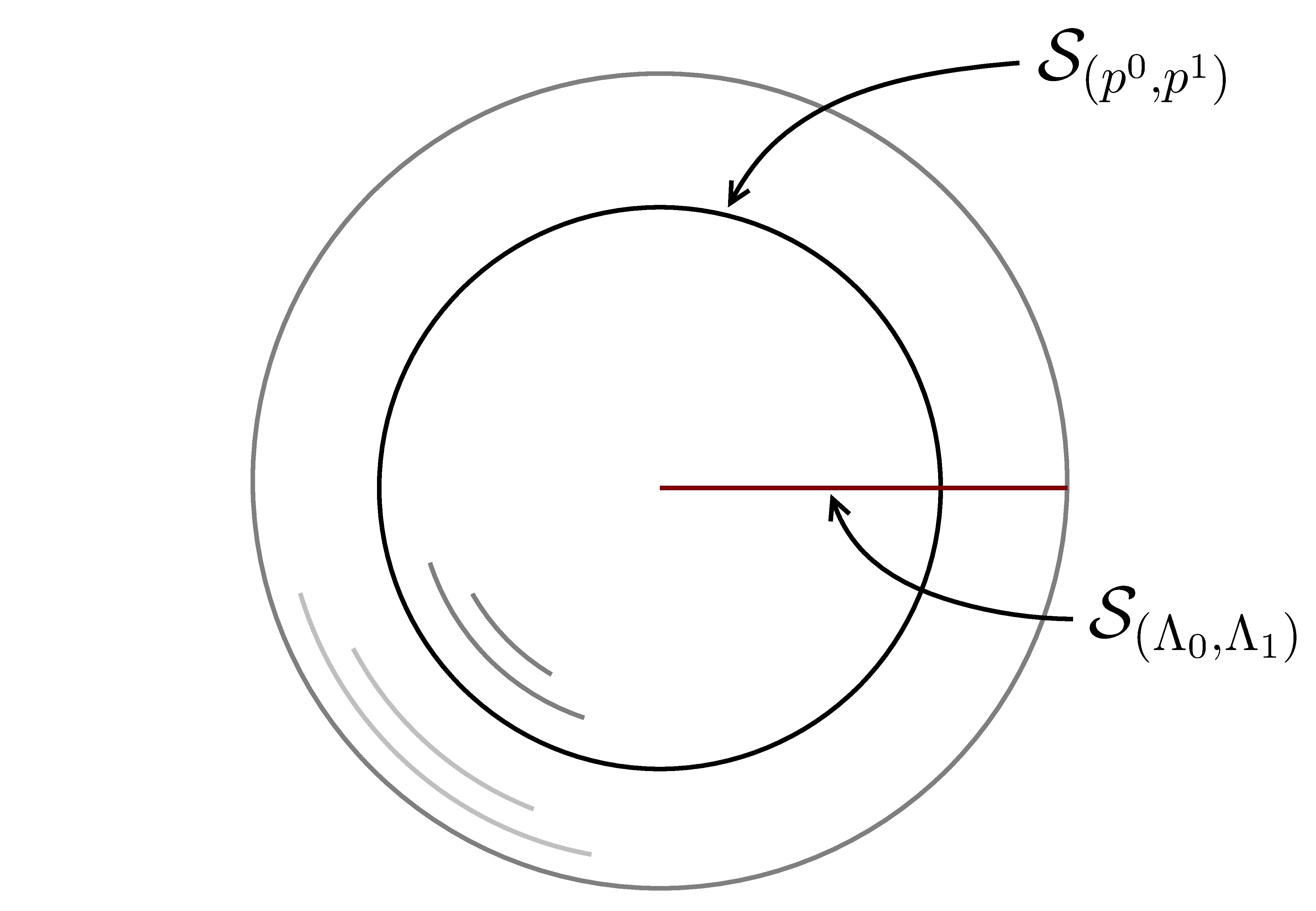}
	\caption{Unitary orbits and classical manifolds in the Bloch ball. The unitary orbits correspond to the round subspheres of the Bloch ball which are centered at the origin, while `most' of the classical manifolds correspond to the open, radial subintervals of the Bloch ball. Only the classical manifolds of the pure states and the maximally mixed state are exceptional as these are single element spaces.}
	\label{fig: spheres}
\end{figure}

The unitary orbits in $\SS^2$ correspond to the nested subspheres in the open Bloch ball which are centered at the origin. Specifically, the unitary orbit $\SS_{(p^0,p^1)}$ correspond to the subsphere of radius $p^0-p^1$, see Figure \ref{fig: spheres}.
The classical manifolds, on the other hand, correspond to the open radial intervals (except for the classical manifold of the maximally mixed state). For instance, if $\Lambda_0=\ketbra{0}{0}$ and 
$\Lambda_1=\ketbra{1}{1}$, then $\SS_{(\Lambda_0,\Lambda_1)}$ corresponds to the intersection of $\SS^2$ and the positive $z$-axis. While if $\Lambda_0=\frac{1}{2}(\ket{0}+\ket{1})(\bra{0}+\bra{1})$ and $\Lambda_1=\frac{1}{2}(\ket{0}-\ket{1})(\bra{0}-\bra{1})$, then $\SS_{(\Lambda_0,\Lambda_1)}$ corresponds to the intersection of $\SS^2$ and the positive $x$-axis, see Figure~\ref{fig: spheres}.

We can use the formulas \eqref{comp metric} and \eqref{invers}
to derive an expression for the complementary metric 
in the Bloch ball; if $\dot{\mathbf{r}}$ is a tangent vector
located at a Euclidean distance $r$ from the origin, then
\begin{equation}
	\gC(\dot{\mathbf{r}},\dot{\mathbf{r}})
	= \frac{1}{4(1-r^2)}\,\dot{\mathbf{r}}\cdot\dot{\mathbf{r}}.
\end{equation}
The ``$\cdot$'' 
is the Euclidean inner product.
Similarly, we can use \eqref{WY pullback} and 
\eqref{rot} to derive an expression for the 
Wigner-Yanase metric. We distinguish between the cases that 
$\dot{\mathbf{r}}$ is tangential to a classical manifold
and that $\dot{\mathbf{r}}$ is tangential to a unitary orbit.
(Remember that the classical and unitary strata are perpendicular
in the Wigner-Yanase geometry.)
In the former case, $\dot{\mathbf{r}}$ is a radial vector, and in the latter case, $\dot{\mathbf{r}}$ is tangential to a subsphere of the Bloch ball. Straightforward calculations yield
\begin{equation}
\gWY(\dot{\mathbf{r}},\dot{\mathbf{r}}) =
\begin{cases}	
	\frac{1}{4(1-r^2)}\,\dot{\mathbf{r}}\cdot\dot{\mathbf{r}}
		&\text{ if $\dot{\mathbf{r}}$ is classical},\\
	\frac{1}{(\sqrt{1+r}+\sqrt{1-r})^2}\,\dot{\mathbf{r}}\cdot\dot{\mathbf{r}}
		&\text{ if $\dot{\mathbf{r}}$ is unitary}.
\end{cases}
\end{equation}
Finally, to derive formulas for the Bures metric in terms of the 
Euclidean metric we use the following formula by Dittman \cite{Di1999} for the Bures metric:
\begin{equation}
	\gB(\dot\rho,\dot\rho)=\frac{1}{4}\Big( \Tr(\dot\rho^2) + \frac{1}{\det\rho}(\dot\rho -\rho\dot\rho)^2\Big).
\end{equation}
Notice that this formula is valid only for qubits.
In Euclidean coordinates,
\begin{equation}
	\gB(\dot{\mathbf{r}},\dot{\mathbf{r}}) =
	\begin{cases}	
	\frac{1}{4(1-r^2)}\,\dot{\mathbf{r}}\cdot\dot{\mathbf{r}}
		&\text{ if $\dot{\mathbf{r}}$ is classical},\\
	\frac{1}{4}\,\dot{\mathbf{r}}\cdot\dot{\mathbf{r}}
		&\text{ if $\dot{\mathbf{r}}$ is unitary}.
\end{cases}
\end{equation}

\subsection{Holonomy of cyclicly evolving mixed qubit states}\label{Uhlmann kubit}
\begin{figure}[t]
	\centering
	\includegraphics[width=0.53\textwidth]{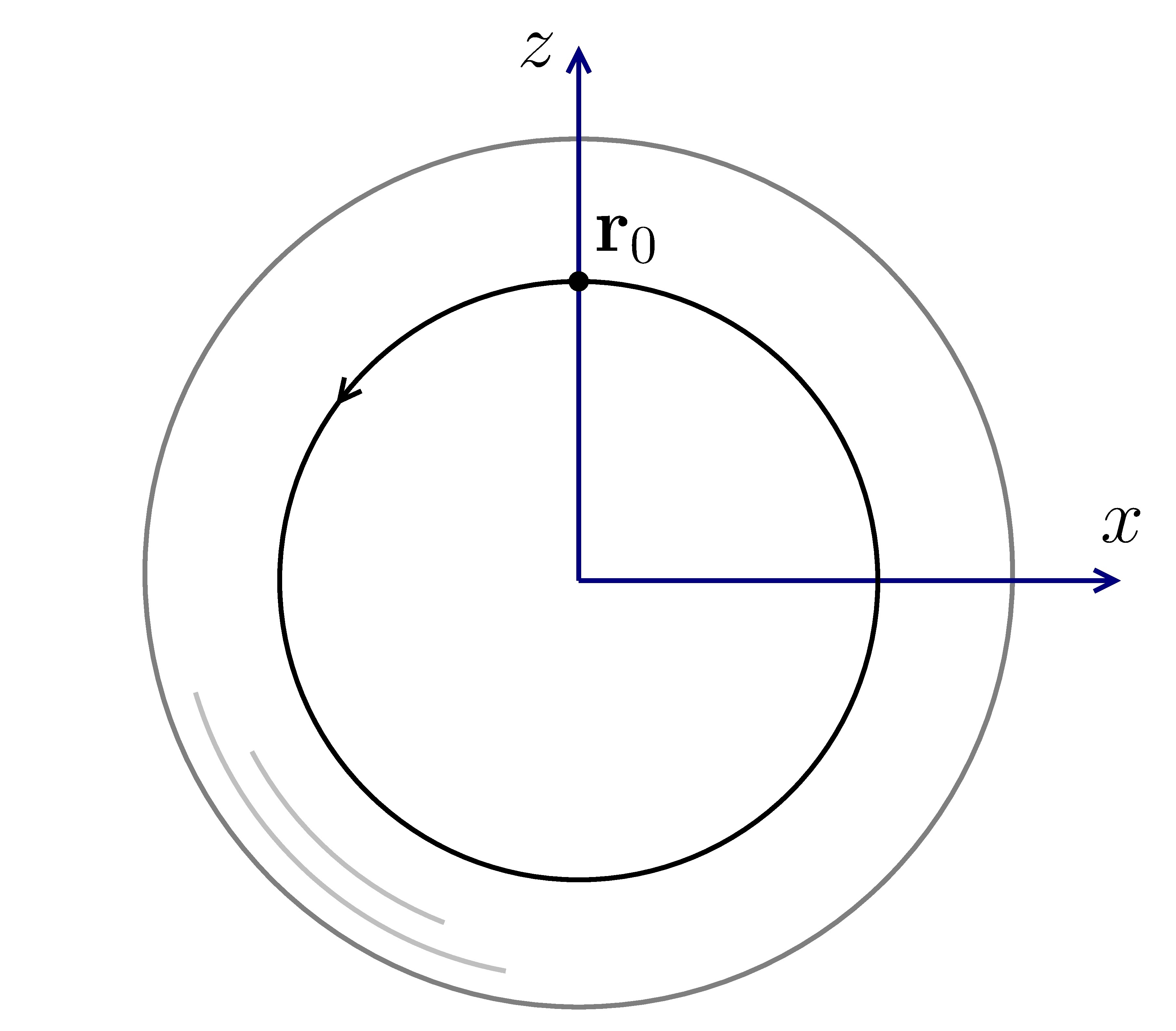}
	\caption{A unitary development in the Bloch ball. The unitary operator propagates the initial state in a round circle in the $xz$-plane.}
	\label{fig: circle}
\end{figure}

In this section we will calculate the Uhlmann and complementary holonomies for a cyclic, unitary development of a mixed qubit state.
(As was mentioned before, the Wigner-Yanase holonomy is always trivial.)
We will take as the initial state 
$\rho_0=p^0\ketbra{0}{0}+p^1\ketbra{1}{1}$, where $p^0>p^1$, corresponding to $\mathbf{r}_0=(0,0,p^0-p^1)$ in the Bloch ball.
Furthermore, to avoid complications arising from the nonabelian character of the holonomy, we will apply a very simple unitary 
propagator to the initial state, namely
\begin{equation}\label{unitaren}
	U_t=\cos(t) \1 + i\sin(t)\sigma_y,
\end{equation}
which causes $\mathbf{r}_0$ to trace out a circle in the $xz$-plane, see Figure \ref{fig: circle}.

Consider the curve $\rho_t=U_t\rho_0 U_t^\dagger$.
The square root lift is $\sqrt{\rho_t}=U_t\sqrt{\rho_0} U_t^\dagger$. A straightforward calculation using 
H\"{u}bner's formula \eqref{hubner} shows that the Bures connection form is constant along the square root lift:
\begin{equation}
	\AB\big(\tfrac{\d}{\d t}\sqrt{\rho_t}\,\big)
	= -i\Big(1-2\sqrt{p^0p^1}\,\Big)\sigma_y.
\end{equation}
An application of the formula \eqref{hubners} shows that 
the same holds for the complementary connection:
\begin{equation}
	\Ac\big(\tfrac{\d}{\d t}\sqrt{\rho_t}\,\big)
	= -i\Big(1-\frac{1}{2\sqrt{p^0p^1}}\,\Big)\sigma_y.
\end{equation}
The curve $\rho_t$ is periodic with a period $\pi$. 
We take the period as the final time, $\tau=\pi$. 
(To calculate the holonomy for a loop over several periods we can apply the group property \eqref{group property}.)
The square root lift is closed and, hence, by Eq.~\eqref{Uh holonomy}
the Uhlmann and complementary holonomies at $\sqrt{\rho_0}$ are 
\begin{equation}
\begin{split}
	\Hol_{\sqrt{\rho_0}}^{\textsc{uh}}
	&=\texp\Big(-\int_0^\pi\dt\, \AB\big(\tfrac{\d}{\d t}\sqrt{\rho_t}\,\big)\Big)\\
	&=\cos\!\Big(\pi\big(1-2\sqrt{p^0p^1}\,\big)\Big) \1 + i\sin\!\Big(\pi\big(1-2\sqrt{p^0p^1}\,\big)\Big)\sigma_y
\end{split}
\end{equation}
and
\begin{equation}
\begin{split}
	\Hol_{\sqrt{\rho_0}}^{\textsc{c}}
	&=\texp\Big(-\int_0^\pi\dt\, \Ac\big(\tfrac{\d}{\d t}\sqrt{\rho_t}\,\big)\Big)\\
	&=\cos\!\Big(\pi\big(1-\frac{1}{2\sqrt{p^0p^1}}\,\big)\Big) \1 + i\sin\!\Big(\pi\big(1-\frac{1}{2\sqrt{p^0p^1}}\,\big)\Big)\sigma_y,
\end{split}
\end{equation}
respectively. 
The corresponding geometric phase factors are 
\begin{align}
	\Geop^{\textsc{uh}}
	&= \Tr\!\big(\rho_0\Hol_{\sqrt{\rho_0}}^{\textsc{uh}}\big)
	= \cos\!\Big(\pi\big(1-2\sqrt{p^0p^1}\,\big)\Big),\\
	\Geop^{\textsc{c}}
	&= \Tr\!\big(\rho_0\Hol_{\sqrt{\rho_0}}^{\textsc{c}}\big)
	= \cos\!\Big(\pi\big(1-\frac{1}{2\sqrt{p^0p^1}}\,\big)\Big),
\end{align}
and, thus, the geometric phases are 
\begin{align}
	\tgeo^{\textsc{uh}}
	&= \arg\cos\!\Big(\pi\big(1-2\sqrt{p^0p^1}\,\big)\Big),\\
	\tgeo^{\textsc{c}}
	&= \arg\cos\!\Big(\pi\big(1-\frac{1}{2\sqrt{p^0p^1}}\,\big)\Big).
\end{align}

The product $p_0p_1$ lies between $0$ and $1/4$ and, consequently, 
the complementary geometric phase will vanish, irrespective of the 
eigenvalue spectrum. However, the Uhlmann geometric phase 
vanishes only for $p^0<(2+\sqrt{3})/4$. If $p^0>(2+\sqrt{3})/4$,
the Uhlmann phase is $\pi$. By varying the spectrum, the initial vector $\mathbf{r}_0$ in the Bloch ball will trace out concentric circles in the $xy$-plane. For small values of the radius of the circles the geometric phase is $0$. But for $r=\sqrt{3}/2$, the phase suddenly jumps 
to $\pi$. This `topological' behavior of the Uhlmann phase for real qubit curves (see Section \ref{sec: real curves}) 
has been used as an argument that the Uhlmann phase can be used as an indicator of a topological phase transition in certain classes of topological insulators and superconductors \cite{ViRiMD2014a,HuAr2014,ViRiMD2014b}.
However, the applicability of the Uhlmann geometric phase to such systems has been questioned \cite{BuDi2015,AnBeErSj2016}. 

\section{The monotone geometries}\label{monotona geometrier}
A distance function $\dist$ on $\SS^n$ is monotone if $\dist(\EE(\rho_1),\EE(\rho_2))\leq \dist(\rho_1,\rho_2)$
for every quantum 
channel $\EE$ on $\HH$.
For a distance function coming from a Riemannian metric $g$,
an equivalent definition is 
$g(d\EE(\dot\rho_1),d\EE(\dot\rho_2))\leq g(\dot\rho_1,\dot\rho_2)$.
Petz \cite{Pe1996} has classified the monotone Riemannian metrics on $\SS^n$.
According to this classification, there is a one-to-one 
correspondence between the symmetric, operator-monotone functions and the 
monotone Riemannian metrics on $\SS^n$:
\begin{equation}
	f\leftrightarrow g_f.
\end{equation}

\subsection{Petz classification theorem}
An operator-monotone function is a real-valued function $f$ on the positive reals $(0,+\infty)$ which is such that $A \preceq B$
implies $f(A) \preceq f(B)$ for every pair of positive operators $A,B$ on $\HH$.
The partial order is defined by $A \preceq B$ if $B-A$ is positive semi-definite. 
We further say that the operator-monotone function is symmetric if
$f(x) = xf(x^{-1})$, and normalized if
$f(1) = 1$.

\subsubsection{The monotone metrics} 
Let $f$ be a symmetric, operator-monotone function and let $g_f$
be the corresponding monotone metric on $\SS^n$.\! This metric is at $\rho$ defined by 
\begin{equation}
	g_f(\dot\rho_1,\dot\rho_2) = \frac{1}{4}\Tr\big(\dot\rho_1 \J^f_\rho(\dot\rho_2)\big),\qquad 
	\J^f_\rho \equiv \R_{\rho^{-1}} f(\L_\rho\R_{\rho^{-1}})^{-1}.
\end{equation}
See \cite{MoCh1990,Pe1996}.
If $f$ is normalized, then $g_f$
is proportional to the quantum Fisher information metric on the classical manifolds:
\begin{equation}
g_f(\dot\rho,\dot\rho) = \frac{1}{4}\gF(\dot\rho,\dot\rho)= \frac{1}{4}\Tr(\dot\rho L_{\dot\rho}).
\end{equation}  
We then say that the metric is Fisher information adjusted.

\subsubsection*{Three examples of monotone metrics}
The symmetric, normalized, operator-monotone functions satisfy \cite{KuAn1980}
\begin{equation}\label{jmf}
	f_{\textsc{c}}(x)
	=\frac{2x}{1+x}\leq f(x)\leq f_{\textsc{b}}(x)=\frac{1+x}{2}.
\end{equation}
For $f=f_{\textsc{b}}$ we have that $\J^f_\rho=2(\L_\rho+\R_\rho)^{-1}$ and, hence, by Eq.~\eqref{alt bures}, the corresponding monotone metric $g_f$ is the Bures metric. At the other end of \eqref{jmf}, if $f=f_{\textsc{c}}$,
then $\J^f_\rho=\frac{1}{2}(\L_\rho^{-1}+\R_\rho^{-1})$ and, by Eq.~\eqref{comp metric}, the monotone metric is the complementary metric. One can show that the greatest function $f_{\textsc{b}}$ defines the smallest Riemannian metric and the smallest function $f_\textsc{c}$ defines the largest Riemannian metric \cite{PeSu1996}. That is, for every normalized $f$,
\begin{equation}
	\gB(\dot\rho,\dot\rho) \leq g_f(\dot\rho,\dot\rho) \leq \gC(\dot\rho,\dot\rho).
\end{equation}
An example of an operator-monotone function in-between $f_\textsc{c}$ and $f_{\textsc{b}}$ 
is
\begin{equation}
	f_\textsc{wy}(x)=\frac{1}{4}(1+\sqrt{x}\,)^2.
\end{equation}
The corresponding monotone metric is the Wigner-Yanase metric.

\subsection[The Dittmann-Rudolph-Uhlmann approach to monotone metrics]{The Dittmann-Rudolph-Uhlmann approach to \\ monotone metrics}
In this section we will describe how every monotone metric on $\SS^n$ can 
be obtained as the projection of the Hilbert-Schmidt Riemannian metric relative to a connection on the standard purification bundle.
More precisely, to every symmetric, operator-monotone function $f$ we will assign a connection form $\A_f$ which is such 
the projection of the Hilbert-Schmidt metric equals $g_f$.
The section consists of a very brief review of the main results in \cite{DiRu1992, DiUh1999}.

\subsubsection{Projected monotone metrics}
Let $f$ be a symmetric, operator-monotone function.
Let $p$ be the unique positive function on the open interval $(0,\infty)$ satisfying
\begin{equation}
	2f(x)(p(x)^2+x)=(p(x) + x)^2,\qquad p(x^{-1})=p(x)^{-1},
\end{equation}
and set
\begin{equation}
	r(x)=\frac{x}{p(x)+x}.
\end{equation}
We define a $\uu(\HH)$-valued connection form $\A_f$ 
on the standard purification bundle
by assigning the skew-Hermitian operator
\begin{equation}\label{fconnection}
	\A_f(\dot\psi)
	=r(\L_{\psi^\dagger\psi}\R_{(\psi^\dagger\psi)^{-1}})(\psi^{-1}\dot\psi)-r(\R_{\psi^\dagger\psi}\L_{(\psi^\dagger\psi)^{-1}})
	((\psi^{-1}\dot\psi)^\dagger).
\end{equation}
to each tangent vector $\dot\psi$ at $\psi$.
One can show, see \cite{DiUh1999}, that  
the projection of the Hilbert-Schmidt metric with respect to $\A_f$ 
is the monotone metric $g_f$.

One can alternatively define $\A_f$ as the mechanical connection of 
a Riemannian metric on $\WW^n$. For if 
\begin{equation}
	k(x)=\frac{p(x)^2+x p(x)}{p(x)^2+x},
\end{equation}
then, for each amplitude $\psi$, the operator $k(\L_{\psi\psi^\dagger}\R_{(\psi^\dagger\psi)^{-1}})$ leaves the tangent space at $\psi$ 
invariant, and $\A_f$ is the mechanical connection of the metric
\begin{equation}\label{fmetric}
	G_f(\dot\psi_1,\dot\psi_2)
	=\GHS\big(\dot\psi_1,k(\L_{\psi\psi^\dagger}\R_{(\psi^\dagger\psi)^{-1}})^{-1}(\dot\psi_2)\big).
\end{equation}
Clearly, the horizontal space at $\psi$ is the image of 
$k(\L_{\psi\psi^\dagger}\R_{(\psi^\dagger\psi)^{-1}})$ of the 
Hilbert-Schmidt horizontal space at $\psi$, i.e., the 
$\GHS$-orthogonal complement of $\V_\psi\WW^n$.
Therefore, a tangent vector $\dot\psi$ is horizontal if, and only if,
\begin{equation} 
	\dot\psi
	=k(\L_{\psi\psi^\dagger}\R_{(\psi^\dagger\psi)^{-1}})\big(\tfrac{1}{2}L\psi\big)
\end{equation}
for some Hermitian $L$. 

\subsubsection{The Bures, Wigner-Yanase and canonical connection forms}
The formulas \eqref{fmetric} and \eqref{fconnection} are somewhat simpler at $\psi=\sqrt{\rho}$. There,
\begin{subequations}
\begin{align}
	G_f(\dot\psi_1,\dot\psi_2)
	&=\GHS\big(\dot\psi_1,k(\L_\rho\R_{\rho^{-1}})^{-1}(\dot\psi_2)\big),\\
	\A_f(\dot\psi)
	&=r(\L_\rho\R_{\rho^{-1}})(\rho^{-1/2}\dot\psi)-r(\R_\rho\L_{\rho^{-1}})
	(\dot\psi^\dagger\rho^{-1/2}).
\end{align}
\end{subequations}

The Bures metric is obtained for $p(x)=1$. Then $k(x)=1$, $r(x)=x/(1+x)$, $f(x)=(1+x)/2$, and the connection assumes the form 
\begin{equation}
\begin{split}
	\A_f(\dot\psi)
	&=(\L_{\rho}+\R_{\rho})^{-1} \big(\L_{\rho} (\rho^{-1/2}\dot\psi) 
		- \R_{\rho}(\dot\psi^\dagger\rho^{-1/2})\big).
\end{split}
\end{equation}
This is the Bures connection form \eqref{Burres}.

If $p(x)=x$, then $k(x)=2x/(1+x)$, $r(x)=1/2$, $f(x)=2x/(1+x)$, and the connection assumes the form 
\begin{equation}
	\A_f(\dot\psi)
	=\frac{1}{2}\big(\rho^{-1/2}\dot\psi 
		- \dot\psi^\dagger\rho^{-1/2}\big).
\end{equation}
We recognize this as the canonical connection form \eqref{Cannes}.

Finally, if $p(x)=\sqrt{x}$, then $k(x)=(1+\sqrt{x})/2$, $r(x)=\sqrt{x}/(1+\sqrt{x})$, $f(x)=(1+\sqrt{x})^2/4$, and  
\begin{equation}
	\A_f(\dot\psi)
	=\frac{ \sqrt{\L_\rho\R_{\rho^{-1}}}}{1+ \sqrt{\L_\rho\R_{\rho^{-1}}}} 
	 \big(\rho^{-1/2}\dot\psi\big) 
	- \frac{ \sqrt{\R_\rho\L_{\rho^{-1}}}}{1+ \sqrt{\R_\rho\L_{\rho^{-1}}}} \big(\dot\psi^\dagger\rho^{-1/2}\big).
\end{equation}
This is the Wigner-Yanase connection form.

\subsection[Extensions of monotone distance functions to nonfaithful density operators]{Extensions of monotone distance functions to\\ nonfaithful density operators}
Whether all monotone distance functions can be extended to distance functions 
on the entire space of density operators on $\HH$ is not known. But both the Bures 
and the Wigner-Yanase distance function do extend to the nonfaithful density operators.
For the formulas \eqref{Buresavstand} and \eqref{Wigneravstand} make sense 
and define proper distance functions on the whole of $\SS$.
Moreover, they agree with the Fubini-Study distance (see Eq.~\eqref{eq: lengths}) 
on the pure, unit-rank density operators. However, the two distance 
functions differ in a fundamental respect. 
Namely, only the Bures distance function is induced by a Riemannian metric 
defined on a neighborhood of $\SS$ in $\hh(\HH)$.
According to a result by Petz \cite{Pe1998} (see also \cite{Uh1995}), the Bures distance function is the only monotone distance function on $\SS$ which is induced by a Riemannian metric which is Fisher information adjusted \emph{and} Fibini-Study metric adjusted (i.e., agrees with the Fubini-Study metric on the tangent bundle of the space of pure states).

\subsubsection{Post-measurement states and the Bures distance}
In quantum mechanics, measurements are modeled 
as actions by positive operator-valued measures, or POVMs for short.
A POVM is a family of positive-semidefinite operators 
$M=\{M_a\}$ labelled by the possible outcomes $a$ of the measurement.
According to the Born rule \cite{Bo1926}, if the system is in the state $\rho$, the outcome $a$ is obtained with the probability $\Tr(\rho M_a)$.
This rule, of course, presupposes that the elements of a POVM add up to the identity operator. Here we will, in addition, assume that the elements of $M$ are projection operators. 
For example, $M$ can consist of the eigenspace projection operators of an observable. 

If the outcome when measuring $M$ on $\rho$ is $a$ we must, for consistency with the repeatability property of projective measurements
(see, e.g., \cite[Sec.\,2.2]{NiCh2000}), assume that 
the state is $\rho_a\equiv M_a\rho M_a/\Tr(\rho M_a)$ immediately after the measurement. We call $\rho_a$ \emph{the} post-measurement state.
However, to just ensure that a second measurement of $M$ gives the outcome $a$ with certainty, it is enough to assume that the support of the post-measurement state is contained in the support of $M_a$.  
We will next show that $\rho_a$ is, in a sense, the state among the `potential post-measurement states' which is closest to the pre-measurement state $\rho$ in the Bures geometry.

By a potential post-measurement state we mean any density operator 
whose support is contained in the support of $M_a$. Now, the support of
every amplitude $\phi$ of a potential post-measurement state $\sigma$ is also contained inside the support 
of $M_a$ and, hence, if $\psi$ is an amplitude for $\rho$,
\begin{equation}\label{ineq}
	|\Tr(\psi^\dagger\phi)|^2
	=|\Tr(\psi^\dagger M_a\phi)|^2
	=|\Tr((M_a\psi)^\dagger \phi)|^2
	\leq \Tr(\rho M_a).
\end{equation}
From this follows that 
$\distB(\rho,\sigma)\geq \arccos\sqrt{\Tr(\rho M_a)}$, according to Uhlmann's theorem. On the other hand, the inequality in \eqref{ineq} is saturated by the amplitude $\phi_a= M_a\psi/\sqrt{\Tr(\rho M_a)}$ of $\rho_a$ and,
hence, 
\begin{equation}
	\distB(\rho,\rho_a) = \arccos F(\rho,\rho_a) = \arccos \sqrt{\Tr(\rho M_a)}\leq \distB(\rho,\sigma).
\end{equation}
We conclude that among the potential post-measurement states, $\rho_a$ 
is the one which is closest to the pre-measurement state $\rho$.

\chapter{Holonomy of isodegenerate mixed states}\label{ch: sjoqvist holonomy}
Consider a curve of density operators $\rho_t$ which have a common, i.e., $t$-independent, nondegenerate weight spectrum.
Letting $p^1,p^2,\dots,p^k$ be the nonzero eigenvalues,  
the density operator $\rho_t$ has a spectral decomposition
\begin{equation}\label{eq: spec dec}
	\rho_t=\sum_{a} p^a\ketbra{\psi_{a;t}}{\psi_{a;t}}.
\end{equation}
The unit rank projectors $\ketbra{\psi_{a;t}}{\psi_{a;t}}$ are unique, being the orthogonal 
projection operators onto the eigenspaces of the $\rho_t$s, but the pure states $\ket{\psi_{a;t}}$ are not unique
as they can be phase shifted without the projection operators being affected.
However, given the initial states $\ket{\psi_{a;0}}$, there are unique curves of pure 
states that are horizontal in the sense of Aharonov and Anandan:
\begin{equation}\label{eq: parallel}
	\braket{\psi_{a;t}}{\dot{\psi}_{a;t}}=0.
\end{equation}
Following Sj\"{o}qvist \emph{et al.}~\cite{SjPaEkAnErOiVe2000} we call
a spectral decomposition in which the pure state curves are horizontal ``parallel transported''.
And we define the geometric phase of a closed curve of density operators 
$\rho_t$ with a parallel transported spectral decomposition as
\begin{equation}\label{eq: ind geo phase}
	\tgeo=\arg\Big(\sum_a p^a \braket{\psi_{a;0}}{\psi_{a;\tau}}\Big).
\end{equation}
The phase factors $e^{i\theta_a}=\braket{\psi_{a;0}}{\psi_{a;\tau}}$ are the Aharonov-Anandan 
holonomies of the projector curves. Below we will see that the diagonal matrix which has these 
phase factors as its diagonal entries can be regarded a holonomy of $\rho_t$.

If the eigenvalues are $t$-independent but degenerate, the parallel transport condition \eqref{eq: parallel} needs to be modified \cite{ToSjKwOh2004}. In this case we say that a spectral decomposition \eqref{eq: spec dec}
is parallel transported provided that
\begin{equation}\label{parat}
    p^a=p^b\implies \braket{\psi_{a;t}}{\dot{\psi}_{b;t}}=0.
\end{equation}
In other words, for each $t$ the velocity vector of $\ket{\psi_{a;t}}$ should be 
perpendicular to the eigenspace to which $\ket{\psi_{a;t}}$ belongs.
Again, we define the geometric phase for a closed curve of density operators with a parallel translated 
spectral decomposition by Eq.~\eqref{eq: ind geo phase}.

Finally, if the eigenvalues change with $t$ but the degeneracies remain constant,
we can generalize the parallel transport condition \eqref{parat} as
\begin{equation}\label{eq: parallel cond}
    p^a_t=p^b_t\implies \braket{\psi_{a;t}}{\dot{\psi}_{b;t}}=0.
\end{equation}
Then we define the geometric phase of a closed curve as
\begin{equation}\label{clgp}
	\tgeo=\arg\Big(\sum_a p^a_0\braket{\psi_{a;0}}{\psi_{b;\tau}}\Big).
\end{equation}
More generally, following Tong \emph{et al.}~\cite{ToSjKwOh2004}, we define the geometric phase of any, not necessarily closed, curve 
of density operators with a parallel transported spectral decomposition as 
\begin{equation}\label{opengp}
	\tgeo=\arg\Big(\sum_a\sqrt{p^a_0p^a_\tau}\braket{\psi_{a;0}}{\psi_{a;\tau}}\Big).
\end{equation}
Clearly, the definition \eqref{opengp} reduces to that in Eq.~\eqref{clgp} for closed curves,
and \eqref{clgp} reduces to the definition \eqref{eq: ind geo phase} if the spectrum is $t$-independent.

In the current chapter we will develop a holonomy theory for isodegenerate 
mixed states, i.e., states which have a common degeneracy spectrum.
 The parallel transport condition 
\eqref{eq: parallel cond} will correspond to a horizontality condition as given by a connection, and 
the geometric phase \eqref{opengp} 
will result from a definition similar to the one in Eq.~\eqref{gmtrc phase}.
The fiber bundle of the holonomy theory can be regarded a reduction of Uhlmanns purification bundle. 
However, the connection is different 
from any of those constructed by Dittmann, Rudolph and Uhlmann \cite{DiRu1992,DiUh1999}, see Section \ref{monotona geometrier}. The theory developed in this chapter provides an answer 
to the question asked in the concluding paragraph in Chapter 5 of \cite{ChJa2004}.
Main references for this chapter are \cite{AnHe2013,AnHe2015}.

\section{Holonomy of isospectral mixed states}\label{sec: isospectral}
In this section we develop a holonomy theory for isospectral mixed states.
It turns out that an appropriate starting point is the symplectic reduction theorem
of Marsden and Weinstein \cite{MaWe1974}.
This theorem of classical mechanics can be used to generate phase spaces for Hamiltonian
dynamical systems with continuous symmetries. We begin with some observations.

\begin{itemize}
\item[(i)] A curve $\rho_t$ of density operators with a common eigenvalue 
	spectrum remains in the unitary orbit of the initial density operator.
	Therefore, the curve is generated by a unitary propagator,
	$\rho_t=U_t\rho_0 U_t^\dagger$. Let $k$ be the common rank of the density operators.
	The propagator $U_t$ generates lifts of $\rho_t$ to $\WW^k$
	when it acts on amplitudes of $\rho_0$.
\item[(ii)] Irrespective of the initial amplitude and the unitary propagator, the lift 
	remains in a fiber of the complementary purification bundle.
	Furthermore, the standard purification projection maps the fiber onto the unitary orbit of 
	$\rho_0$.
\item[(iii)] The fibers of the complementary purification bundle are submanifolds 
	of $\LL^k$.
	The space $\LL^k$ is a symplectic manifold with a symplectic form 
	$\OHS$, the Hilbert-Schmidt symplectic form \eqref{ohs}.
	Now, the right action by $\UU(k)$ on $\LL^k$ is symplectic, i.e.,
	$R_U^*\OHS=\OHS$ for all $U$ in $\UU(k)$.
	Moreover, the action has an associated coadjoint-equivariant
	momentum map whose level sets in $\WW^k$ agree with the fibers of the 
	complementary purification projection.
\end{itemize}
The observations suggest an application of the Marsden-Weinstein symplectic reduction theorem to $(\LL^k,\UU(k),\OHS)$. 

\subsection{The symplectic reduction theorem}
Let $\GG$ be a real Lie group with a Lie algebra $\gg$.
Write $\gg^*$ for the space of real-valued linear functions on $\gg$
and let $g\to \Ad_g^*$ be the right coadjoint representation of $\GG$ on $\gg^*$,
$\Ad_g^*(\mu)=\mu\circ\Ad_g$.

Let $M$ be a manifold with a symplectic form $\Omega$.
Assume that $\GG$ acts on $M$ from the right via transformations $R_g$.
We say that the action is Hamiltonian if the transformations are symplectic, i.e., if $R_g^*\Omega=\Omega$, and 
the action has an associated coadjoint-equivariant momentum map $\J$. The momentum map is a $\gg^*$-valued map on $M$ and the coadjoint-equivariance requirement is that $\J\circ R_g=\Ad_g^*\circ\,\J$ for all $g$ in $\GG$.
The symplectic reduction theorem by Marsden and Weinstein \cite{MaWe1974} reads:

\vspace{5pt}
\noindent \emph{Assume that a Lie group $\GG$ acts from the right on a manifold $M$ with a symplectic form $\Omega$.
Furthermore, assume that the action is Hamiltonian, that $\mu$ is a regular value of the momentum map $\J$,
and that the isotropy group $\GG_\mu$ of $\mu$ acts freely and properly discontinuously on the level set $\J^{-1}(\mu)$. 
Then the projection $\pi$ from $\J^{-1}(\mu)$ onto the orbit space $\J^{-1}(\mu)/\GG_\mu$ is a principal $\GG_\mu$-bundle.
Moreover, the orbit space has a unique symplectic form $\omega$ characterized by 
$\pi^*\omega=\Omega|_{\J^{-1}(\mu)}$.}

\vspace{5pt}
 Recall that $\mu$ is a regular value of $\J$
if the differential of $\J$ maps the tangent spaces at every 
point in the level set $\J^{-1}(\mu)$ onto $\g^*$.
The level set is then a smooth manifold according to the inverse function theorem \cite[Ch.\,I, Th.\,2.1]{Sa1996}.
If the isotropy group $\GG_\mu$, i.e., the group of all $g$ in $\GG$ for which $\Ad_g^*(\mu)=\mu$,
acts freely and properly discontinuously on $\J^{-1}(\mu)$, then also the orbit space $\J^{-1}(\mu)/\GG_\mu$ is a smooth manifold
\cite[Sec.\,I, Prop.\,4.3]{KoNo1996}.

\subsection{Application of the symplectic reduction theorem}
Define a $\uu(k)^*$ valued map $\J$ on $\LL^k$
by declaring the action of $\J(\psi)$ on $\xi$ in $\uu(k)$ to be $\J(\psi)\xi=-i\Tr(\psi^\dagger\psi\xi)$.
The map $\J$ is coadjoint-equivariant since
\begin{equation}
	\J(\psi U)\xi 
	= -i \Tr( U^\dagger\psi^\dagger\psi U\xi)
	= -i \Tr( \psi^\dagger\psi U\xi U^\dagger)
	= \J(\psi)\Ad_U\!\xi.
\end{equation}
Furthermore, every $\J(\psi)$ is a regular value.
To see this assume that the differential of $\J$
maps the tangent space at some $\psi$ into but not onto $\uu(k)^*$.
According to Riesz' lemma \cite[Ch.\,II, Th.\,II.4]{ReSi1980},
there then exists a nonzero $\xi$ in $\uu(k)$ such that $d\J(\dot\psi)\xi=0$ for all tangent vectors $\dot\psi$ at $\psi$. 
But since $d\J(\dot\psi)\xi=\OHS(\psi\xi,\dot\psi)$ and $\OHS$ is nondegenerate, this implies the contradictory conclusion that $\xi=0$.

Let $\UU_{\J(\psi)}$ be the right isotropy group of $\J(\psi)$:
\begin{equation}
	\begin{split}
		\UU_{\J(\psi)}
		&=\{U\in\UU(k):\Ad_U^*\J(\psi)=\J(\psi)\}\\
		&=\{U\in\UU(k): U\psi^\dagger\psi=\psi^\dagger\psi U\}.
	\end{split}
\end{equation}
This group acts freely on the manifold $\J^{-1}(\J(\psi))$ from the right, and properly discontinuously because it is compact.
Thus, the coset space $\J^{-1}(\J(\psi))/\UU_{\J(\psi)}$ is a manifold.
By the Marsden-Weinstein reduction theorem, the quotient map $\phi\to \phi\UU_{\J(\psi)}$ 
is a principal fiber bundle with symmetry group $\UU_{\J(\psi)}$.
Furthermore, the coset space comes equipped with a unique symplectic structure 
whose pullback to $\J^{-1}(\J(\psi))$ equals the restriction of $\OHS$.

\subsection{Bundles over isospectral mixed states}
Let $\bfp$ be a weight spectrum of length $k$,
let $\psi$ be an amplitude for a density operator in $\SS_{\bfp}$,
and let $\bfl$ be the eigenprojector spectrum of $\psi^\dagger\psi$.
In Section \ref{comp pur} we introduced the notation $\WW_{\bfp;\bfl}$
for the level set $\J^{-1}(\J(\psi))$, which is also the fiber of the complementary 
purification bundle over $\psi^\dagger\psi$. Moreover, we introduced the notation $\UU_{\bfl}$ for the isotropy group of $\J(\psi)$. 
The standard purification bundle projection $\Pi$ maps $\WW_{\bfp;\bfl}$ onto the unitary orbit $\SS_{\bfp}$,
and the following diagram is commutative:
\begin{equation}
	\begin{tikzcd}
	{} & \WW_{\bfp;\bfl} \arrow{dl}[swap]{\phi\to \phi\UU_{\bfl}} \arrow{dr}{\Pi:\phi\to\phi\phi^\dagger} & {} \\
	\WW_{\bfp;\bfl}/\UU_{\bfl} \arrow{rr}{\phi\UU_{\bfl} \to \phi\phi^\dagger} & {} & \SS_{\bfp}
	\end{tikzcd}
\end{equation}
The horizontal map in the diagram is a diffeomorphism and, consequently, the 
restriction of $\p$ to $\WW_{\bfp;\bfl}$ is a principal fiber bundle with 
right acting symmetry group $\UU_{\bfl}$. 
From now on we will write $\p_{\bfp;\bfl}$ for the restriction of $\p$ to $\WW_{\bfp;\bfl}$.

An important observation is that the isomorphism class of 
$\p_{\bfp;\bfl}$ does not depend on the choice of the
eigenprojector spectrum $\bfl$. Indeed, if $\bfl'$ is another eigenprojector spectrum compatible with $\bfp$, there is a unitary operator $U$ on $\HH^k$ such that $\bfl'= U^\dagger \bfl U$. The isotropy groups $\UU_{\bfl'}$ and $\UU_{\bfl}$
are $U$-conjugated, i.e., $\UU_{\bfl'}=U^\dagger \UU_{\bfl} U$, and 
$\R_U$ restricts to an equivariant bundle map from $\WW_{\bfp;\bfl}$ onto $\WW_{\bfp;\bfl'}$. 
In other words, the diagram 
\begin{equation}
	\begin{tikzcd}\label{diagrammet}
	\WW_{\bfp;\bfl} \arrow{dr}[swap]{\p_{\bfp;\bfl}} \arrow{rr}{\R_U} & {} & \WW_{\bfp;\bfl'} \arrow{dl}{\p_{\bfp;\bfl'}} \\
	{} & \SS_{\bfp} & {}
	\end{tikzcd}
\end{equation}
is commutative and $\R_{U^\dagger V U}\circ \R_U = \R_U\circ \R_V$ for every $V$ in $\UU_{\bfl}$.

\subsubsection{The connection}
We consider the bundle $\p_{\bfp;\bfl}$ from $\WW_{\bfp;\bfl}$ onto $\SS_{\bfp}$.
The vertical bundle $\V\WW_{\bfp;\bfl}$ of $\p_{\bfp;\bfl}$ is the kernel bundle of the differential 
of $\p_{\bfp;\bfl}$. The vertical space at $\psi$ in $\WW_{\bfp;\bfl}$ is, hence,
$\V_\psi\WW_{\bfp;\bfl}=\{\psi\xi: \xi\in\uu_{\bfl}\}$, where $\uu_{\bfl}$ is the Lie algebra of the symmetry group.
The Lie algebra consists of the skew-Hermitian
operators on $\HH^k$ which commute with all the $\lambda_a$s.

Define $\H\WW_{\bfp;\bfl}$ as the orthogonal complement of 
$\V\WW_{\bfp;\bfl}$ with respect to the restriction of the Hilbert-Schmidt Riemannian metric $\GHS$ to $\WW_{\bfp;\bfl}$.
The horizontal bundle $\H\WW_{\bfp;\bfl}$ is then the kernel bundle of the mechanical 
connection form of the restriction of $\GHS$, see Section~\ref{sec: horiz tang vect}.
We denote the mechanical connection form simply by $\A$. 

An explicit formula for the connection form is
\begin{equation}\label{eq: connection formen}
	\A_{\psi}(\dot\psi)=(\psi^\dagger\psi)^{-1}\sum_a\lambda_a\psi^\dagger \dot\psi\lambda_a=\sum_a(P^a)^{-1}\lambda_a\psi^\dagger \dot\psi\lambda_a.
\end{equation}
The $P^a$s are the distinct eigenvalues in $\bfp$.
Letting $\II$ and $\JJ$ be the $\uu_{\bfl}^*$-valued 
moment of inertia tensor and metric momentum, respectively, 
\begin{subequations}
\begin{align}
\II_{\psi}(\xi)\eta=\GHS(\psi\xi,\psi\eta),\\
\JJ_{\psi}(\dot\psi)\eta=\GHS(\dot\psi,\psi\eta),
\end{align}
\end{subequations}
the formula \eqref{eq: connection formen} follows from the calculation
\begin{equation}
	\begin{split}
		\II_{\psi}\big(\sum_a (P^a)^{-1}  \lambda_a\psi^\dagger \dot\psi \lambda_a\big)\eta
		&=\sum_a\Tr\left(\lambda_a \dot\psi^\dagger \psi \lambda_a \eta 
			- \eta \lambda_a \psi^\dagger \dot\psi \lambda_a\right)\\
		&=\Tr\left(\dot\psi^\dagger\psi\eta - \eta\psi^\dagger \dot\psi\right)\\
		&=\JJ_\psi(\dot\psi)\eta,
	\end{split}
\end{equation}
As expected, the connection form transforms covariantly under an application of the bundle map in the diagram \eqref{diagrammet}: 
$\R_U^*\A_{\psi U}=\Ad_{U^\dagger}\circ \A_\psi$.

\subsection{Holonomy and geometric phase}
Let $\rho_t$ be a closed curve in $\SS_{\bfp}$
and let $\Gamma$ be the parallel transport operator of $\rho_t$.
The holonomy of $\rho_t$ based at the amplitude $\psi_0$ of $\rho_0$ 
in $\WW_{\bfp;\bfl}$ is the
unitary $\Hol_{\psi_0}$ in $\UU_{\bfl}$ defined by 
\begin{equation}
	\Gamma(\psi_0) = \psi_0 \Hol_{\psi_0}.
\end{equation}
As usual, we define the geometric phase factor and geometric phase of $\rho_t$ as
\begin{subequations}\label{geometriphase}
\begin{align}	
	&\Geop
	 = \Tr \left( \psi_0^\dagger \Gamma(\psi_0) \right)
	 = \Tr \left( \psi_0^\dagger \psi_0 \Hol_{\psi_0}\right),\label{gpf}\\
	&\tgeo = \arg \Geop.\label{gp}
\end{align}
\end{subequations}
Also, we take the first identity in Eq.~\eqref{gpf} as the definition of the geometric phase factor for a nonclosed curve.
Clearly, the geometric phase factor and the geometric phase are invariant under the action of $\UU_{\bfl}$. Moreover, they do not depend on 
the choice of the eigenprojector spectrum $\bfl$ because every $\R_U$ maps $\WW_{\bfp;\bfl}$ isometrically 
onto $\WW_{\bfp;U^\dagger \bfl U}$ with respect to $\GHS$ and, therefore, $\R_U$ preserves 
horizontality of curves. 

We can write down an 
explicit expression for the holonomy and the geometric phase factor
in terms of the propagator.
Suppose that $\rho_t=U_t\rho_0 U_t^\dagger$
and let $\psi_t=U_t\psi_0$.
The horizontal lift of $\rho_t$ extending from $\psi_0$ is
\begin{equation}\label{hurru}
	\psi_{t}^h=\psi_t\texp\Big(\!-\!\int_0^t \dt\, \A(\dot\psi_t)\Big).
\end{equation}
Then, by Eq.~\eqref{eq: connection formen}, 
\begin{equation}
	\Gamma(\psi_0)
	= U_\tau\psi_0\texp\Big(\!-\!\int_0^\tau\dt \sum_a(P^a)^{-1}\lambda_a\psi_0^\dagger U_t^\dagger\dot{U}_t\psi_0\lambda_a\Big).
\end{equation}
Consequently,
\begin{equation}\label{eq: explicit holonomy}
	\Hol_{\psi_0} = 
	(\psi_0^\dagger\psi_0)^{-1}\psi_0^\dagger U_\tau\psi_0 \texp\Big(\!-\!\int_0^\tau\dt \sum_a (P^a)^{-1}\lambda_a\psi_0^\dagger U_t^\dagger\dot{U}_t\psi_0\lambda_a\Big).
\end{equation}
This formula simplifies somewhat for faithful density operators, in which case we can choose $\HH^k=\HH$ and $\psi_0=\sqrt{\rho_0}$. Then
\begin{equation}\label{femtjugo}
	\Hol_{\sqrt{\rho_0}}
	= U_\tau \texp\Big(\!-\!\int_0^\tau \dt \sum_a \lambda_a U_t^\dagger\dot{U}_t\lambda_a\Big).
\end{equation}
Equation \eqref{femtjugo} is a counterpart to \eqref{eq: cocycle} for faithful density operators.

We will next show that the definition of geometric phase \eqref{gp} agrees 
with the definition \eqref{eq: ind geo phase} by Sj\"oqvist \emph{et al.}
To this end let $p^1,p^2,\dots,p^k$ be the common nonzero eigenvalues of the $\rho_t$s
and let $\psi_t$ be a horizontal lift of $\rho_t$.
The operator $\psi_0^\dagger\psi_0$ also has the weight spectrum $p^1,p^2,\dots,p^k$
and we fix an orthonormal basis $\ket{1},\ket{2},\dots,\ket{k}$ for $\HH^k$ such that $\psi_0^\dagger\psi_0\ket{a}=p^a\ket{a}$. 
Set $\ket{\psi_{a;t}}=\psi_t\ket{a}/\sqrt{p^a}$. Then
\begin{equation}\label{kurva}
	\rho_t=\sum_a p^a \ketbra{\psi_{a;t}}{\psi_{a;t}}
\end{equation}
is a parallel transported spectral decomposition: 
\begin{equation}
	p^a=p^b 
	\implies
	\braket{\psi_{a;t}}{\dot{\psi}_{b;t}}
	=\bra{a}\A(\dot\psi_t)\ket{b}
	= 0.
\end{equation}
Conversely, for a parallel transported spectral decomposition \eqref{kurva} 
we can define $\psi_t=\sum_a\sqrt{p^a}\ketbra{\psi_{a;t}}{a}$ where $\ket{1},\ket{2},\dots,\ket{k}$ 
is \emph{any} orthonormal basis for $\HH^k$. Then $\psi_t$ is horizontal as
\begin{equation}
	p^a=p^b 
	\implies
	\bra{a}\A(\dot\psi_t)\ket{b}
	= \braket{\psi_{a;t}}{\dot{\psi}_{b;t}}
	= 0.
\end{equation}
The conformity of the definitions \eqref{gp} and \eqref{eq: ind geo phase}
now follows from the calculation
\begin{equation}
	 \Geop
	  = \Tr\left(\psi_0^\dagger\psi_\tau\right)
	  = \Tr\Big(\sum_{a,b}\sqrt{p^ap^b}\ket{a}\braket{\psi_{a;0}}{\psi_{b;\tau}}\bra{b}\Big)
	  = \sum_{a}p^a\braket{\psi_{a;0}}{\psi_{a;\tau}}.
\end{equation}

\section{Holonomy of isodegenerate mixed states}\label{sec: isodegenerate}
In this section we will extend the fiber bundle and the connection over the unitary orbit $\SS_{\bfp}$ to a fiber bundle with a connection over $\SS_{\bfm}$,
where $\bfm$ is the degeneracy spectrum of $\bfp$. Recall that $\SS_{\bfm}$ is
the space of all and only those density operators on $\HH$ which have degeneracy spectrum $\bfm$.

\subsection{The bundle}
Let $k$ be the rank of the density operators in $\SS_{\bfm}$.
In Section \ref{comp pur} we introduced the notation 
$\SS_{\bfl}^c$ for the space of all density operators on $\HH^k$
which have the eigenprojection spectrum $\bfl$
and we defined $\WW_{\bfl}$ as the preimage of $\SS_{\bfl}^c$
under the complementary purification bundle.
We assume that $\bfl$ is compatible with $\bfm$.
Then, if $l$ is the common length of $\bfm$ and $\bfl$,
the space $\WW_{\bfl}$ is a manifold of dimension 
$2nk-k^2+l-1$. This follows from $\p^c$ being transversal to $\SS_{\bfl}^c$, 
which is manifold of dimension $l-1$.

The standard purification projection 
restricts to a principal fiber bundle $\p_{\bfl}$ from $\WW_{\bfl}$
onto $\SS_{\bfm}$ having symmetry group $\UU_{\bfl}$.
The vertical space at $\psi$ is $\V_\psi\WW_{\bfl}=\V_\psi\WW_{\bfp;\bfl}$, where $\bfp$ 
is the weight spectrum of $\psi\psi^\dagger$, and we define $\H_\psi\WW_{\bfl}$ 
as the orthogonal complement of $\V_\psi\WW_{\bfl}$ in $\T_\psi\WW_{\bfl}$ with respect to the metric $\GHS$. The union of the horizontal spaces is a connection on $\p_{\bfl}$.

The tangent space of $\SS_{\bfm}$ at $\rho$ splits into the tangent spaces
$\T_\rho\SS_{\bfp}$ and $\T_\rho\SS_{\bfL}$, where $\bfp$ and $\bfL$ are 
the weight and eigenprojector spectra of $\rho$.
Horizontal lifting of tangent vectors
induces a corresponding splitting of the horizontal spaces.
For a curve $\rho_t$ in $\SS_{\bfm}$ let $\bfL_{t}=(\Lambda_{1;t},\Lambda_{2;t},\dots,\Lambda_{l;t})$
be the eigenprojector spectrum of $\rho_t$, and let $P^a_t$ be the eigenvalue corresponding to $\Lambda_{a;t}$.
We can define a lift of $\rho_t$ to $\WW_{\bfl}$ as follows. Let  $\Lambda_{a;t}=\sum_{b_a}\ketbra{\psi_{ab_{a};t}}{\psi_{ab_{a};t}}$ and 
$\lambda_{a}=\sum_{b_{a}}\ketbra{ab_{a}}{ab_{a}}$ be orthonormal expansions. Then 
\begin{equation}\label{lyft}
	\psi_t=\sum_{a,b_{a}}\sqrt{ P^a_t}\ketbra{\psi_{ab_{a};t}}{ab_{a}}
\end{equation}
is a lift of $\rho_t$.
The velocity vectors of $\psi_t$ naturally split into two components: 
\begin{equation}
	\dot{\psi}_t 
	= \dot{\psi}^c_t + \dot{\psi}^u_t
	= \sum_{a,b_{a}} \dd{t}\Big(\!\sqrt{P^a_t}\,\Big)\ketbra{\psi_{ab_{a};t}}{ab_{a}} 
		+ \sum_{a,b_{a}} \sqrt{P^a_t}\ketbra{\dot{\psi}_{ab_{a};t}}{ab_{a}}.
\end{equation}
The component $\dot{\psi}^c_t$ is horizontal, since 
$\dot{\psi}^{c\dagger}_t\psi_t=\psi_t^\dagger\dot{\psi}^c_t$ 
implies that $\dot{\psi}^c_t$ and $\psi_t\xi$ are perpendicular for every $\xi\in\uu_{\bfl}$,
and its projection is tangential to $\SS_{\bfL_t}$.
The component $\dot{\psi}^u_t$, on the other hand, is horizontal if, and only if, the parallel transport condition
\begin{equation}\label{para}
  	\braket{\psi_{ab_a;t}}{\dot{\psi}_{a\beta_a;t}}=0
\end{equation}
is satisfied. The projection of $\dot{\psi}^u_t$ is tangential to the unitary orbit of $\rho_t$.

\subsection{Holonomy and geometric phase}
According to Eq.~\eqref{eq: connection formen},
\begin{equation}\label{A}
	\A(\dot\psi_t)
	= \A(\dot\psi^u_t)
	= \sum_a\sum_{b_a,\beta_a}\ket{ab_a}\braket{\psi_{ab_a;t}}{\dot{\psi}_{a\beta_a;t}}\bra{a\beta_a}.
\end{equation}
By adjusting the phases of the curves $\ket{\psi_{ab_a;t}}$, if necessary, we can make $\psi_t$ a closed curve.
The holonomy of $\rho_t$ at $\psi_0$ is, then, 
\begin{equation}
	\Hol_{\psi_0}
	= \texp\Big(\!-\!\int_0^\tau \dt \sum_a\sum_{b_a,\beta_a}\ket{ab_a}\braket{\psi_{ab_a;t}}{\dot{\psi}_{a\beta_a;t}}\bra{a\beta_a}\Big).
\end{equation}
As before, we define the geometric phase factor and the geometric phase by
\begin{subequations}\label{gffgf}
\begin{align}
	\Geop &= \Tr\big(\psi_0^\dagger\Gamma(\psi_0)\big),\\
	\tgeo &= \arg\Tr\big(\psi_0^\dagger\Gamma(\psi_0)\big).
\end{align}
\end{subequations}
Here, as usual, $\Gamma$ is the parallel transport operator associated with $\rho_t$. Similar arguments as in the isospectral case show that 
the geometric phase agrees with the geometric phase defined in Eq.~\eqref{opengp}:
The lift \eqref{lyft} is horizontal if, and only if, the parallel transport condition \eqref{para}
is satisfied and
\begin{equation}
\begin{split}
	\Geop	&=\Tr\Big(\big(\sum_{a,b_a}\sqrt{P^a_0}\ketbra{ab_a}{\psi_{ab_a;0}}\big)\big(\sum_{\alpha,\beta_{\alpha}}\sqrt{P^{\alpha}_t}\ketbra{\psi_{\alpha \beta_\alpha;\tau}}{\alpha\beta_\alpha}\big)\Big)\\
	&=\Tr\Big(\sum_{a,b_a}\sqrt{P^a_0P^a_\tau}\ket{ab_a}\braket{\psi_{ab_a;0}}{\psi_{ab_a;\tau}}\bra{ab_a}\Big)\\
	&=\sum_{a,b_a}\sqrt{P^a_0P^a_\tau}\braket{\psi_{ab_a;0}}{\psi_{ab_a;\tau}}.
\end{split}
\end{equation}

\section{Holonomy for mixed qubits}
The density operators which represent mixed qubit states having 
degeneracy spectrum $(1,1)$ all belong to $\SS_{(1,1)}$.
The space $\SS_{(1,1)}$ can be identified with the interior of the Bloch ball with the 
origin removed. Given an orthonormal basis $\{\ket{0},\ket{1}\}$ for the Hilbert space, an explicit diffeomorphism from $\SS_{(1,1)}$ onto
the `punctured' interior of the Bloch ball   
is given by the map in Eq.~\eqref{diffeomorphism}.

\subsection{Holonomy of cyclicly evolving mixed qubit states}
Consider a closed curve of mixed qubits
\begin{equation}\label{mixed kub}
	\rho_t=p^0_t\ketbra{\psi_{0;t}}{\psi_{0;t}}+p^1_t\ketbra{\psi_{1;t}}{\psi_{1;t}}.
\end{equation}
We define a curve of amplitudes of the $\rho_t$s by
\begin{equation}\label{plain}
	\psi_t = \sqrt{p^0_t}\ketbra{\psi_{0;t}}{0}+\sqrt{p^1_t}\ketbra{\psi_{1;t}}{1}.
\end{equation}
According to Eq.~\eqref{A},
\begin{equation}
	\A(\dot{\psi}_t) 
	= \ket{0}\braket{\psi_{0;t}}{\dot{\psi}_{0;t}}\bra{0}
	+ \ket{1}\braket{\psi_{1;t}}{\dot{\psi}_{1;t}}\bra{1}
\end{equation}
and, hence, by Eq.~\eqref{hurru}, a horizontal lift of $\rho_t$ is
\begin{equation}\label{straight}
	\psi_{t}^h 
	= \sqrt{p^0_t}\ketbra{\psi_{0;t}}{0}e^{-\int_0^t\dt\braket{\psi_{0;t}}{\dot{\psi}_{0;t}}}
	+\sqrt{p^1_t}\ketbra{\psi_{1;t}}{1}e^{-\int_0^t\dt\braket{\psi_{1;t}}{\dot{\psi}_{1;t}}}.
\end{equation}
The holonomy of $\rho_t$ at $\psi_0$ is
\begin{equation}\label{qubit holonomy}
	\Hol_{\psi_0}=\ket{0} \braket{\psi_{0;0}}{\psi_{0;\tau}}\bra{0}e^{-\int_0^\tau\dt\braket{\psi_{0;t}}{\dot{\psi}_{0;t}}}
	+\ket{1}\braket{\psi_{1;0}}{\psi_{1;\tau}}\bra{1}e^{-\int_0^\tau\dt\braket{\psi_{1;t}}{\dot{\psi}_{1;t}}},
\end{equation}
and the geometric phase factor is
\begin{equation}\label{qubit gpf}
	\Geop
	= p^0_0\braket{\psi_{0;0}}{\psi_{0;\tau}}e^{-\int_0^t\dt\braket{\psi_{0;t}}{\dot{\psi}_{0;t}}}
		+ p^1_0\braket{\psi_{1;0}}{\psi_{1;\tau}}e^{-\int_0^t\dt\braket{\psi_{1;t}}{\dot{\psi}_{1;t}}}.
\end{equation}

\subsubsection{Holonomy of isospectral mixed qubit states}
Consider a unitarily developed curve of density operators $	\rho_t=U_t\rho_0 U_t^\dagger$.
We take the initial state to be diagonal in the orthonormal basis $\{\ket{0},\ket{1}\}$,
\begin{equation}\label{initial}
	\rho_0=p^0\ketbra{0}{0}+p^1\ketbra{1}{1},
\end{equation}
and we define the propagator as   
\begin{equation}\label{up}
	U_t=\cos(t)\1-i\sin(t)\mathbf{n}\cdot\bfsigma,
\end{equation}
where $\mathbf{n}=(n^1,n^2,n^3)$ is a unit vector.
The curve is periodic with period $\tau=\pi$, which we take as the final time.

\begin{figure}[t]
	\centering
	\includegraphics[width=0.56\textwidth]{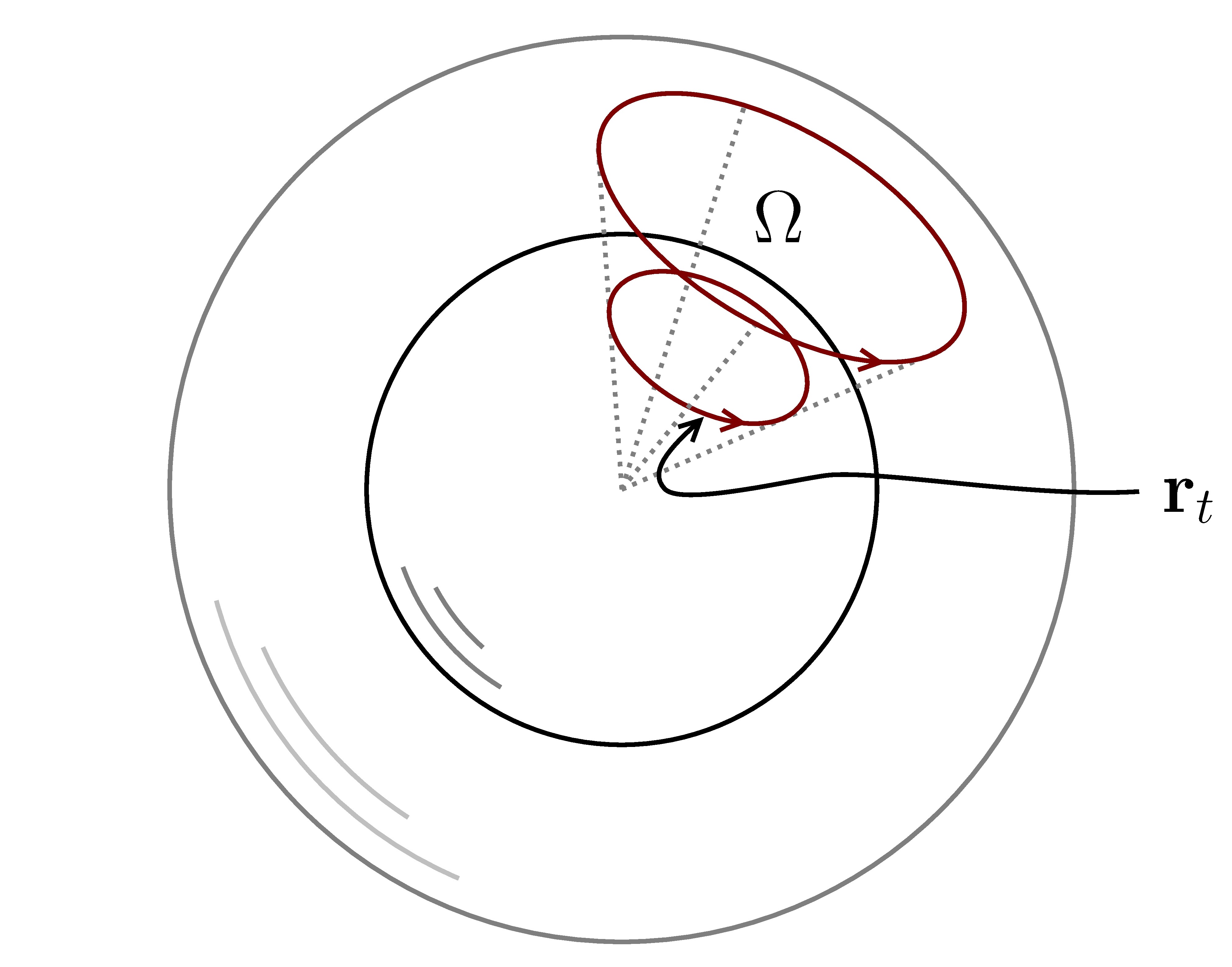}
	\caption{The solid angle of the curve $\mathbf{r}_t$ in the Bloch ball. The radial projection of $\mathbf{r}_t$ on the Bloch sphere separates the Bloch sphere into two regions.
The solid angle is the area of the region `lying to the left' of the projection.}
	\label{fig: solid angle}
\end{figure}

Define $\ket{\psi_{0;t}}=U_t\ket{0}$ and $\ket{\psi_{1;t}}=U_t\ket{1}$.
By Eqs.~\eqref{qubit holonomy} and \eqref{qubit gpf}, the holonomy of $\rho_t$
at $\sqrt{\rho_0}$ and the geometric phase factor is 
\begin{subequations} 
\begin{align}
	\Hol_{\sqrt{\rho_0}}
	&= \ketbra{0}{0}e^{i(1+n^3)\pi}+\ketbra{1}{1}e^{i(1-n^3)\pi},\\
	\Geop 
	&= p^0e^{i(1+n^3)\pi} + p^1e^{i(1-n^3)\pi}.
\end{align}
\end{subequations}
We can interpret these results geometrically.
Let $\mathbf{r}_t$ be the curve in the Bloch ball corresponding to 
the curve $\rho_t$. 
The curve $\mathbf{r}_t$ extends from $\mathbf{r}_0=(0,0,p^0-p^1)$
and traces out a circle in a plane perpendicular to $\mathbf{n}$.
Let $\Omega=2\pi(1-\cos\theta)$ be the solid angle of the circle traced out by $\mathbf{r}_t$, see Figure \ref{fig: solid angle}. 
Then 
\begin{subequations} 
\begin{align}
	i(1+n^3)\pi &= i(1+\cos\theta)\pi=i2\pi-i(\Omega/2),\\
	i(1-n^3)\pi &= i(1-\cos\theta)\pi=i(\Omega/2).
\end{align}
\end{subequations}
Consequently,
\begin{subequations} 
\begin{align}
	&\Hol_{\sqrt{\rho_0}}
	= \ketbra{0}{0}e^{-i(\Omega/2)}+\ketbra{1}{1}e^{i(\Omega/2)},\label{h}\\
	&\Geop = p^0e^{-i(\Omega/2)} + p^1e^{i(\Omega/2)}.\label{g}
\end{align}
\end{subequations}

Finally, if we take $\mathbf{n}=(0,-1,0)$, then $\rho_t$ 
is identical to the curve considered in Section \ref{Uhlmann kubit}.
In this case, $\Omega=2\pi$ and, hence, $\Geop = -1$.
Thus, irre\-spective of the initial spectrum, the geometric phase of $\rho_t$
is $\pi$. This shows that the geometric phase developed here is indeed different from the 
geometric phase of Uhlmann as well as the Wigner-Yanase and complementary geometric phases.
 
\subsubsection{Holonomy of isodegenerate mixed qubit states}
Formulas \eqref{qubit holonomy} and \eqref{qubit gpf} show that the holonomy
of a curve of isodegenerate mixed qubit states
depends only on the Aharonov-Anandan geometric phase factors of
the pure state curves represented by the evolving eigenprojectors 
and on the spectrum of the initial 
state. We can use this observation to determine the holonomy and the geometric phase. For example, if $\mathbf{r}_t$ runs once around the origin in the
$xz$-plane, so that the radial projection of $\mathbf{r}_t$ onto the Bloch sphere parameterizes a great circle, then
$\Hol_{\sqrt{\rho_0}}= -\1$ and $\Geop=-1$. And if $\mathbf{r}_t$ is confined to the $xz$-plane 
but does not encircle the origin, then $\Hol_{\sqrt{\rho_0}}= \1$ and $\Geop=1$. 
This follows immediately from Eqs.~\eqref{h} and \eqref{g}.
In Section \ref{sec: real curves} we will derive a topological result which for mixed qubit
states implies that the holonomy and the geometric phase factor of a curve in the $xz$-plane are homotopy invariants. From this follows that if $\mathbf{r}_t$ 
encircles the origin an even number of times, then the holonomy is $\1$ and the geometric phase factor is $1$.
And if $\mathbf{r}_t$ 
encircles the origin an odd number of times, then the holonomy is $-\1$ and the geometric phase factor is $-1$.

We finish this section with some formulas using which the 
holonomy and the geometric phase can be derived in the general case.
Thus, consider a curve of density operators $\rho_t=(\1+\mathbf{r}_t\cdot\bfsigma)/2$, where $\mathbf{r}_t = (x_t,y_t,z_t)$.
Writing $r_t=\sqrt{x_t^2+y_t^2+z_t^2}$, the eigenvalues and eigenvectors
of $\rho_t$ are
\begin{subequations}
\begin{align}
	p^0&=\frac{1}{2}(1+r_t), & \ket{\psi_{0;t}}
	= \frac{(r_t+z_t)e^{i\theta^0_t}}{\sqrt{2r_t(r_t+z_t)}}\ket{0}
	+ \frac{(x_t+iy_t)e^{i\theta^0_t}}{\sqrt{2r_t(r_t+z_t)}}\ket{1},\label{nollanden}\\
	p^1&=\frac{1}{2}(1-r_t), & \ket{\psi_{1;t}}
	= \frac{(iy_t-x_t)e^{i\theta^1_t}}{\sqrt{2r_t(r_t+z_t)}}\ket{0}
	+ \frac{(r_t+z_t)e^{i\theta^1_t}}{\sqrt{2r_t(r_t+z_t)}}\ket{1}.\label{ettanden}
\end{align}
\end{subequations}
The phase factors $e^{i\theta^0_t}$ and $e^{i\theta^1_t}$ are arbitrary 
and can be chosen so that $\ket{\psi_{0;t}}$ and $\ket{\psi_{1;t}}$
satisfies the parallel transport condition \eqref{para}.
Clearly, the formulas \eqref{nollande} and \eqref{ettande} are valid only if $z_t\ne -r_t$.
For $z_t=-r_t$,
\begin{subequations}
\begin{align}
	&\ket{\psi_{0;t}} = \ket{1}e^{i\theta^0_t},\\
	&\ket{\psi_{1;t}} = \ket{0}e^{i\theta^1_t}.
\end{align}
\end{subequations}

\subsection{Nonextendability of the connection}
We have constructed fiber bundles over the spaces
of isodegenerate mixed states and we have equipped these bundles with connections.
A natural question is if these bundles and connections can be extended to a larger class of density operators. 
Here, we show that this is in general not possible.

\begin{figure}[t]
	\centering
	\includegraphics[width=0.75\textwidth]{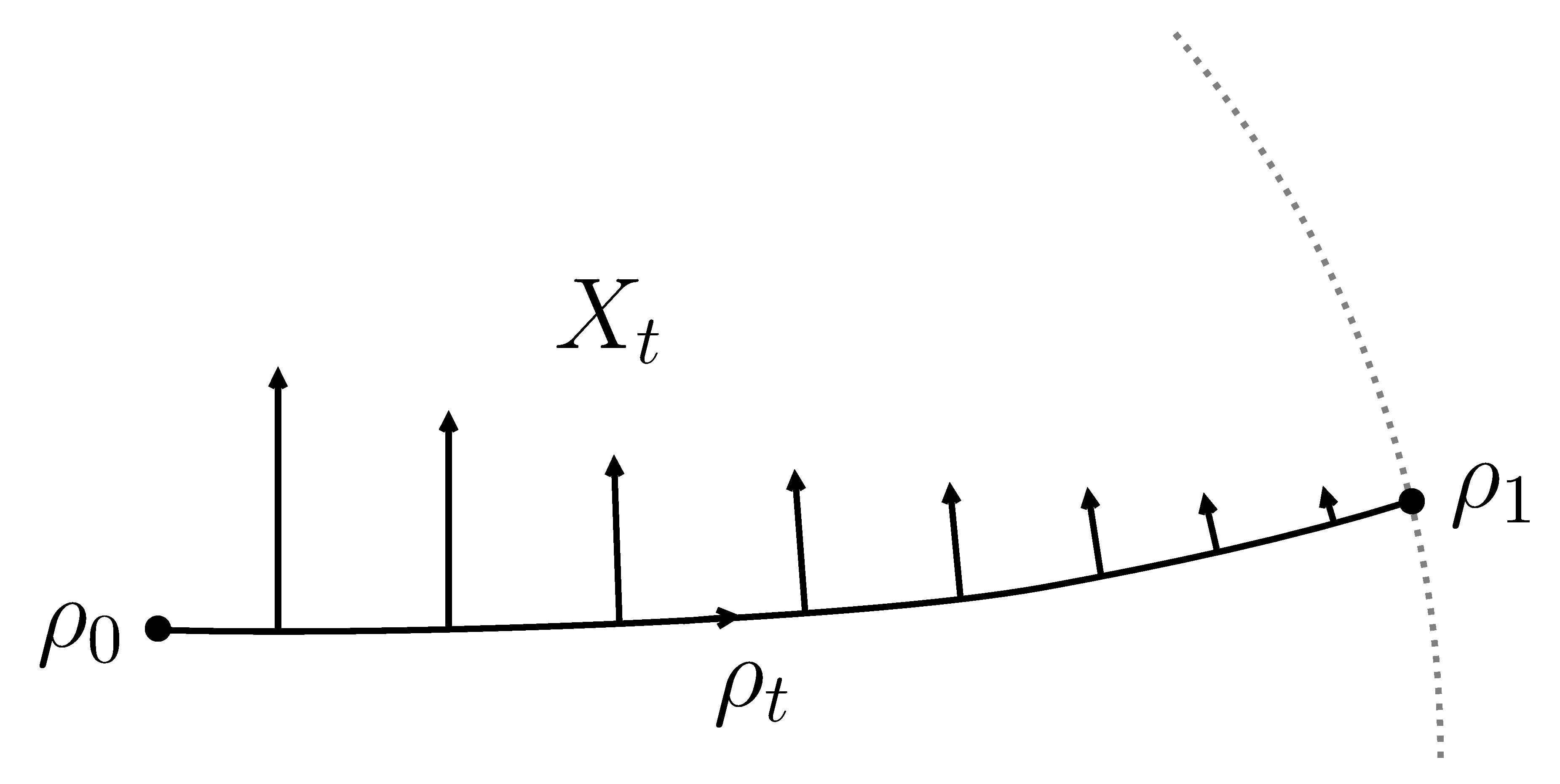}
	\caption{A vector field along $\rho_t$ which converges to the zero-vector at the boundary of $\SS_{\bfm}$. Since the vectors converge to the zero-vector but the lengths of the vectors do not converge to zero, the metric cannot be extended to a neighborhood of the limiting density operator.}
	\label{fig: vector field}
\end{figure}

Since the symmetry group of $\p_{\bfl}$ acts by isometries we can define a metric $g$ on $\SS_{\bfm}$ by projecting down the metric $\GHS$.
We prove that $g$ does not extend over the boundary of the closure of $\SS_{\bfm}$.
To this end, consider an eigenvalue spectrum $\bfP$ and an eigenprojector spectrum $\bfL$  compatible with $\bfm$.
Define a curve of density operators in $\SS_{\bfm}$
by $\rho_t=\sum_a P^a_t\Lambda_a$, $0\leq t< 1$, where
\begin{equation}
	P^a_t=
	\begin{cases}
		t(P^2-P^1)/m_1+P^1  &\text{ if $a=1$},\\
		t(P^1-P^2)/m_2+P^2  &\text{ if $a=2$},\\
		P^a                       &\text{ otherwise}.
	\end{cases}
\end{equation}
Let $l$ be the common length of the spectra $\bfP$ and $\bfL$
and set $m_{l+1}=n-k$, where $k$ is the rank of the density operators in $\SS_{\bfm}$. 
Fix an orthonormal basis $\{\ket{a;b_a}:a=1,2,\dots,l+1; b_a=1,2,\dots,m_a\}$
for $\HH$ such that $\Lambda_a=\sum_{b_a}\ketbra{a;b_a}{a;b_a}$, and define unitary operators $U_\eps$ as
\begin{equation}
\begin{split}
	U_\eps=\big(\cos \eps\ket{1;1} &+ \sin \eps\ket{2;1}\big)\bra{1;1} -
		 \big(\sin \eps\ket{1;1} - \cos \eps\ket{2;1}\big)\bra{2;1} \\
		 &+\sum_{a=1}^2 \sum_{b_a=2}^{m_a} \ketbra{a;b_a}{a;b_a} + 
		 \sum_{a=3}^{l+1} \sum_{b_a=1}^{m_a} \ketbra{a;b_a}{a;b_a}.
\end{split}
\end{equation}
Furthermore, define a vector field $X_t$ along $\rho_t$ by
\begin{equation}
	X_t=\dd{\eps} U_\eps\rho_t U_\eps^\dagger\big|_{\eps=0}.
\end{equation}
See Figure~\ref{fig: vector field}. 
As $t\to 0$, the curve $\rho_t$ converges to a 
density operator $\rho_1$ on the boundary of $\SS_{\bfm}$,
and the vector field $X_t$ converges to the zero-vector at $\rho_1$.
The following bound on the squared lengths of the vectors $X_t$ shows that $g$ cannot be extended over a neighborhood of $\rho_1$:
 \begin{equation}
	g(X_t,X_t)=P^1_t+P^2_t \geq \frac{P^1}{m_2}+\frac{P^2}{m_1}>0.
\end{equation}
We conclude that the connection cannot be extended to a fiber over $\rho^1$.

\subsubsection{Nonextendability of the connection in the case of qubits}
We denote the metric in the interior of the Bloch ball obtained by pulling back 
the metric $g$ in $\SS_{(1,1)}$ by the diffeomorphism \eqref{diffeomorphismen} also by $g$. Consider a curve of qubit density operators $\rho_t=U_t\rho_0 U_t^\dagger$ where
$\rho_0 = p^0\ketbra{0}{0} + p^1\ketbra{1}{1}$ and $U_t=\cos(t)\1+i\sin(t)\sigma_x$.
The corresponding curve in the Bloch ball is a circle
of Euclidean radius $r=p^0-p^1$: $\mathbf{r}_t=(0,r\sin(2t),r\cos(2t))$.
To calculate the squared speed of $\mathbf{r}_t$,
we first note that 
$\psi_t = \sqrt{p^0} U_t\ketbra{0}{0}+\sqrt{p^1}U_t\ketbra{1}{1}$
is a horizontal lift of $\rho_t$.  Consequently,
\begin{equation}
\begin{split}
	g(\dot{\mathbf{r}}_t,\dot{\mathbf{r}}_t)
	= g(\dot{\rho}_t,\dot{\rho}_t)
	= \GHS(\dot{\psi}_t,\dot{\psi}_t)
	= \Tr(\dot{\psi}_t^\dagger\dot{\psi}_t)
	= 1.
\end{split}
\end{equation}
A peculiar feature of this result is that it does not depend on the spectrum and, hence, not on the radius $r$. Consequently, irrespective of the spectrum,  if traversed once, the circle $\mathbf{r}_t$ has the length
\begin{equation}
	\length{\mathbf{r}_t} 
	= \int_0^{\pi}\dt \sqrt{g(\dot{\mathbf{r}}_t,\dot{\mathbf{r}}_t)}
	= \pi.
\end{equation}
Moreover, by rotational symmetry of $g$, every unitary orbit $\SS_{(p^0,p^1)}$ has the area $\pi$. 
We can now consider the limit when $r\to 0$.
In the Euclidean picture of the Bloch ball, the circles $\mathbf{r}_t$ then converge to a 
constant curve at the origin, which corresponds to the maximally mixed qubit state. 
But since the lengths of these circles do not converge to $0$ according to $g$,
we conclude that $g$ cannot be extended over the maximally mixed state.

\begin{remark*}
Here we have only proven that the 
connection cannot be extended over density operators obtained by the `merging' of two nonzero eigenvalues.
One can ask if it is possible to extend the connection over density operators obtained 
by putting one, or more, eigenvalues to zero. For example, can the connection on $\SS_{(1,1)}$ 
be extended to the pure states? The answer is not known.
\end{remark*}

\section{Curves of real density operators}\label{sec: real curves}
Each density operator can be described by a real matrix 
relative to some basis for $\HH$.
But it is not true, in general, that all the density operators in a curve can be described by real matrices relative to one and the same basis. In this section we consider curves of density operators
for which this \emph{is} the case.
We will call such curves ``real curves'' or ``real developments''.
Notice that, without loss of generality, we can take the basis 
to be an eigenbasis of the initial density operator.

\subsubsection{Real developments of qubits}
A curve of mixed qubits is real if, and only if, the corresponding curve in the Bloch ball is confined to a plane through the origin.

\subsubsection{Curves with anti-unitary, involutive symmetries}
Suppose that all the density operators in a curve commute 
with an anti-unitary involution $T$. Then the curve is real.
To see this, inductively define a $T$-invariant basis for $\HH$ by letting
$\ket{k}=a_k\ket{k'}+a_k^*T\ket{k'}$,
where $\ket{1'}$ is any nonzero vector, $\ket{k'+1}$ is a vector perpendicular 
to the linear span of $\{\ket{1},\ket{2},\dots,\ket{k}\}$, and 
the $a_k$s are chosen so that the vectors $\ket{k}$ are normalized.
The so obtained basis $\{\ket{1},\ket{2},\dots,\ket{n}\}$ is orthogonal since if $l>k$, then
\begin{equation}
	\braket{k}{l}
	= a_l\braket{k}{l'} + a_l^* \big{\langle} k\,\big| \,T\ket{l'}\big\rangle
	= a_l^*\braket{k}{l'}^*=0.
\end{equation}
Relative to this basis,
\begin{equation}
	\bra{k}\rho_t\ket{l}
	=\big{\langle} T\ket{k}\, \big{|}\, \rho_t T\ket{l} \big{\rangle}
	=\big{\langle} T\ket{k}\, \big{|}\, T\rho_t \ket{l} \big{\rangle}
	=\bra{k}\rho_t\ket{l}^*.
\end{equation}
Thus, $\rho_t$ is real. Notice also that we can assume that the initial density operator is diagonal in 
the constructed basis. For if $\ket{k'}$ is an eigenvector of $\rho_0$ with eigenvalue $p^k$, then so is $\ket{k}$. 
We can now consistently choose the $\ket{k'}$s as eigenvectors of $\rho_0$.

\subsection{Holonomy of real developments}
Consider a closed curve $\rho_t$ in $\SS_{\bfm}$.
Let $\bfL$ be the eigenprojector spectrum of the initial density operator $\rho_0$. We can expand $\rho_t$ as 
\begin{equation}\label{real curve}
	\rho_t = \sum_a P^a_t \Lambda_{a;t} = \sum_a P^a_t S_t\Lambda_a S_t^\dagger,
\end{equation}
where $S_t$ is a curve of unitary operators which, 
without loss of generality, we can assume extends from the identity operator.
We say that $\rho_t$ is a real development, or a real curve, if
the $S_t$s can be chosen such that they are represented by special orthogonal matrices relative an
eigenbasis for the initial operator $\rho_0$.
The assumption that \eqref{real curve} is a real curve is
equivalent to the assumption that there is an orthonormal basis 
\begin{equation}\label{big basis}
	\{\ket{a;b_a}:a=1,2,\dots l+1; b_a=1,2,\dots,m_a\}
\end{equation}
for $\HH$, where $l$ is the length of $\bfm$ and $m_{l+1}\equiv n-k$, such that
\begin{subequations}
\begin{align}
	&\Lambda_a=\sum_{b_a}\ketbra{a;b_a}{a;b_a},\\
	&\1-\sum_a\Lambda_a=\sum_{b_{l+1}}\ketbra{l+1;b_{l+1}}{l+1;b_{l+1}},
\end{align}
\end{subequations}
and such that the matrix with entries $\bra{b_a;a}S_t\ket{{b'}_{a'};a'}$ is special orthogonal.
Let $\bfl$ be an eigenprojector spectrum for density operators on $\HH^k$ compatible with $\bfm$, and 
let 
\begin{equation}\label{little basis}
	\{\ket{ab_a}:a=1,2,\dots,l; b_a=1,2,\dots,m_a\}
\end{equation}
be an orthonormal basis for $\HH^k$ such that 
\begin{equation}
	\lambda_a=\sum_{b_a=1}^{m_a}\ketbra{ab_a}{ab_a}.
\end{equation}
The curve $\psi_t=\sum_{a,{b_a}}\sqrt{P^a_t}S_t\ketbra{a;b_a}{ab_a}$ 
is a curve of amplitudes for the $\rho_t$s, and according to Eq.~\eqref{the holonomy}, the holonomy of $\rho_t$ 
at $\psi_0$ is 
\begin{equation}
	\Hol_{\psi_0}
	=(\psi_0^\dagger\psi_0)^{-1}\psi_0^\dagger\psi_\tau\texp\Big(\!-\!\int_0^\tau\dt\,\A(\dot{\psi}_t^u)\Big),
\end{equation}
where $\dot{\psi}^u_t=\sum_{a,{b_a}}\sqrt{P^a_t}\dot{S}_t\ketbra{a;b_a}{ab_a}$.
The connection applied to $\dot{\psi}^u_t$ is
\begin{equation}
	\A(\dot{\psi}_t^u)
	=\sum_a\sum_{b_a}\sum_{b'_a}\ketbra{ab_a}{a;b_a}S_t^T\dot{S}_t\ketbra{a;b'_a}{ab'_a},
\end{equation}
and 
\begin{equation}
	(\psi_0^\dagger\psi_0)^{-1}\psi_0^\dagger\psi_\tau
	=\sum_a\sum_{b_a}\sum_{b'_a} \ketbra{ab_a}{a;b_a} S_{\tau} \ketbra{a;b'_a}{ab'_a}.
\end{equation}
Apparently, the holonomy, when expanded as a matrix relative to the basis~\eqref{little basis}, belongs to the block-diagonal group
\begin{equation}
	\mathbb{SO}(\bfl)\equiv
	\{\diag(O^1,O^2,\dots,O^l)\in\mathbb{SO}(k):O^a\in\mathbb{O}(m_a)\}.
\end{equation}
Here, $\mathbb{O}(m_a)$ is the group of orthogonal $m_a\times m_a$-matrices and 
$\mathbb{SO}(k)$ is the group of special orthogonal $k\times k$-matrices.
 
The geometric phase factor of $\rho_t$,
\begin{equation}
	\Geop
	=\sum_a\sum_{b_a}P^a_0\bra{ab_a}\Hol_{\psi_0}\ket{ab_a},
\end{equation}
is real and, hence, the geometric phase 
\begin{equation}
	\tgeo=\arg\Big(\sum_a\sum_{b_a}P^a_0\bra{ab_a}\Hol_{\psi_0}\ket{ab_a}\Big)
\end{equation}
is either $0$ or $\pi$ modulo integer multiples of $2\pi$.
Interestingly, if the initial state is `almost pure', i.e., if $m_1=1$
and $p^1_0>1/2$, the geometric phase will agree with the Aharonov-Anandan phase of the eigenprojector curve extending from the projection onto the $p^1_0$-eigenspace of the initial state.

\subsubsection{Real purification bundles}
There is a quicker route to the result above.
Define $\SS_{\bfm}^\reals$ as the intersection 
of $\SS_{\bfm}$ and the real span of the operators of the form 
$\ketbra{a;b_a}{a';b'_{a'}}$, i.e., the outer products of all pairs of vectors from the basis \eqref{big basis}.
Furthermore, let $\HH_{\reals}$ be the real span of the basis \eqref{big basis},
and let $\HH^k_{\reals}$ be the real span of the basis \eqref{little basis}.
In complete analogy with the theory developed in Sections \ref{sec: isospectral} and \ref{sec: isodegenerate},
we can define fiber bundles $\p_{\bfl}^\reals$ from $\WW_{\bfl}^\reals$ onto $\SS_{\bfm}^\reals$,
where the elements of $\WW_{\bfl}^\reals$ are $\reals$-linear maps from 
$\HH^k_{\reals}$ to $\HH_{\reals}$.
The symmetry group of $\p_{\bfl}^\reals$ is $\OO_{\bfl}$,
which is the group of orthogonal operators on $\HH^k_\reals$ which commute
with all the projectors $\lambda_a$. 
The holonomies belong to the group $\SO_{\bfl}$ of the special orthogonal operators 
on $\HH^k_\reals$ which commute
with the $\lambda_a$s.

\subsubsection{Real, nondegenerate developments}
We have seen that the holonomy of a closed curve of 
nondegenerate density operators is a diagonal matrix
with the diagonal entries being the Aharonov-Anandan phase factors for the eigenprojector curves of the density operators.
The product of the phase factors is $1$ since the holonomy is special unitary and, hence, the geometric phases of the eigenprojector curves sum up to $0\mod 2\pi$.
If the curve is real, these phase factors are $+1$ or $-1$.
Consequently, the geometric phase factor is
\begin{equation}
	\Geop=\pm p^1_0 \pm p^2_0 \pm \dots \pm p^k_0,
\end{equation}
where the signs correspond to the Aharonov-Anandan phase factors.

\subsection{Holonomy and homotopy}
A one-parameter deformation of a curve $\rho_t$ 
is a continuous one-parameter family of curves 
$\rho_{t;s}$, $0\leq s\leq 1$, such that 
$\rho_{t;0}=\rho_t$.
The deformation is called a homotopy if the endpoints remain fixed
during the deformation, i.e., if $\rho_{0;s}=\rho_0$ and $\rho_{\tau;s}=\rho_\tau$ for all $s$. 
Thus, if $\rho_t$ is a closed curve, a homotopy is a one-parameter deformation of $\rho_t$ through closed curves at $\rho_0$. 

Homotopy is an equivalence relation on the space of closed curves at $\rho_0$.
The set of equivalence classes is denoted $\pi_1(\SS_{\bfm},\rho_0)$.
We can equip this set with a group structure by defining the product of two 
equivalence classes by
\begin{equation}
	[\rho_t][\sigma_t]=[(\rho*\sigma)_t].
\end{equation}
The group $\pi_1(\SS_{\bfm},\rho_0)$ is the fundamental group of $\SS_{\bfm}$ at $\rho_0$.

\begin{figure}[t]
	\centering
	\includegraphics[width=0.9\textwidth]{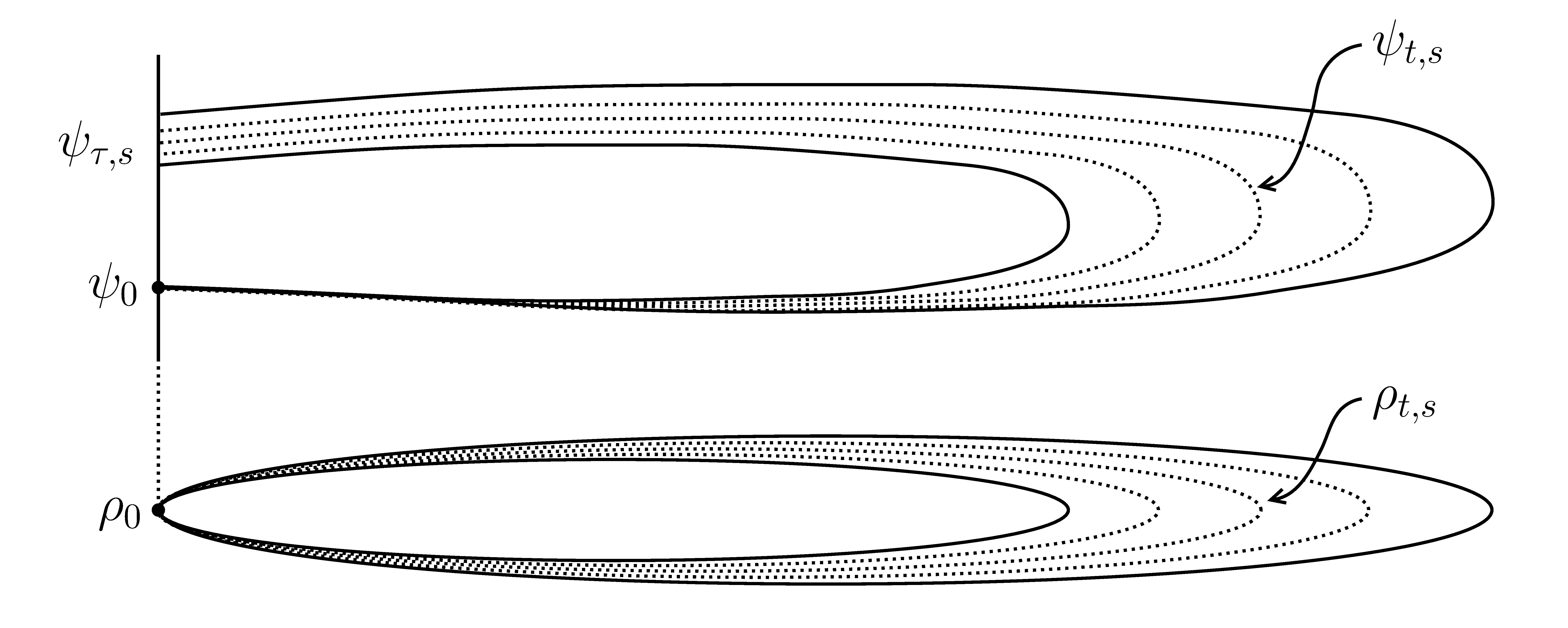}
	\caption{A lift of a homotopy of a curve of density operators.
	The lift can be chosen such that it is via horizontal curves.
	Therefore, the holonomy, the geometric phase factor, and the geometric phase vary continuously with the deformation.}
	\label{fig: homotopy lift}
\end{figure}

The bundle $\p_{\bfl}$ has the homotopy lifting property \cite[Ch.\,1, Sec.\,5]{Hu1994}: 
If $\psi_0$ is an amplitude for $\rho_0$, the homotopy 
$\rho_{t,s}$ lifts to a one-parameter deformation 
$\psi_{t,s}$ such that $\psi_{0,s}=\psi_0$ for all $s$.
See Figure \ref{fig: homotopy lift}.
We can further assume that the curves $t\to \psi_{t,s}$
are horizontal for each $s$. From this follows that the holonomy and the geometric phase factor varies continuously under homotopies.
In the particular case when the homotopies are by real curves of density operators, the geometric phase remains fixed. 
This is so because the geometric phase factor remains positive or negative during the homotopy. For real, nondegenerate curves also the holonomy remains fixed since in this case $SO_{\bfl}$ is a discrete group.

\chapter{Applications}\label{apps}
In this final chapter we have included three applications of the theory developed in Chapter \ref{ch: sjoqvist holonomy}.
In Section \ref{sec: nodes} we introduce higher order geometric phases,
in Section \ref{sec: anur} we derive an uncertainty relation, and in Section \ref{Oqsl} we extend the quantum speed limits of Mandelstam-Tamm and Margolus-Levitin to mixed quantum states. References for these applications are \cite{AnHe2013, AnHe2014a, AnHe2014b}.

\section{Nodes and higher order geometric phases}\label{sec: nodes}
The geometric phase of a curve of density operators is defined
as the argument of the geometric phase factor, see Eq.~\eqref{gffgf}.
Of course, this is a proper definition only if the geometric phase factor does not vanish; if the geometric phase factor \emph{does} vanish, the geometric phase is undefined.
In this section we introduce ``higher order geometric phases'' and
prove that under a generically satisfied condition there will always be a well-defined geometric phase of some order.

\subsection{Nodes}
Let $\bfm$ be a degeneracy spectrum of rank $k$ density operators on $\HH$.
Consider a curve $\rho_t$ in $\SS_{\bfm}$\ and let $\bfl$ be a degeneracy spectrum for density operators on $\HH^k$.
The geometric phase factor of $\rho_t$ is 
$\Geop=\Tr(\psi_0^\dagger\Gamma(\psi_0))$,
where $\psi_0$ is any amplitude in $\WW_{\bfl}$ for the initial density operator 
and $\Gamma$ is the parallel transport operator associated with $\rho_t$.
We call the final density operator $\rho_\tau$ a node if $\Geop=0$.
If the final state is \emph{not} a node, the geometric phase is defined as the argument of
the geometric phase factor. But if the final state \emph{is} a node, the geometric phase is not 
defined.

\subsubsection{Nodal structure for qubits}
\begin{figure}[t]
	\centering
	\includegraphics[width=0.45\textwidth]{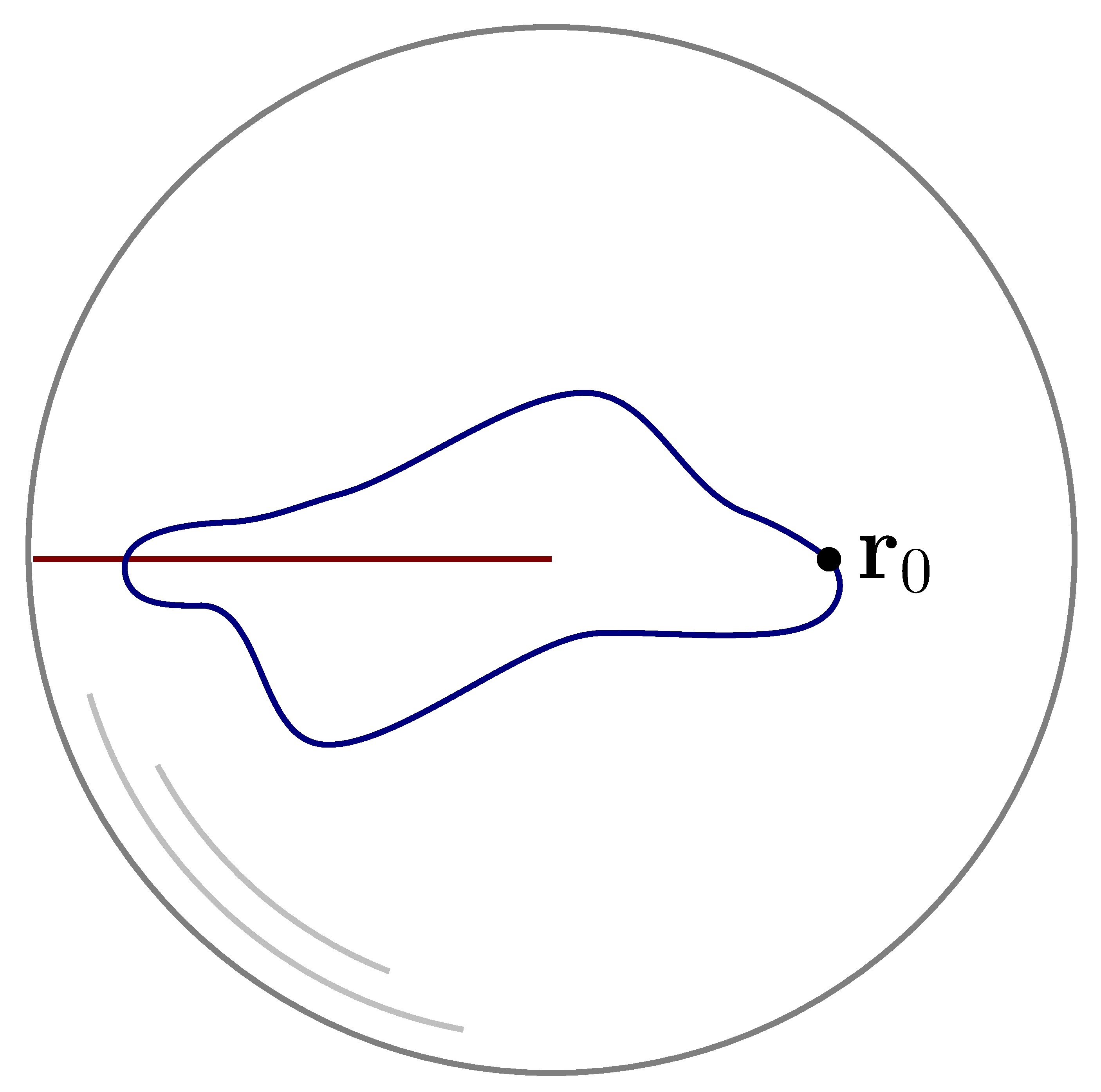}
	\caption{The nodes of a qubit system. 
	The nodes of a curve in the Bloch ball are located on the classical manifold opposite to the initial state. }
	\label{fig: qubit nodes}
\end{figure}

By varying the final time $\tau$, we can speak of the nodes for a developing system.
For a qubit system it is particularly easy to locate the nodes;
the nodes correspond to all the points on the radial segment opposite to the inital state in the Bloch ball, see Figure \ref{fig: qubit nodes}.
To see this consider a curve $\rho_t$ as in Eq.~\eqref{mixed kub}.
We assume that the initial state is diagonal in the computational basis $\{\ket{0},\ket{1}\}$,
so that the initial state in the Bloch ball is $\mathbf{r}_0=(0,0,p^0_0-p^1_0)$.
Define a lift $\psi_t$ of $\rho_t$ by Eq.~\eqref{plain}, where the 
vectors $\ket{\psi_{0:t}}$ and $\ket{\psi_{1;t}}$ are given by 
Eqs.~\eqref{nollanden} and \eqref{ettanden}, respectively, 
and adjust the phases $\theta^0_t$ and $\theta^1_t$ so that 
$\psi_t$ is horizontal. Write $r_t$ for the Euclidean 
length of the position vector $\mathbf{r}_t$.
The geometric phase factor is, then,
\begin{equation}
	\Geop=\sqrt{\frac{r_t+z_t}{2r_t}}\Big(\sqrt{p^0_0\,p^0_\tau}e^{i\theta^0_t}+\sqrt{p^1_0\,p^1_\tau}e^{i\theta^1_t}\Big).
\end{equation}
The parenthesis does not vanish because $p^0_0\,p^0_\tau> 1/4$ and $p^1_0\,p^1_\tau<1/4$. Thus, the final state is a node if, and only if, $z_\tau=-r_\tau$. This means that $\mathbf{r}_\tau$ is on the radial segment in the Bloch ball opposite to the initial state.

\subsection{Higher order geometric phases}
Let $l$ be the common length of $\bfm$ and $\bfl$. That is,
assume that the density operators $\rho_t$ have $l$ different nonzero eigenvalues.
Let $\psi_0$ be an amplitude for $\rho_0$ in $\WW_{\bfl}$.
For each sequence $\alpha=(a_1,a_2,\dots,a_d)$ of integers $1\leq a_i\leq l$ 
define the geometric phase factor and geometric phase for $\rho$ corresponding to $\alpha$ by
\begin{subequations}
\begin{align}
	\Geop^\alpha
	&=\Tr\Big(\prod_{i} \lambda_{a_i}\psi_0^\dagger\Gamma(\psi_0)\lambda_{a_{i+1}}\Big)
	\text{ where } \lambda_{a_{d+1}}\equiv\lambda_{a_1}, \label{higher geophase factor}\\
	\tgeo^\alpha&=\arg \Geop^\alpha.\label{higher geophase}
\end{align}
\end{subequations}
These quantities are independent of the initial amplitude since the unitaries in the symmetry group $\UU_{\bfl}$ commute with the 
projectors $\lambda_a$. 
The usual geometric phase factor can be recovered from the geometric phase factors corresponding to single element sequences:
\begin{equation}
	\Geop=\sum_{a=1}^l\Geop^{(a)}.
\end{equation}
More generally, if $d$ is any positive integer,  
\begin{equation}
	\Geop^d\equiv
\Tr\Big(\big(\psi_0\Gamma(\psi_0)\big)^d\Big)=\sum_\alpha\Geop^\alpha
\end{equation}
where the sum is over all the different sequences 
$\alpha$ of length $d$ that can be formed from integers $1\leq a_i\leq l$.
The quantity $\Geop^d$ is the 
$d$th order geometric phase factor of $\rho_t$, and
we define the $d$th order geometric phase as
\begin{equation}\label{eq:high}
	\tgeo^d=\arg\Geop^d.
\end{equation}

\subsubsection{Properties of the $d$th order geometric phase}
For a closed curve, the geometric phase factor corresponding to 
a sequence $\alpha$ vanishes if $\alpha$ is not constant;
if $\alpha=(a_1,a_2,\dots,a_d)$, then 
\begin{equation}\label{property}
	\prod_{i} \lambda_{a_i}\psi_0^\dagger\Gamma(\psi_0)\lambda_{a_{i+1}}
	=\begin{cases}
		(P^a)^d\lambda_a\Hol_{\psi_0}^d \lambda_{a}
		\text{ if $a_i=a$ for all $i$,}\\
		0\text{ otherwise.}
\end{cases}
\end{equation}
Here, $P^a$ is the $a$th distinct nonzero eigenvalue of $\rho_0=\rho_\tau$.
Equation \eqref{property} is a consequence of the fact that the holonomy of $\rho_t$ commutes with the projectors $\lambda_a$. 
It is apparent from this equation that $\Geop^\alpha$ is a  
homogeneous polynomial of degree $d$ in the entries of $\Hol_{\psi_0}$.
Hence the name ``higher order'' geometric phase factors.
For a periodic evolution, $\Hol_{\psi_0}^d$ 
is the holonomy of the system when it has gone trough $d$ cycles.
This follows from the group property \eqref{group property}.

\subsection{The existence of higher order geometric phases}
If $\rho_0$ and $\rho_\tau$ are fully distinguishable \cite{En1996,MaMiPuZy2008}, then $\psi_0^\dagger\Gamma(\psi_0)=0$ and, hence, $\Geop^\alpha=0$ for every sequence $\alpha$. But this situation is very nongeneric and, e.g., cannot appear if the rank of the $\rho_t$s is greater than half of the the dimension of $\HH$.
If $\psi_0^\dagger\Gamma(\psi_0)$ has at least one nonzero eigenvalue, then 
$\rho_t$ has a well-defined geometric phase of some order less than or equal to the rank $k$.
To see this, we first observe that the 
 characteristic polynomial of $\psi_0^\dagger\Gamma(\psi_0)$ is
\begin{equation}
	\det\big(x-\psi_0^\dagger\Gamma(\psi_0)\big)
	=\sum_{a=0}^{k}(-1)^as_a\,x^{k-a},
\end{equation}
where
\begin{equation}
	s_0=1,\quad s_a=\frac{1}{a}\sum_{d=1}^{a}(-1)^{d-1}s_{a-d}\Geop^d.
\end{equation}
It follows that 
if $\psi_0^\dagger\Gamma(\psi_0)$ has a nonzero eigenvalue, 
then $\Geop^d$ is nonzero for some $d\leq k$.

\subsubsection{Higher order geometric phases for qubits}
Consider a unitary evolution $\rho_t=U_t\rho_0U_t^\dagger$.
Assume that the initial density operator is given by \eqref{initial} and the evolution operator by
\eqref{up}. If we set the final time to $\tau=\pi/2$, 
then $\rho_\tau$ is a node and, hence, $\Geop=0$.
Consequently, the geometric phase of $\rho_t$ is \emph{not} defined.
But $\Geop^2=2p_1p_2>0$ and, hence, the second order geometric phase of $\rho_t$ \emph{is} defined. 

\subsection{Off-diagonal geometric phase factors}
Consider a curve of nondegenerate, isospectral mixed states. Such a curve has a unique spectral representation 
$\rho_t=\sum_{a}p^a\ketbra{\psi_{a;t}}{\psi_{a;t}}$.
If the spectral decomposition is parallel transported, the geometric 
phase factor is the expected value of the relative phase factors 
$\braket{\psi_{a;0}}{\psi_{a;\tau}}$.

If all the relative phase factors vanish, the geometric phase is not defined.
Manini and Pistolesi \cite{MaPi2000} then proposed to study the off-diagonal phase factors
\begin{equation}\label{offisar}
	\gamma_{a_1,a_2,\dots,a_d}^{(d)}
	=\prod_{i=1}^d \braket{\psi_{a_i;0}}{\psi_{a_{i+1};\tau}},\qquad \big(\ket{\psi_{a_{d+1};\tau}}\equiv\ket{\psi_{a_1;\tau}}\big).
\end{equation} 
These factors are invariant under the action of the symmetry group.
Mukunda \emph{et al.} \cite{MuArChSi2001} extended the work of Manini and Pistolesi. 
They expressed the off-diagonal phase factors in terms of Bargmann invariants, and, 
furthermore, used the fact that
the Bargmann invariants satisfy certain relations to give a full description of how the off-diagonal phase
factors are interrelated.

For degenerate mixed states, the off-diagonal phase factors need not 
be invariant under the action of the symmetry group.
But appropriate combinations of them are. For example,
\begin{equation}\label{exp}
	\Geop^{(a_1,a_2,\dots,a_d)}
	=\sum_{b_1}\sum_{b_2}\dots\sum_{b_d}p_{b_1}p_{b_2}\dots p_{b_d}
	\gamma_{b_1,b_2,\dots,b_d}^{(d)},
\end{equation}
where each index $b_s$ run through all integers satisfying
$0<b_s-\sum_{i=1}^{a_s-1}m_i\leq m_{a_s}$.
Thus, using \eqref{exp} and the results of Mukunda \emph{et al.}\ one can, in principle, describe 
all the relations between the higher order geometric phases. 

\section{An uncertainty relation}\label{sec: anur}
In quantum theory, observables of systems are represented by Hermitian operators.
A characteristic of quantum theory is the 
impossibility to fully predict the outcome of a measurement of an observable---the actual value of an observable cannot in general be known prior to measurement.
Only expectation values and variances of values of observables can be calculated.

Another characteristic of quantum theory is that there is a limit to the precision with which the values of a pair of observables can be known simultaneously.
This is the famous quantum uncertainty principle.
The classic uncertainty relations are the Heisenberg-Kennard-Weyl \cite{He1927,Ke1927,We1927} and Robertson-Schr{\"o}dinger \cite{Ro1929, 
Sc1930} uncertainty relations. Several generalizations of these uncertainty relations exist. Examples include relations involving the Wigner-Yanase skew-information \cite{Lu2003,Lu2005, Pa2005}, purity-bounded relations \cite{Do2002} and entropic uncertainty relations \cite{WeWi2010}.
In this section, we will derive a symplectic-geometric uncertainty relation.

\subsection{The Robertson-Schr\"{o}dinger uncertainty relation}
Let $A$ be an observable on $\HH$.
The precision with which the value of $A$ can be known 
is quantified by the variance,
$\var_\rho\!A = \Tr(\rho A^2)-\Tr(\rho A)^2$.
If $B$ is another observable, the precision with which the values of $A$ and $B$ can be known simultaneously is limited by the Robertson-Schr\"{o}dinger uncertainty relation \cite{Ro1929, Sc1930}:
\begin{equation}\label{Sur}
4\var_\rho\!A\var_\rho\!B
\geq
\big(\Tr(\rho\{A,B\})-2\Tr(\rho A)\Tr(\rho B)\big)^2-\big(\Tr(\rho[A,B])\big)^2.
\end{equation}

\subsection{Hamiltonian vector fields of observables}\label{Ham vec fields}
Consider the fiber bundle $\p_{\bfp;\bfl}$ over the unitary orbit $\SS_{\bfp}$
defined in Section \ref{sec: isospectral}. By the Marsden-Weinstein symplectic reduction theorem,
$\SS_{\bfp}$ admits a unique symplectic structure $\omega$ whose pullback to
$\WW_{\bfp;\bfl}$ equals the restriction of the Hilbert-Schmidt symplectic form $\OHS$. 
For an observable $A$, let $X_A$ be the Hamiltonian vector field of
the expectation value function $E_\rho(A)=\Tr(\rho A)$.
A straightforward calculation shows that $X_{A}(\rho)= -i[A,\rho]$.
Let $B$ be another observable and let $\psi$ be an amplitude of $\rho$.
The vectors $Y_A(\psi)=-iA\psi$ and $Y_B(\psi)=-iB\psi$ are lifts of 
$X_{A}(\rho)$ and $X_{B}(\rho)$ to $\psi$.
Consequently,
\begin{equation}\label{symp}
	\omega(X_A(\rho),X_B(\rho))
	=\OHS(Y_A(\psi),Y_B(\psi))
	=-i\Tr(\rho[A,B]).
\end{equation}
Notice that the lifts $Y_A(\psi)$ and $Y_B(\psi)$ need not be horizontal.
We say that an observable $A$ is parallel at $\rho$
if for some, hence every, amplitude $\psi$ of $\rho$ the lift $Y_A(\psi)$
\emph{is} horizontal. 

As we have seen in Section \ref{sec: isospectral}, we can define a metric $g$ 
on $\SS_{\bfp}$ by projecting down the the Hilbert-Schmidt Riemannian metric $\GHS$. 
\emph{In this application it turns out to be convenient to redefine $\GHS$ 
 as twice the real part of the Hilbert-Schmidt Hermitian product}. See Section \ref{sec: the HS metric}.
For observables $A$ and $B$ which are parallel at $\rho$, we then have that
\begin{equation}
	g(X_A(\rho),X_B(\rho))
	= \GHS(Y_A(\psi),Y_B(\psi))
	= \Tr(\rho\{A,B\}).
\end{equation}

\subsection{A geometric uncertainty relation}
Let $Y_A^h(\psi)$ and $Y_B^h(\psi)$ be the 
horizontal lifts of $X_A(\rho)$ and $X_B(\rho)$
or, equivalently, the horizontal projections 
of $Y_A(\psi)$ and $Y_B(\psi)$, respectively.
Furthermore, let $Y_A^v(\psi)$ and $Y_B^v(\psi)$ be the 
vertical projections of $Y_A(\psi)$ and $Y_B(\psi)$.
A key observation is that 
$\sqrt{2} E_\rho(A)$ equals the length of the projection 
of $Y_A^v(\psi)$ on the vertical line spanned by the unit vector $\chi(\psi)=-i\psi/\sqrt{2}$:
\begin{equation}
	\sqrt{2} E_\rho(A)
	= \GHS(\chi(\psi),Y_A(\psi))
	= \GHS(\chi(\psi),Y_A^v(\psi)).
\end{equation}
It follows that the variance of $A$ at $\rho$ is greater than half of the squared 
length the Hamiltonian vector field of $A$ at $\rho$:
\begin{equation}\label{pregur}
\begin{split}
	2\var_\rho\!A
	&= \Tr(\rho\{A,A\})-2 E_\rho(A)^2\\
	&= \GHS(Y_A^h(\psi),Y_A^h(\psi))+
			\GHS(Y_A^v(\psi),Y_A^v(\psi))-2 E_\rho(A)^2\\
	&\geq g(X_A(\rho),X_A(\rho)).
\end{split}
\end{equation}
Then, by the Cauchy-Schwarz inequality,
\begin{equation}\label{gur}
\begin{split}
	4\var_\rho\!A\var_\rho\!B
	&\geq g(X_A(\rho),X_A(\rho))\,g(X_B(\rho),X_B(\rho))\\
	&\geq g(X_A(\rho),X_B(\rho))^2+\omega(X_A(\rho),X_B(\rho))^2.
\end{split}
\end{equation}
We call the inequality of the left hand side and the right hand side of \eqref{gur} the geometric uncertainty relation.

\subsection[Comparison with the Robertson-Schr\"{o}dinger uncertainty relation]{Comparison with the Robertson-Schr\"{o}dinger \\ uncertainty relation}
For systems in pure states the uncertainty relation
\eqref{gur} is a geometric version of the Robertson-Schr\"{o}dinger uncertainty 
relation \eqref{Sur}, see \cite{Sc1996,AsSc1998}.
But for general mixed states, the two uncertainty relations are not equivalent. The 
difference is due to the multiple dimensions of the vertical bundles of the 
$\p_{\bfp;\bfl}$s. Before we compare the two uncertainty relations we
give a short geometric proof of the Robertson-Schr\"{o}dinger uncertainty relation. 
To this end, we denote the projection of $Y_A^v(\psi)$
on the hyperplane in $\V_{\psi}\WW_{\bfp;\bfl}$ perpendicular to $\chi(\psi)$
by $Z_A(\psi)$. By Eq.~\eqref{pregur}, the left hand side of \eqref{Sur} equals
\begin{equation}
\begin{split}
	g(X_A,X_A)&\,g(X_B,X_B)+\GHS(Z_A,Z_A)\,\GHS(Z_B,Z_B)\\
	+&\,g(X_A,X_A)\,\GHS(Z_B,Z_B)+g(X_B,X_B)\,\GHS(Z_A,Z_A).
\end{split}
\end{equation}
(Here we have omitted the references to the arguments $\rho$ and $\psi$
to increase the readability.)
By Eqs.~\eqref{symp} and \eqref{pregur}, the right hand side of \eqref{Sur} equals
\begin{equation}\label{nyckel}
	g(X_A,X_B)^2+\omega(X_A,X_B)^2+\GHS(Z_A,Z_B)^2
	+2g(X_A,X_B)\,\GHS(Z_A,Z_B).
\end{equation}
Equation \eqref{Sur} now follows from the three inequalities
\begin{subequations}
\begin{align}
	&g(X_A,X_A)\,g(X_B,X_B)\geq g(X_A,X_B)^2+\omega(X_A,X_B)^2,\\
	&\GHS(Z_A,Z_A)\,\GHS(Z_B,Z_B)\geq \GHS(Z_A,Z_B)^2,\\
	&g(X_A,X_A)\,\GHS(Z_B,Z_B)+g(X_B,\,X_B)\,\GHS(Z_A,Z_A)\nonumber\\
	&\hspace{155pt}\geq 2g(X_A,X_B)\,\GHS(Z_A,Z_B).
\end{align}
\end{subequations}
Apparently, the difference between uncertainty relations \eqref{Sur} and \eqref{gur} lies in the sum $\GHS(Z_A,Z_B)^2+2g(X_A,X_B)\,\GHS(Z_A,Z_B)$.
The relation \eqref{gur} is a geometric equivalent of 
\eqref{Sur} if this sum vanishes.
This is, e.g., the case if $A$ or $B$ is parallel at $\rho$, or if $\rho$ is pure, in which case the vertical bundle of $\p_{\bfp;\bfl}$ has
one-dimensional fibers. In general, however, the two relations are not equivalent.

\subsubsection{Uncertainty of qubit observables}
Consider a system in the mixed qubit state $\rho=p^0\ketbra{0}{0}+p^1\ketbra{1}{1}$. Let $A=\sigma_x+\eps\sigma_z$ and $B=\sigma_x-\eps\sigma_z$
where $\eps$ is some real, positive constant.
An amplitude for $\rho$ is $\psi=\sqrt{p^0}\ketbra{0}{0}+\sqrt{p^1}\ketbra{1}{1}$.
We have that
\begin{subequations}
\begin{align}
	\hspace{-5pt}Y_A(\psi) &= -i\sqrt{p^0}\ketbra{1}{0}-i\sqrt{p^1}\ketbra{0}{1}-i\eps\sqrt{p^0}\ketbra{0}{0}+i\eps\sqrt{p^1}\ketbra{1}{1},\\
	\hspace{-5pt}Y_B(\psi) &= -i\sqrt{p^0}\ketbra{1}{0}-i\sqrt{p^1}\ketbra{0}{1}+i\eps\sqrt{p^0}\ketbra{0}{0}-i\eps\sqrt{p^1}\ketbra{1}{1}.
\end{align}
\end{subequations}
The horizontal and vertical components of these vectors are given by
\begin{subequations}
\begin{align}
	Y_A^h(\psi) &= Y_B^h(\psi) = -i\sqrt{p^0}\ketbra{1}{0}-i\sqrt{p^1}\ketbra{0}{1},\\
	Y_A^v(\psi) &= -Y_B^v(\psi) = -i\eps\sqrt{p^0}\ketbra{0}{0}+i\eps\sqrt{p^1}\ketbra{1}{1}.
\end{align}
\end{subequations}
Moreover, the projections of $Y_A^v(\psi)$ and $Y_B^v(\psi)$
onto the vertical hyperplane perpendicular to $\chi(\psi)$ are given by
\begin{equation}
	Z_A(\psi) = - Z_B(\psi) = -2i\eps p^1\sqrt{p^0}\ketbra{0}{0}+2i\eps p^0\sqrt{p^1}\ketbra{1}{1}.
\end{equation}
Then
\begin{equation}
\begin{split}
\GHS(Z_A,Z_B)^2&+2g(X_A,X_B)\,\GHS(Z_A,Z_B)=\\
& = \GHS(Z_A,Z_A)(\GHS(Z_A,Z_A)-2g(X_A,X_A))\\
& = 32\eps^2p^0p^1(2\eps^2p^0p^1-1)
\end{split}
\end{equation}
We conclude from Eq.~\eqref{nyckel} that 
if $\eps>\sqrt{1/2p^0p^1}$, the Robertson-Schr\"{o}dinger uncertainty relation gives a stronger bound on the product of the variances of $A$ and $B$.
But if $\eps<\sqrt{1/2p^0p^1}$, the geometric uncertainty relation \eqref{gur}
gives a stronger bound. Interestingly, for large $\eps$ the state is close to being a mixture of common eigenstates of $A$ and $B$, while for small values of $\eps$
the observables $A$ and $B$ are almost parallel at $\rho$.

\subsection{The Kirillov-Kostant-Souriau metric}\label{sec: KKS}
The symplectic form $\omega$
is the Kirillov-Kostant-Souriau symplectic form, see, e.g., \cite[Ch.\,22]{CdS2001}.
Let $g$ be the metric obtained by projecting down $\GHS$.
We can define an almost complex structure $J$ by the requirement that
\begin{equation}\label{compatible}
	g(\dot\rho_1,\dot\rho_2)=\omega(\dot\rho_1,J\dot\rho_2).
\end{equation}
Recall that an almost complex structure on $\SS_{\bfp}$
is a bundle map $J$ of the tangent bundle of $\SS_{\bfp}$ such that $J^2=-\1$.
The identity \eqref{compatible} says that $J$ is compatible with the pair $(\omega,g)$. Given \emph{any} almost complex structure $J$ on $\SS_{\bfp}$
we can always define a metric $g_J$ so that $J$ is compatible with $(\omega,g_J)$,
by simply \emph{defining} $g_J$ by the compatibility requirement \eqref{compatible}.
A theorem of Newlander and Nirenberg \cite[Ch.\,14]{CdS2001}
states that $\SS_{\bfp}$ admits a holomorphic atlas such that $h=g_J+i\omega$
is a Hermitian metric 
if, and only if, the Nijenhuis tensor vanishes identically on $\SS_{\bfp}$:
\begin{equation}\label{Nijenhuis}
	\mathcal{N}_J(\dot\rho_1,\dot\rho_2)=[J\dot\rho_1,J\dot\rho_2]-J[\dot\rho_1,J\dot\rho_2]-J[J\dot\rho_1,\dot\rho_2]-[\dot\rho_1,\dot\rho_2] = 0.
\end{equation} 

There is an almost complex structure $J$ on $\SS_{\bfp}$ 
for which the Nijenhuis tensor vanishes identically.
It can be defined as follows. 
According to \eqref{unitary tangent}, a tangent vector $\dot\rho$ at $\rho$
has a representation $\dot\rho=-i[A,\rho]$ for some Hermitian $A$.
We can take $A$ to be parallel at $\rho$ which means that $\Lambda_aA\Lambda_a=0$
for all the eigenprojectors $\Lambda_a$ of $\rho$.
This also means that $A$ can be decomposed as $A=A^u+A^l$ where $A^u$ has an upper triangular matrix representation, and $A^l$ has a lower triangular matrix representation, relative to every eigenbasis of $\rho$. 
We define $J$ by
\begin{equation}\label{ac}
	J\dot\rho=-i[iA^u-iA^l,\rho].
\end{equation}
Notice that $iA^u-iA^l$ is Hermitian since $A^l$ is the adjoint of $A^u$.
It is straightforward to check that $J^2=-\1$.
The metric $g_J$ is the Kirillov-Kostant-Souriau metric \cite{Au1994} on $\SS_{\bfp}$. It turns out that this metric does not agree with $g$, i.e., the projection of $\GHS$. To see this we evaluate $g$ and $g_J$ on $\dot\rho=-i[A,\rho]$,
where $A$ is parallel at $\rho$. Thus let $\psi$ be an amplitude for $\rho$. Then
\begin{equation}
	g(\dot\rho,\dot\rho)
	=\GHS(-iA\psi,-iA\psi)=2\Tr(\rho A^2)=2\Tr(\rho\{A^l,A^u\}).
\end{equation}
For the Kirillov-Kostant-Souriau metric we have that
\begin{equation}
	g_J(\dot\rho,\dot\rho)
	=\OHS(-iA\psi,-i(iA^u-iA^l)\psi)=2\Tr(\rho[A^l,A^u]).
\end{equation}
The length of vectors is shorter in the Kirillov-Kostant-Souriau geometry than in the geometry of $g$:
\begin{equation}
	g(\dot\rho,\dot\rho)-g_J(\dot\rho,\dot\rho)=4\Tr(\rho A^uA^l)\geq 0.
\end{equation}
One can also derive an uncertainty relation involving the Kirillov-Kostant-Souriau metric \cite{He2014}:
\begin{equation}
	4\var_\rho A\var_\rho B\geq g_J(X_A(\rho),X_B(\rho))^2+\omega(X_A(\rho),X_B(\rho))^2.
\end{equation}
However, this relation is weaker than the uncertainty relation \eqref{gur}.

It is known that the Nijenhuis tensor \eqref{Nijenhuis}
vanishes for the almost complex structure defined by Eq.~\eqref{ac}.
But it is not known if this is also true for the almost complex structure 
$J$ compatible with $(\omega,g)$. Let us finish this section with a more explicit definition of $J$. The tensors $\omega^\flat$ and $g^\flat$, assigning 
the one-forms $\omega^\flat(X)$ and $g^\flat(X)$ defined by
\begin{align}
	\omega^\flat(X)\dot\rho&=\omega(X,\dot\rho),\\
	g^\flat(X)\dot\rho&=g(X,\dot\rho),
\end{align}
to every vector field $X$ on $\SS_{\bfp}$, are isomorphisms from the tangent bundle 
of $\SS_{\bfp}$ onto the cotangent bundle of $\SS_{\bfp}$.
It follows that there is a bundle map $Q$ of the tangent bundle 
of $\SS_{\bfp}$ such that $\omega^\flat=g^\flat\circ Q$.
Recall from Section \ref{hermitian rep} that each tangent space of $\SS_{\bfp}$ can be identified with a subspace of the special Hermitian operators.
The restriction of $Q$ to each tangent space  is a
skew-symmetric linear map, and the almost complex structure is 
\begin{equation}
	J=\left(\sqrt{QQ^T}\,\right)^{-1}Q.
\end{equation}

\section{Two quantum speed limits}\label{Oqsl}
In Chapter \ref{ch: introduction} we gave examples of two quantum speed limits. One was the Mandelstam-Tamm quantum speed limit which bounds the evolution-time in terms of the energy uncertainty, the other was the Margolus-Levitin quantum speed limit which bounds the evolution-time in terms of the energy.
In this section we extend these quantum speed limits to mixed quantum states.

\subsection{Parallel and perpendicular Hamiltonians}

Consider the bundle $\Pi_{\bfp;\bfl}$ from $\WW_{\bfp;\bfl}$ onto $\SS_{\bfp}$
equipped with the mechanical connection $\A$ of the restriction of the Hilbert-Schmidt metric to $\WW_{\bfp;\bfl}$. 
(Here we use the `standard' definition \eqref{ghs} for the Hilbert-Schmidt metric.)
Write $g$ for the projection of the Hilbert-Schmidt metric on 
$\SS_{\bfp}$ and write 
$\dist_g$ for the associated distance function.
Furthermore, let $q$ be the common image operator of all the amplitudes in $\WW_{\bfp;\bfl}$ under the complementary purification projection, $q=\sum_a P^a\lambda_a$.

Let $H$ be a Hamiltonian on $\HH$.
In Section \ref{Ham vec fields} we introduced the notation $X_H$ for the Hamiltonian vector field on $\SS_{\bfp}$ associated with the 
average energy function $E_{\rho}(H)=\Tr(\rho H)$. 
The vector field is defined by $X_H(\rho)=-i[H,\rho]$. 
We then called $H$ parallel at $\rho$ if its canonical lift $Y_H$, defined by $Y_H(\psi)=-iH\psi$,
is horizontal along the fiber of $\rho$.
Notice that this condintion needs to be checked only at one amplitude of $\rho$. This since the composition $\A\circ Y_H$ is constant along the fibers of $\Pi_{\bfp;\bfl}$.

The inertia tensor can be used to measure deviation from parallelism.
When restricted to $\WW_{\bfp;\bfl}$, the inertia tensor is of ``constant bi-invariant type''. That is, 
$\II_{\psi}$ is an adjoint-invariant form on $\uu_{\bfl}$ which is independent 
of $\psi$. Accordingly, all the $\II_{\psi}$:s define the same metric on $\uu_{\bfl}$, namely 
\begin{equation}\label{beta}
(\xi|\eta)=-\frac{1}{2}\Tr\big((\xi \eta+\eta \xi)q\big).
\end{equation}
We define a $\uu_{\bfl}$-valued field $\xi_H$ on $\SS_{\bfp}$ by $\xi_H(\rho)=\A(Y_{H}(\psi))$,
where $\psi$ is any amplitude of $\rho$. 
Then $(\xi_H|\xi_H)$ equals the squared length of the vertical part of $X_{H}$. 
We furthermore say that the Hamiltonian $H$ is 
perpendicular at $\rho$ if $Y_{H}$ is vertical along the 
fiber over $\rho$, or, equivalently, if $Y_H(\psi)=\psi\xi_H(\rho)$ for every amplitude $\psi$ of $\rho$. In this case $X_H(\rho)=0$, and $\rho$ represents a mixture of eigenstates of ${H}$.

\subsection{A Mandelstam-Tamm quantum speed limit}\label{evolution time}
Let $\zeta_H$ be the projection of $\xi_H$ on the 
orthogonal complement of the unit vector $-i\1$ in $\uu_{\bfl}$.
According to Eq.~\eqref{pregur}, the vector field of $H$ satisfies 
\begin{equation}\label{c}
g(X_H ,X_H)=\var H - (\zeta_H|\zeta_H).
\end{equation}
We thus have that $ g(X_H(\rho),X_H(\rho))=\var_\rho H$ if $H$ is parallel at $\rho$.
Now, suppose that the curve $\rho_t$ satisfies $i\dot \rho_t=[H,\rho_t]$.
By Eq.~\eqref{pregur}, the distance between the initial state $\rho_0$ and the final state $\rho_\tau$
is upper bounded by the final time $\tau$ times the time-average of the uncertainty of $H$ along $\rho_t$:
\begin{equation}\label{qslen}
	\dist_{g}(\rho_0,\rho_\tau)
	\leq \length{\rho_t}
	\leq\tau \Delta E.
\end{equation}
Notice that for pure states, the vertical bundle is one-dimensional. Then $\zeta_H=0$ for any Hamiltonian and $g(X_H,X_H)=\var H$. The inequality~\eqref{qslen} is, hence, an extension of \eqref{eq: jost}
from pure to mixed states.

\subsubsection{A comparison with the Uhlmann quantum speed limit}
The distance between two density operators in $\SS_{\bfp}$, as measured by $\dist_{g}$,
is not less than their Bures distance. This is obvious from the observations that
$\dist_{g}(\rho_1,\rho_2)$ is the Hilbert-Schmidt distance between the fibers of $\Pi_{\bfp;\bfl}$
over $\rho_1$ and $\rho_2$, that $\distB(\rho_1,\rho_2)$ is the Hilbert-Schmidt distance between the fibers of the full standard purification bundle $\Pi$ over $\rho_1$ and $\rho_2$, and that the fibers of  $\Pi_{\bfp;\bfl}$ are contained in the fibers of $\Pi$.
Accordingly, Eq.~\eqref{qslen} provides a stronger quantum speed limit than the Uhlmann quantum speed limit \eqref{Uhqsl}.

One might ask what is the relation between the distance function 
$\dist_{g}$ and the distance function associated with the \emph{restriction} of the
Bures metric to $\SS_{\bfp}$. (The latter is not equal to $\distB$ since the geodesic distance
is now defined as an infimum of lengths of curves constrained 
to lie \emph{in} $\SS_{\bfp}$,
and geodesics of $\gB$ are in general not isospectral.)
The inequality in \eqref{uppskattning} is also valid for the restricted Bures metric, here denoted 
$\gB|_{\SS_{\bfp}}$, and hence we have that
\begin{equation}\label{tjopp}
	\dist_{\gB|_{\SS_{\bfp}}}(\rho_0,\rho_\tau)
	\leq\tau \Delta E.
\end{equation}
On the other hand, 
\begin{equation}\label{infHHH}
	\dist_{g}(\rho_0,\rho_\tau)=\inf_H \int_{0}^{\tau} \dt\,\sqrt{\var_{\rho_t}(H)},
\end{equation}
where the infimum is taken over all Hamiltonians $ H$ for which the boundary value von Neumann equation 
\begin{equation}\label{HHH}
	i\dot\rho_t=[H,\rho_t],\qquad \rho_{t=0}=\rho_0,\quad \rho_{t=\tau}=\rho_\tau,
\end{equation}
is solvable. To see this, we first note that the set of Hamiltonians for which Eq.~\eqref{HHH}
is solvable is never empty. This is so because $\UU(\HH)$ acts transitively on $\SS_{\bfp}$.
Secondly, if $\rho_t$ satisfies \eqref{HHH} for \emph{some} Hamiltonian $H$,
then, by Eq.~\eqref{pregur}, the length of $\rho_t$ is a lower bound for the   
integral of the energy uncertainty along $\rho_t$: 
\begin{equation}
	\length{\rho_t}\leq\int_{0}^{\tau}\dt\sqrt{\var_{\rho_t}(H)}.
\label{enekvation}
\end{equation}
Finally, for any curve $\rho_t$ there is a Hamiltonian $H$ that 
generates a horizontal lift of $\rho_t$ because the unitary group of $\HH$ acts transitively on $\WW_{\bfp;\bfl}$.
For such a Hamiltonian we have equality in \eqref{enekvation}. Then, if $\rho_t$ is a shortest geodesic,
\begin{equation}
	\dist_{g}(\rho_0,\rho_\tau)=\int_{0}^{\tau}\dt\sqrt{\var_{\rho_t}(H)}.
\label{finalen}
\end{equation}
This proves Eq.~\eqref{infHHH}, and that 
$\dist_{\gB|_{\SS_{\bfp}}}(\rho_0,\rho_\tau)\leq \dist_{g}(\rho_0,\rho_\tau)$ by Eq.~\eqref{tjopp}. 
We conclude that
\begin{equation}\label{estimat}
	\tau \Delta E
	\geq \dist_{g}(\rho_0,\rho_\tau)
	\geq \dist_{\gB|_{\SS_{\bfp}}}(\rho_0,\rho_\tau)
	\geq \distB(\rho_0,\rho_\tau).
\end{equation}

In Section \ref{comparison} we will construct examples of density operators between which the $g$-distance is strictly greater than the Bures distance. 
Thus, the quantum speed limit given by the leftmost inequality in \eqref{estimat} is sometimes greater than the quantum speed limit derived by Uhlmann. However, for fully distinguishable states they are the same.
For if $\rho_1$ and $\rho_2$ in $\SS_{\bfp}$ are fully distinguishable, then 
\begin{equation}\label{joru}
	\dist_{g}(\rho_1,\rho_2)=\distB(\rho_1,\rho_2)=\frac{\pi}{2}.
\end{equation}
This is a consequence of the fact that purification amplitudes of fully distinguishable 
mixed states have orthogonal supports, and, hence, are Hilbert-Schmidt orthogonal.
(An amplitude has the same support as the density operator it purifies,
and two fully distinguishable density operators have orthogonal supports \cite{En1996,MaMiPuZy2008}.)
That two amplitudes $\psi_1$ and $\psi_2$ have orthogonal 
supports can concisely be expressed as $\psi_1^\dagger\psi_2=0$.
The second equality in \eqref{joru} follows immediately from this observation. 
And since $\dist_g$ is bounded from below by $\distB$, it is sufficient to observe that the $\Pi_{\bfp;\bfl}$-fibers of $\rho_1$ and $\rho_2$ can be connected by a horizontal curve inside $\WW_{\bfp;\bfl}$ having length $\pi/2$. The curve $\psi_t=\cos(t)\psi_1+\sin(t)\psi_2$, 
where $\psi_1$ and $\psi_2$ are any amplitudes of $\rho_1$ and $\rho_2$ in $\WW_{\bfp;\bfl}$ and $0\leq t\leq \pi/2$, is such a curve.
For direct computations show that $\psi_t^\dagger\psi_t=q$ and 
$\A(\dot\psi_t)=0$. We conclude that also the first equality 
in \eqref{joru} holds.

\subsection{A Margolus-Levitin quantum speed limit}\label{margolus}
Equation \eqref{estimat} generalizes the Mandelstam-Tamm quantum speed limit \eqref{eq: MT speed limit} to mixed states.
In this section we will generalize the Margolus-Levitin quantum speed limit \eqref{eq: Margolus-Levitin quantum speed limit} to mixed states.
We thus consider a system whose dynamics is governed by a, possibly time-dependent, Hamiltonian $H_t$ which is such that $H_{t_1}$ and $H_{t_2}$ 
commute for any $0\leq t_1,t_2\leq\tau$.
The Hamiltonian then has an instantaneous spectral decomposition $H_t=\sum_aE^a_t\ketbra{E_a}{E_a}$, where the orthonormalized energy eigenstates $\ket{E_a}$ are time-independent.

For a solution $\rho_t$ of the von Neumann equation \eqref{HHH},
consider the lift 
\begin{equation}
	\psi_t=\sum_a e^{-it\bar{E}^a_t}\ketbra{E_a}{E_a}\psi_0,
	\qquad 
	\bar{E}^a_t\equiv\frac{1}{t}\int_0^t\dt\big(E^a_t-E^0_t\big),
\end{equation}
where $\psi_0$ is any amplitude for the initial density operator
and $E^0_t$ is the smallest energy eigenvalue of $H_t$.
Using that $\cos x\geq 2(x+\sin x)/\pi$ for $x\geq 0$,
\begin{equation}\label{ett}
\begin{split}
	\Re\Tr\big(\psi_0^\dagger \psi_\tau\big)
	&=\sum_{a} \Re\bra{E_a}\psi_\tau\psi_0^\dagger\ket{E_a}\\
	&=\sum_{a} \bra{E_a}\rho_0\ket{E_a}\cos\big(\tau\bar{E}^a_\tau\big)\\
&\geq \sum_{a} \bra{E_a}\rho_0\ket{E_a}\left(1-\frac{2}{\pi}\left(\tau\bar{E}^a_\tau+\sin\big(\tau\bar{E}^a_\tau\big)\right)\right)\\
&=1-\frac{2\tau}{\pi}\sum_{a} \bra{E_a}\rho_0\ket{E_a}\bar{E}^a_\tau+\frac{2}{\pi}\Im\Tr\big(\psi_0^\dagger \psi_\tau\big).
\end{split}
\end{equation}
Then, if the initial an final amplitudes are perpendicular, which is, e.g, the case when the initial and final density operators are fully distinguishable, we find that
\begin{equation}	
	\tau\geq\frac{\pi}{2\bar{E}},\qquad \bar{E}\equiv\sum_{a} \bra{E_a}\rho_0\ket{E_a}\bar{E}^a_\tau
\end{equation}
This, clearly, is a generalization of the Margolus-Levitin quantum speed limit. 

\begin{figure}[t]
	\centering
	\includegraphics[width=0.75\textwidth]{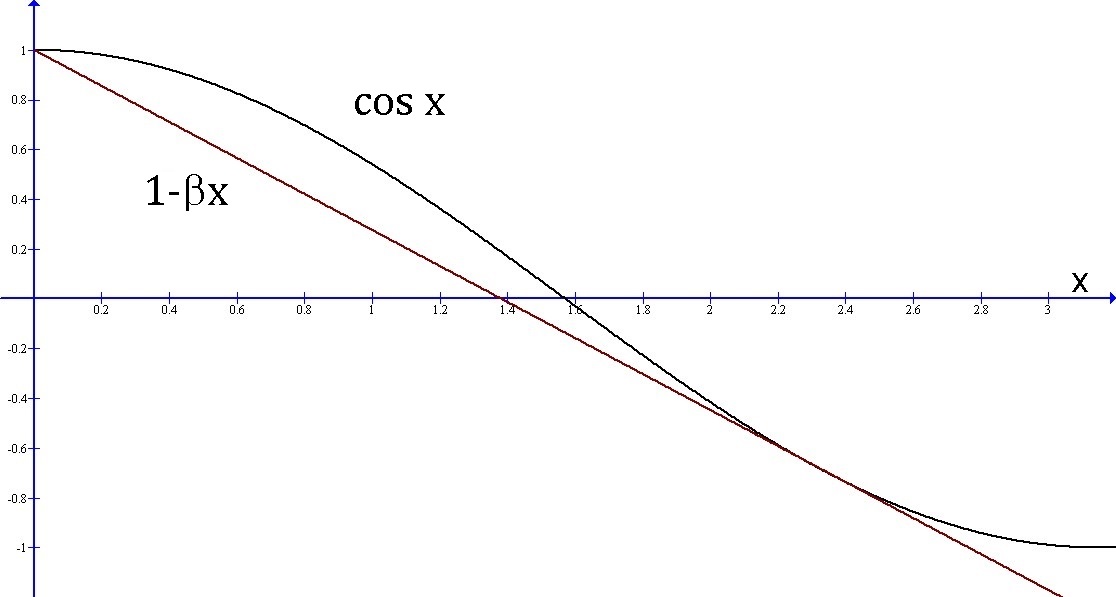}
	\caption{A tangent line to the cosine function.}
	\label{fig: graphs}
\end{figure}

Another quantum speed limit, which does not rely on the perpendicularity of the initial and final amplitudes, can be obtained from a minor modification of the calculation \eqref{ett}.
To this end let $\beta\approx 0.724$ be such that $1-\beta x$ is a tangent line to $\cos x$, see Figure \ref{fig: graphs}. 
Then, 
\begin{equation}
	\big|\Tr(\psi_0^\dagger\psi_\tau)\big|
	\geq \Re\Tr(\psi_0^\dagger\psi_\tau)
	\geq \sum_a \bra{E_a}\rho_0\ket{E_a}(1-\beta\tau\bar{E}^a_\tau)
	=1-\beta\tau\bar{E}.
\end{equation}
Since $1-x\geq 4\arccos^2(x)/\pi^2$ for $0\leq x\leq 1$,
\begin{equation}\label{DL-improved}
	\tau\geq \frac{1}{\beta\bar{E}}\big(1-|\Tr(\psi_0^\dagger\psi_\tau)|\big)
	\geq\frac{4}{\beta\pi^2\bar{E}}\arccos^2\big(|\Tr(\psi_0^\dagger\psi_\tau)|\big).
	\end{equation}
By the Uhlmann theorem \cite{Uh1976}, $|\Tr(\psi_0^\dagger\psi_\tau)|\leq F(\rho_0,\rho_\tau)$ and, hence, since arccos is decreasing,
$\arccos(|\Tr(\psi_0^\dagger\psi_\tau)|)\geq \distB(\rho_0,\rho_\tau)$.
We conclude that
\begin{equation}
	\tau\geq\frac{4\distB^2(\rho_0,\rho_\tau)}{\beta\pi^2\bar{E}}.	\end{equation}
(Despite the resemblance, this quantum speed limit is different from that derived in the appendix of \cite{DeLu2013}.)

\subsection{Geodesic generating Hamiltonians}\label{optimal hamiltonians}
In this last section, we return to the leftmost inequality in \eqref{estimat}
and we ask under what conditions on the Hamiltonian $H$ this inequality 
is tight. According to Eqs.~\eqref{c} and \eqref{qslen}, a sufficient condition is that the solutions to 
the Schr\"{o}dinger equation of $H$ that extend from the 
fiber over the initial density operator are horizontal geodesics.\footnote{Strictly speaking, the condition is sufficient only for small enough values on the final time. If the distance between the initial and final states is greater 
than the injectivity radius it is not certain that the geodesic
generated by $H$ is the shortest curve between the initial and 
final states.}
We thus ask for a characterization of those Hamiltonians that generate horizontal geodesics.

For simplicity, we assume that the initial state $\rho_0$ is faithful. Let $\bfp$ and $\bfl$ be the weight and eigenprojector spectra of $\rho_0$, and let $\psi_0$ in $\WW_{\bfp;\bfl}$
be an amplitude for $\rho_0$. Consider a curve  $\psi_t$ 
which extends from $\psi_0$. Since the left action of the unitary group of $\HH$ on $\WW_{\bfp;\bfl}$ is free and transitive, there is a unique curve of unitary operators $U_t$ such that $\psi_t=U_t\psi_0$.
We define the curve $\xi_t$ in $\uu(\HH)$ by $\dot U_t=U_t\xi_t$.
Then we define $H=iU_t\xi_t U_t^\dagger$ or, more explicitly,
\begin{equation}\label{Hfromxi}
	H=i\texp\Big(\!-\!\int_0^t\dt\,\xi_t\Big)^\dagger\xi_t \texp\Big(\!-\!\int_0^t\dt\,\xi_t\Big).
\end{equation}
The curve $\psi_t$ then satisfies the Schr\"{o}dinger equation $i\dot\psi_t=H\psi_t$,
 and our task now is to 
formulate conditions on $\xi_t$ 
which guarantees that $\psi_t$ is a horizontal geodesic. 

The tangent space at $\psi_0$ can be identified with $\uu(\HH)$.
An explicit identification is given by $\zeta\to\zeta\psi_0$.
We equip $\uu(\HH)$ with the metric that makes this isomorphism an isometry:
\begin{equation}
	\zeta\ast\eta=-\frac{1}{2}\Tr\big((\zeta\eta+\eta\zeta)\rho_0\big).
\end{equation}
Furthermore,
we write $X_\zeta$ for the left invariant vector field on $\WW_{\bfp;\bfl}$
which at $\psi_0$ equals $\zeta\psi_0$. At $\psi=U\psi_0$, then,
\begin{equation}\label{vanster}
	X_\zeta(\psi) = X_\zeta(\L_U(\psi_0)) = d\L_U(\zeta\psi_0) = U\zeta U^\dagger\psi.
\end{equation}
Notice that $\psi_t$ is an integral curve of $X_{\xi_t}$. Now, 
$\nabla_{X_\zeta} X_\zeta = - X_{\ad_\zeta^*\zeta}$, where $\nabla$ is the Levi-Civita connection of the Hilbert-Schmidt metric and $\ad_\zeta^*\zeta$ is the skew-Hermitian operator defined by $\ad_\zeta^*\zeta\ast\eta=\zeta\ast[\zeta,\eta]$. 
Indeed, by the Kozul formula \cite[Ch.\,IV, Prop.\,2.3]{KoNo1996},
\begin{equation}
\begin{split}
	2\GHS(\nabla_{X_\zeta}X_\zeta,X_\eta)
	& = X_{\zeta}\GHS(X_{\zeta},X_{\eta})
		+ X_{\zeta}\GHS(X_{\eta},X_{\zeta})\\
	&\hspace{13pt} - X_{\eta}\GHS(X_{\zeta},X_{\zeta})
		- \GHS(X_\zeta,[X_\zeta,X_\eta])\\
	&\hspace{13pt} + \GHS(X_\zeta,[X_\eta,X_\zeta])
		+ \GHS(X_\eta,[X_\zeta,X_\zeta])\\
	& = -\, 2\ad_\zeta^*\zeta\ast\eta\\
&=-\,2\,\GHS(X_{\ad_\zeta^*\zeta},X_\eta).
\end{split}
\end{equation}
The covariant derivative of the velocity field of $\psi_t$ is, thus,
\begin{equation}
	\nabla_{t}\dot \psi
	= X_{\dot\xi_t}(\psi_t)+\nabla_{X_{\xi_t}}X_{\xi_t}(\psi_t)
	= X_{\dot\xi_t-\ad_{\xi_t}^*\xi_t}(\psi_t).
\end{equation}
We conclude that $\psi_t$ is a geodesic
if, and only if, $\xi_t$ satisfies 
\begin{equation}\label{EA eq}
	\dot{\xi}_t=\ad_{\xi_t}^*\xi_t.
\end{equation}
Notice that this equation, which is called the Euler-Poincar\'{e} equation \cite[Ch.\,13, Eq.\,(13.5.17)]{MaRa1999}, is of first order (unlike the geodesic equation which is of second order).
From the discussion in Section \ref{le mech con} we conclude that $\psi_t$ is a horizontal geodesic if $\xi_t$, in addition to satisfying Eq.~\eqref{EA eq},
is such that the initial velocity $\dot\psi_0=\xi_0\psi_0$ is horizontal. That is, if $\A(\xi_0\psi_0)=0$.

\subsubsection{Density operators with two distinct eigenvalues}\label{comparison}
A geodesic orbit space is a Riemannian homogeneous space 
in which each geodesic is an orbit of a one-parameter subgroup of its isometry group. If the density operators in $\SS_{\bfp}$
are faithful and have precisely two different, possibly degenerate eigenvalues, then $\SS_{\bfp}$ is a geodesic orbit space.
This is so because then every geodesic in $\SS_{\bfp}$
is generated by a time-independent Hamiltonian.

To see this, let $P^1$ and $P^2$ be the common eigenvalues of the operators in $\SS_{\bfp}$ and let $m_1$ and $m_2$ be the corresponding 
degeneracies. 
Furthermore, let $\xi$ be a skew-Hermitian operator on $\HH$ such that $X_\xi$ is horizontal along the fiber over $\rho_0$,
and let $\zeta$ be any skew-Hermitian operator on $\HH$.
Express $\xi$ and $\zeta$ as matrices with respect to
an eigenbasis for $\rho_0$:
\begin{equation}
	\xi = \begin{bmatrix} 0 & \xi_{12}\\ -\xi_{12}^\dagger & 0 	\end{bmatrix},
	\qquad
	\zeta = \begin{bmatrix} \zeta_{11} & \zeta_{12}\\ -\zeta_{12}^\dagger & \zeta_{22}\end{bmatrix}.
\end{equation}
Here, $\xi_{12}$ and $\zeta_{12}$ have dimensions $m_1\times m_2$, and $\zeta_{11}$ and $\zeta_{22}$ have dimensions $m_1\times m_1$ and $m_2\times m_2$, respectively. Now $\ad_\xi^*\xi=0$ because
\begin{equation}
	\xi\ast [\xi,\zeta]
	=\frac{1}{2}\left(P^1\Tr[\xi_{12}\xi_{12}^\dagger,\zeta_{11}] + P^1\Tr[\xi_{12}^\dagger\xi_{12},\zeta_{22}]\right)=0.
\end{equation}
(As commutators of matrices have vanishing trace.)
This in turn implies that every curve of skew-Hermitian operators $\xi_t$ which satisfies Eq.~\eqref{EA eq} is stationary, and, hence, that the Hamiltonian defined by \eqref{Hfromxi} is time-independent.
Next we use this observation to produce curves of qubit density operators for 
which the quantum speed limit given by the first inequality in \eqref{estimat} is 
strictly stronger than the Uhlmann quantum speed limit.
\begin{figure}[t]
	\centering
	\includegraphics[width=0.5\textheight]{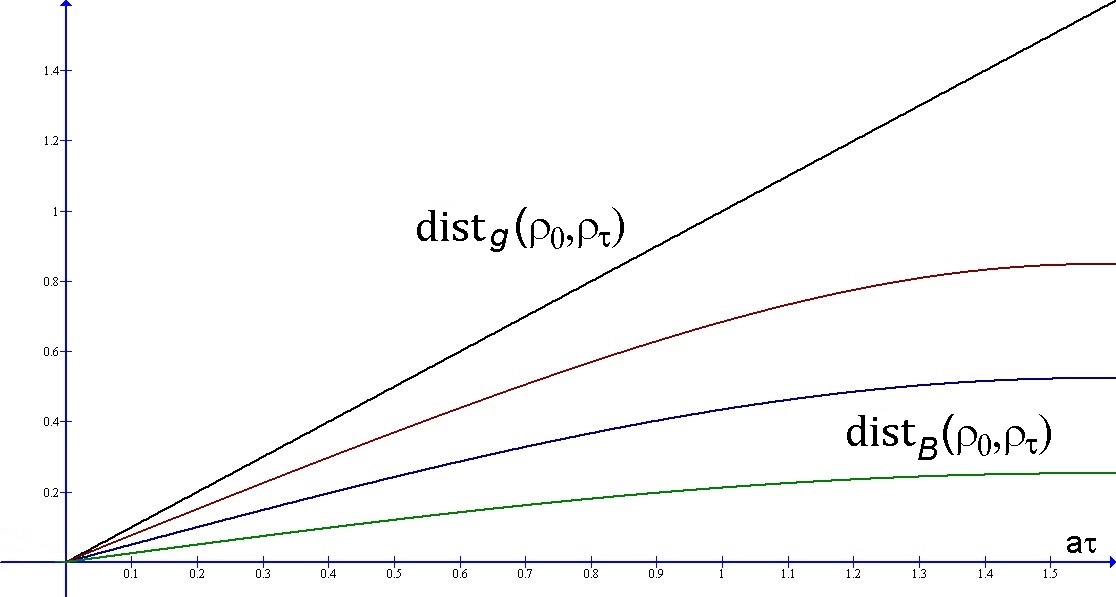}
	\caption{A comparison of the distance between $\rho_0$ and $\rho_\tau$ as measured by $\dist_g$ and $\distB$. We have plotted the distances for the cases $(p_0,p_1)=(7/8,1/8)$ (red), $(p_0,p_1)=(3/4,1/4)$ (blue), and $(p_0,p_1)=(5/8,3/8)$ (green).}
	\label{fig: Sj vs Uh}
\end{figure}
\subsubsection{Geodesics of mixed qubits}
For a qubit, $\xi_{12}= ae^{i\theta}$, where we can take $a>0$.
The solution to the Schr\"odinger equation of $H=i\xi$ which extends from $\sqrt{\rho_0}$ is
\begin{equation}
	\psi_t = \begin{bmatrix} 
				\sqrt{p^0}\cos\,(at) & \sqrt{p^1}e^{i\theta}\sin\,(at)\\ -\sqrt{p^0}e^{-i\theta}\sin\,(at) & \sqrt{p^1}\cos\,(at)\end{bmatrix}.
\end{equation}
This curve is a horizontal geodesic, and, hence, its projection is a geodesic extending from $\rho_0$: 
\begin{equation}
	\rho_t=\begin{bmatrix} p^0\cos^2{(at)}+p^1\sin^2{(at)} & e^{i\theta}(p^1-p^0)\cos {(at)}\sin {(at)}\\ e^{-i\theta}(p^1-p^0)\cos{(at)}\sin{(at)} & p^0\sin^2{(at)}+p^1\cos^2{(at)}\end{bmatrix}.
\end{equation}
Let $d>0$ be the injectivity radius of $\SS_{\bfp}$, and assume that $0<\tau<d/a$.
Then $\rho_t$ is a shortest geodesic between $\rho_0$ and $\rho_\tau$, and
\begin{equation}
	\dist_{g}(\rho_0,\rho_\tau)
	=\length{\rho_t}=a\tau.
\end{equation}
A direct calculation shows that the energy uncertainty 
along $\rho_t$ equals $a$. Therefore, the first inequality in 
\eqref{estimat} is saturated.

Next, we will show that the Bures distance between $\rho_0$ and $\rho_\tau$ is strictly less than $ a\tau$. From this follows that 
there is at least one case for which the first inequality in \eqref{estimat} is a stronger quantum speed limit than the Uhlmann quantum speed limit.

The fidelity of $\rho_0$ and $\rho_\tau$ is 
\begin{equation}\label{trogen}
	F(\rho_0,\rho_\tau)
	= \Tr(\rho_0\rho_\tau)+2\sqrt{\det\rho_0\det\rho_\tau}
	= (p^0-p^1)^2\cos^2 (a\tau)+4p^0p^1,
\end{equation}
and, hence, the Bures distance between $\rho_0$ and $\rho_\tau$ is 
\begin{equation}
	\distB(\rho_0,\rho_\tau)
	=\arccos\sqrt{(p^0-p^1)^2\cos^2(a\tau)+4p^0p^1}.
\end{equation}
Since the difference $a\tau-\arccos\sqrt{(p^0-p^1)^2\cos^2(a\tau)+4p^0p^1}$ is strictly positive for $\tau>0$, see Figure \ref{fig: Sj vs Uh}, we have that
$\dist_g(\rho_0,\rho_\tau)>\distB(\rho_0,\rho_\tau)$.
\vfill
\cleardoublepage
\pagestyle{plain}

\end{document}